\documentclass{article}

\usepackage{style/arxiv}

\usepackage[utf8]{inputenc} 
\usepackage[T1]{fontenc}    
\usepackage{hyperref}       
\usepackage{url}            
\usepackage{booktabs}       
\usepackage{amsfonts}       
\usepackage{nicefrac}       
\usepackage{microtype}      
\usepackage{graphicx}
\usepackage{float}
\usepackage{multirow}
\usepackage[round]{natbib}

\title{Filtering in tractography using autoencoders (FINTA)}

\author{
Jon Haitz Legarreta \\
Sherbrooke Connectivity Imaging Laboratory (SCIL) \\
Videos \& Images Theory and Analytics Laboratory (VITAL) \\
Department of Computer Science \\
Universit\'{e} de Sherbrooke, Canada \\
\And
Laurent Petit \\
Groupe d'Imagerie Neurofonctionnelle (GIN) \\
Univ. Bordeaux, CNRS, CEA, IMN, UMR 5293, France \\
\And
Fran\c{c}ois Rheault \\
Sherbrooke Connectivity Imaging Laboratory (SCIL) \\
Department of Computer Science \\
Universit\'{e} de Sherbrooke, Canada \\
\And
Guillaume Theaud \\
Sherbrooke Connectivity Imaging Laboratory (SCIL) \\
Department of Computer Science \\
Universit\'{e} de Sherbrooke, Canada \\
\And
Carl Lemaire \\
Centre de Calcul Scientifique \\
Universit\'{e} de Sherbrooke, Canada \\
\And
Maxime Descoteaux* \\
Sherbrooke Connectivity Imaging Laboratory (SCIL) \\
Department of Computer Science \\
Universit\'{e} de Sherbrooke, Canada \\
\And
Pierre-Marc Jodoin* \\
Videos \& Images Theory and Analytics Laboratory (VITAL) \\
Department of Computer Science \\
Universit\'{e} de Sherbrooke, Canada \\

\\
{*}{Co-senior author. These authors contributed equally.}
}

\begin{document}
\maketitle
\begin{abstract}
Current brain white matter fiber tracking techniques show a number of problems, including: generating large proportions of streamlines that do not accurately describe the underlying anatomy; extracting streamlines that are not supported by the underlying diffusion signal; and under-representing some fiber populations, among others. In this paper, we describe a novel autoencoder-based learning method to filter streamlines from diffusion MRI tractography, and hence, to obtain more reliable tractograms. Our method, dubbed FINTA ({\em Filtering in Tractography using Autoencoders}) uses raw, unlabeled tractograms to train the autoencoder, and to learn a robust representation of brain streamlines. Such an embedding is then used to filter undesired streamline samples using a nearest neighbor algorithm. Our experiments on both synthetic and in vivo human brain diffusion MRI tractography data obtain accuracy scores exceeding the $90\%$ threshold on the test set. Results reveal that FINTA has a superior filtering performance compared to conventional, anatomy-based methods, and the RecoBundles state-of-the-art method. Additionally, we demonstrate that FINTA can be applied to partial tractograms without requiring changes to the framework. We also show that the proposed method generalizes well across different tracking methods and datasets, and shortens significantly the computation time for large ($>1$ M streamlines) tractograms. Together, this work brings forward a new deep learning framework in tractography based on autoencoders, which offers a flexible and powerful method for white matter filtering and bundling that could enhance tractometry and connectivity analyses.
\end{abstract}


\section{Introduction}
\label{sec:introduction}
\begin{figure*}[!th]
\centering
\begin{tabular}{cccccc}
\includegraphics[scale=0.95, trim=0.9in 0.2in 0.9in 0.2in, clip=true, width=0.14\linewidth, keepaspectratio=true]{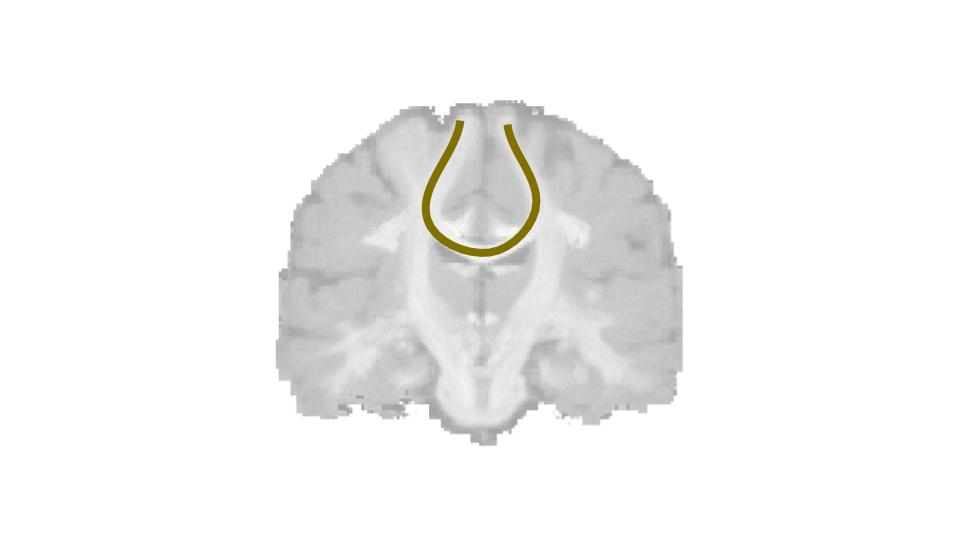} &
\includegraphics[scale=0.95, trim=0.9in 0.2in 0.9in 0.2in, clip=true, width=0.14\linewidth, keepaspectratio=true]{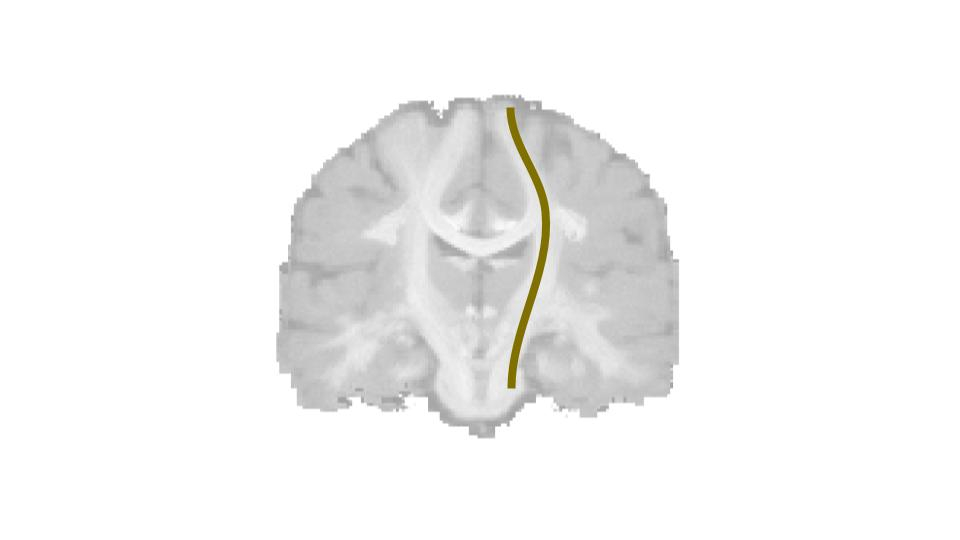} &
\includegraphics[scale=0.95, trim=0.9in 0.2in 0.9in 0.2in, clip=true, width=0.14\linewidth, keepaspectratio=true]{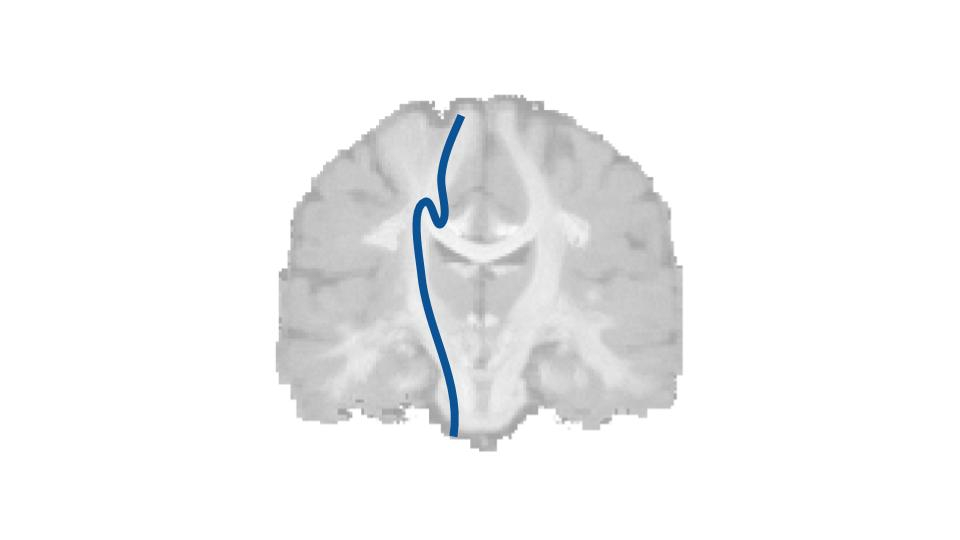} &
\includegraphics[scale=0.95, trim=0.9in 0.1in 0.9in 0.1in, clip=true, width=0.14\linewidth, keepaspectratio=true]{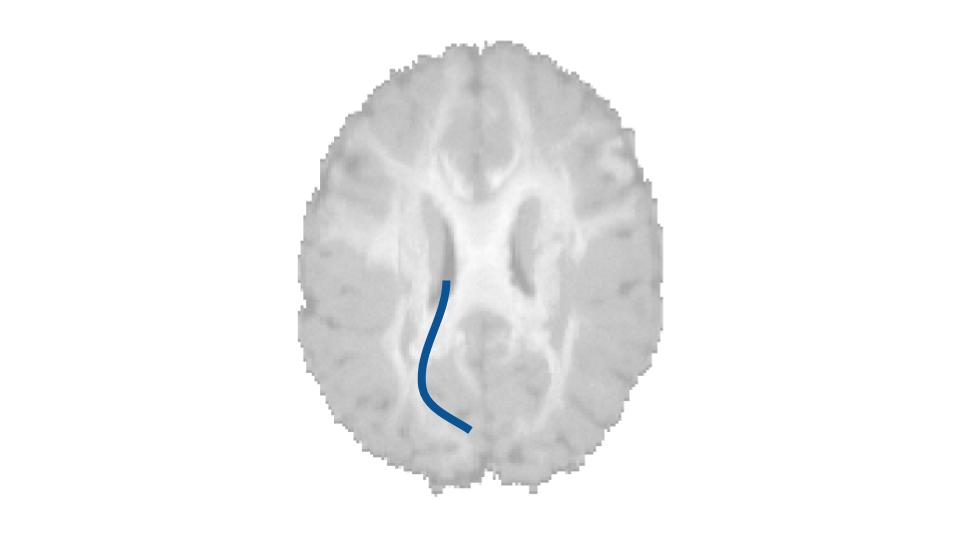} &
\includegraphics[scale=0.95, trim=0.9in 0.2in 0.9in 0.2in, clip=true, width=0.15\linewidth, keepaspectratio=true]{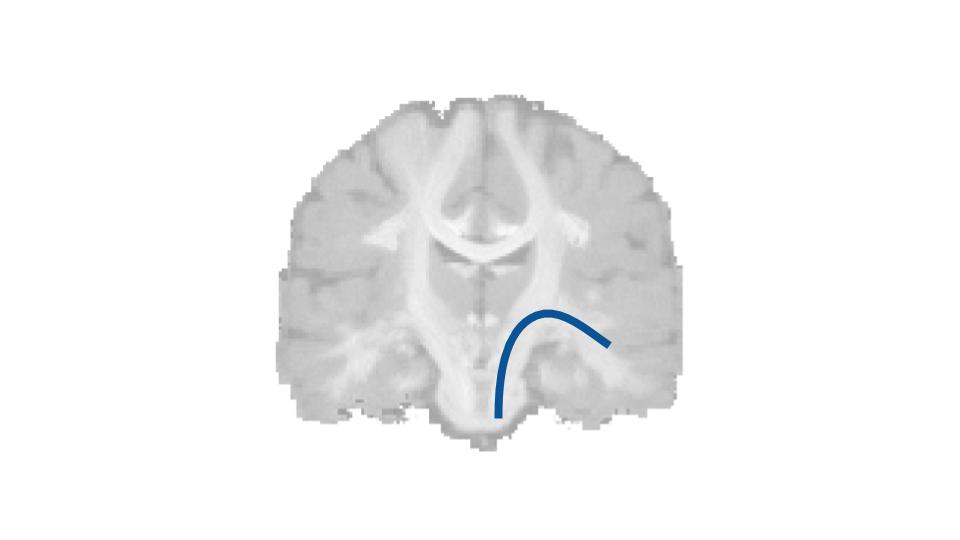} &
\includegraphics[scale=0.95, trim=0.9in 0.2in 0.9in 0.2in, clip=true, width=0.14\linewidth, keepaspectratio=true]{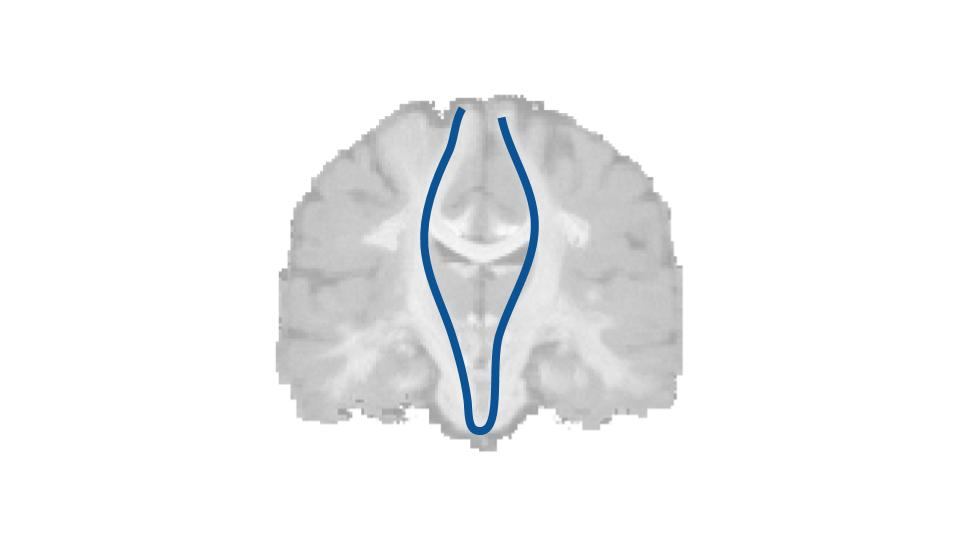}\\
\textbf{(a)} & \textbf{(b)} & \textbf{(c)} & \textbf{(d)} & \textbf{(e)} & \textbf{(f)} \\
\end{tabular}
\caption{\label{fig:plausible_vs_implausible}Schematic representation of plausible and implausible streamlines: (a) plausible streamlines belonging to the corpus callosum; (b) plausible streamlines belonging to the corticospinal tract; (c) implausible streamline containing a loop or a sharp bend; (d) implausible streamline stopped at the cerebrospinal fluid; (e) implausible streamline stopped in the white matter (or hitting a white matter or atlas boundary); (f): implausible streamline describing an invalid bundle.}
\end{figure*}

Tractography uses local orientation information from diffusion Magnetic Resonance Imaging (dMRI) data to delineate the brain white matter fiber pathways, and hence, to provide information about its structural connectivity. These connectivity pathways in tractograms are composed of streamlines virtually representing fascicles of white matter fibers. Tractography methods face a number of challenges when propagating streamlines, including avoiding the early termination of the tracking procedure \citep{Smith:Neuroimage:2012, Girard:Neuroimage:2014}, providing streamlines between gray matter regions that are known to be connected while avoiding spurious streamlines, ensuring the full occupancy of the white matter volume by the streamlines \citep{Rheault:Neuroimage:2019}, and a complete gray matter surface coverage when streamlines reach the cortex \citep{StOnge:Neuroimage:2018}. As a result, and despite the efforts to address these issues, white matter tracking procedures are known to produce a disproportionately large amount of invalid streamlines (referred to as ``implausibles'' in this work, as opposed to ``plausible'' streamlines). The category of invalid streamlines spans a broad group of streamlines that are known not to satisfy accepted anatomical constraints. These include streamlines that contain loops or sharp bends; streamlines that stop in non-gray matter tissues, such as the cerebrospinal fluid (CSF); streamlines that prematurely stop in the white matter; or streamlines describing trajectories between gray matter regions that are not connected structurally. An illustration of plausible and implausible streamlines can be seen in figure \ref{fig:plausible_vs_implausible}. According to the ISMRM 2015 Tractography Challenge benchmark \citep{Maier-Hein:NatureComm:2017}, tracking methods produce a non-negligible proportion of invalid streamlines as a trade-off between sensitivity and specificity. The study reported that non-existing, false positive connections were present in as many as $64\%$ of the submissions. Hence, improved filtering strategies are required to detect and remove anatomically unfeasible streamlines and mitigate some of the limitations derived from current streamline propagation methods.

Filtering implausible streamlines generated by tracking methods is a necessary step to avoid biologically unrealistic connectivity patterns and biases in reconstructing the brain structural connectome from diffusion MRI data \citep{Sotiropoulos:NMRBiomed:2019, Yeh:Neuroimage:2019} (see section \ref{subsubsec:plausible_implausible_filtering}). The quality of the tractograms influence various aspects of the connectome analysis, including the connectivity density, the clustering degree or the existence of connections themselves, among others \citep{Yeh:Neuroimage:2016}. Hence, tractogram filtering, in its various forms, is essential towards providing a reliable imaging evidence using diffusion MRI of how the white matter fascicles are arranged in the brain.

Filtering in tractography can also be regarded as a procedure that separates streamlines that are useful to serve a certain task from those that are not. As such, some filtering methods involve a clustering operation which may be used to extract a predefined set of bundles (see section \ref{subsubsec:clustering}). While connectomics applications~\citep{Sotiropoulos:NMRBiomed:2019} require every anatomically plausible fiber to be recovered, other applications often need a more specialized set of fibers. For example, in the human \textit{versus} macaque study by~\cite{Takemura:CerebralCortex:2017}, only streamlines from the occipital white matter are considered. In~\citet{Chenot:BrainStructFunc:2019} only fibers from the pyramidal tract are retained, while in ~\citet{Bullock:BrainStructFunc:2019} the brain fibers connecting the dorsal and ventral posterior cortex are the ones of interest.

Streamline filtering can be seen as an application of choice for deep learning classification methods. Unfortunately, neural networks are not easily applicable to brain tractography, mostly because of the fundamental difficulty of building a labeled dataset. In fact, finding irreproachable ground truth streamlines is a very difficult endeavour, with significant differences in reproducibility measures even among well-trained expert neuroanatomists, as shown in \citet{Rheault:HBM:2020}. Also, depending on the task at hand, the labeling of the streamlines might vary considerably, and hence may require time-consuming manual verification and/or relabeling. Classification neural networks are trained to predict a fixed number of classes, and thus cannot be used to predict a different set of classes without being retrained on a newly labeled set of data.

This work proposes a deep autoencoder for streamline-based tractography filtering. The method, named FINTA ({\em Filtering in Tractography using Autoencoders}), is trained on non-curated, raw tractograms. Once training is over, the learned latent space is a low-dimensional representation of the input streamlines where similar streamlines are located next to each other. The filtering process is then carried out by projecting to the latent space examples of reference streamlines. Then, the to-be-filtered streamlines are projected to the latent space and labeled according to a nearest neighbor test. We demonstrate the filtering ability of FINTA on both synthetic and human brain datasets. Additional quantitative evaluations show its superior performance compared to both anatomy-based tractography filtering methods and the state-of-the-art RecoBundles \citep{Garyfallidis:Neuroimage:2018, Rheault:PhD:2020} streamline similarity filtering method. Similarly, FINTA is linear in terms of the streamline count at test time, allowing for a faster tractogram filtering.

\subsection{Related work}
\label{subsec:related_work}
\subsubsection{Filtering plausible \textit{versus} implausible streamlines}
\label{subsubsec:plausible_implausible_filtering}
Diffusion MRI tractography methods generate tractograms that may contain several million candidate streamlines representing white matter  fiber populations. Yet, a large proportion of such streamlines are anatomically implausible, meaning that they do not comply with neuroanatomical constraints attributed to fiber populations. A regular tractogram including only anatomically plausible streamlines can contain in the order of $500\,000$ to $3\,000\,000$ streamlines \citep{Presseau:Neuroimage:2015}. Hence, having an automatic procedure to filter out implausible streamlines is critical. Tractography filtering is currently approached based on five main patterns: (i) streamline geometry features; (ii) region-of-interest-driven streamline inclusion and exclusion; (iii) clustering; (iv) connectivity-based; and (v) diffusion signal mapping or forward models \citep{Jeurissen:NMRBiomed:2019, Sotiropoulos:NMRBiomed:2019, Jorgens:Springer:2020}. The first category of criteria is based on anatomical features of individual streamlines, such as unfeasible streamline length or local curvature indices. Streamline inclusion and exclusion criteria, adopted, for example, in \citet{Catani:Neuroimage:2002, Wakana:Neuroimage:2007, Yendiki:FrontiersNeuroinf:2011, Wassermann:BrainStructFunc:2016, Warrington:Neuroimage:2020}, incorporate white matter and/or tissue local and connectivity constraints for streamline traversal \citep{Schilling:BrainStructFunc:2020}. Such principles are usually concatenated into a sequence of restrictions to form a succession of step-wise refinement rules \citep{Maier-Hein:NatureComm:2017, Sarubbo:BrainStructFunc:2019, Jeurissen:NMRBiomed:2019, Jorgens:Springer:2020}. Clustering approaches, such as QuickBundles \citep{Garyfallidis:FrontiersNeurosc:2012}, use a similarity measure to remove non-meaningful data issued by a tracking method. These methods are based on the assumption that a properly defined distance measure is able to split apart the relevant streamlines from the rest. These methods usually require a representative set of streamlines of the population that is being studied to be built so that they can be used as reference models to exclude the extraneous streamlines. Connectivity-based approaches, such as the one presented by \citet{Wang:Neuroimage:2018}, propose to remove undesired streamlines usually imposing some regularization constraint on the tractogram-derived connectivity matrices. The forward models, such as SIFT/SIFT2 \citep{Smith:Neuroimage:2013, Smith:Neuroimage:2015}, LiFE \citep{Pestilli:NatureMethods:2014} and COMMIT/COMMIT2 \citep{Daducci:TMI:2015, Schiavi:ScienceAdv:2020}, posit that only a subset of streamlines in a whole-brain tractogram are essential or relevant to explain the underlying diffusion signal. SIFT/SIFT2 is based on modifying the local fiber Orientation Distribution Function (fODF) based on the diffusion signal, whereas LiFE and COMMIT/COMMIT2 use specific local models to weigh the contribution of the signal to the streamline representation. Anatomical \textit{a priori} constraints can be incorporated to improve the accuracy of the reconstructed pathways. Some of these methods have large computational demands in terms of the time required to filter the tractogram of a single subject (in the order of hours) \citep{ODonnell:2019:NMRBiomed}.

A number of works use deep learning-based methods for tractography filtering. \citet{Lucena:MSc:2018} used an autoencoder to learn features about the input tractogram, and then used a residual network to regress on those features. However, their results were only limited to the uncinate fascicle and no comparison to other filtering methods was provided. More recently \citet{Astolfi:arxiv:2020} proposed a geometrical deep learning method to filter tractograms. The method implements a supervised geometric model which builds on streamline graph structures.

\subsubsection{Streamline clustering}
\label{subsubsec:clustering}
White matter streamline clustering (also referred to as \textit{bundling}) allows to assign every streamline in a tractogram into anatomically coherent bundles \citep{Maddah:MIA:2008, Li:Neuroimage:2010, Wassermann:Neuroimage:2010, Guevara:Neuroimage:2011, Garyfallidis:FrontiersNeurosc:2012, Jin:Neuroimage:2014, Wassermann:BrainStructFunc:2016, Siless:Neuroimage:2018, Sharmin:FrontiersNeurosc:2018, Garyfallidis:Neuroimage:2018}. It is a procedure where the streamlines of interest need to be identified from the rest of the streamlines (the rejection class). The task of attributing a streamline to a bundle needs to address the fundamental question of how streamlines are similar to, or different from, each other. One of the challenges lies in finding the appropriate measure to quantify the affinity of streamlines, the other major difficulty being computing such affinities or distances in a reasonable time.

Probabilistic model-based methods (e.g. \citet{Dayan:ICIP:2018}) or traditional machine learning-based frameworks (e.g. \citet{Brun:MICCAI:2004, Wang:MICCAI-CNI:2018, Kumar:PatternRecog:2019, Berto:biorxiv:2020}) have also been proposed to solve the clustering task. However, some of the findings in these works are limited to clustering a single bundle, require separate learning models for each bundle, are still affected by the streamline-wise distance computational complexity, or suffer from the trade-off of the granularity of the arrangement.

Some works \citep{Brun:MICCAI:2004, ODonnell:AJNR:2006, ODonnell:TMI:2007, Sydnor:Neuroimage:2018} have investigated the feasibility of feature spaces to perform the streamline clustering, and have highlighted the ability of such embeddings to overcome some of the limitations of region-of-interest-based methods.

A number of different strategies have been developed recently to address the streamline clustering task using deep learning networks, including stacked, bi-directional siamese networks (FS2Net) \citep{Patil:arxiv:2017}, regular convolutional neural networks using different streamline re-parameterizations (FiberNet/FiberNet2.0) \citep{Gupta:MICCAI:2017, Gupta:ISBI:2018}, or custom feature vectors (TRAFIC) \citep{Lam:SPIE:2018}, (DeepWMA) \citep{Zhang:MIA:2020}, ensemble models using fully connected neural networks \citep{Ugurlu:MICCAI-CDMRI:2018}, or representation learning methods \citep{Zhong:ISMRM:2020} based on recurrent autoencoders. Yet, many of these methods still work in a supervised fashion, and thus do not allow to generalize to a variable number of classes, require a distinctly trained network for each bundle, or present results limited to a reduced number of bundles.

As opposed to the previous deep learning approaches, a few other ones (\citet{Wasserthal:Neuroimage:2018, Pomiecko:ISBI:2019, Li:Neuroimage:2020}) provide voxel-wise segmentations representing the locations traversed by a given bundle's streamlines. Most of these approaches are based on a U-Net \citep{Ronneberger:MICCAI:2015} or another form of classification convolutional neural network using a specialized feature vector or parameterization. The major drawback of voxel-based bundle segmentation methods is the need of a separate model (including the need to train it) for each bundle, and hence, a variable number of models depending on the target classes.

In this work, we describe a novel unsupervised learning framework based on deep autoencoders to filter streamlines in diffusion MRI tractography. The proposed autoencoder is trained in an unsupervised fashion, and an optimal filtering threshold is computed in the learned representation space using a set of labeled streamlines. We demonstrate its performance on both synthetic and in vivo human brain tractography data, including partial tractograms. We provide evidence that the described deep learning setting can be applied to filter plausible and implausible streamlines as well as to multi-label classification problems without requiring to be retrained.

\section{Material and methods}
\label{sec:materials_methods}
\begin{figure*}[!htp]
\centering
\begin{tabular}{cc}
\textbf{(a)} &
\raisebox{-0.5\height}{\includegraphics[scale=0.95, trim=0.45in 0.15in 0.45in 0.2in, clip=true, width=0.7\linewidth, keepaspectratio=true]{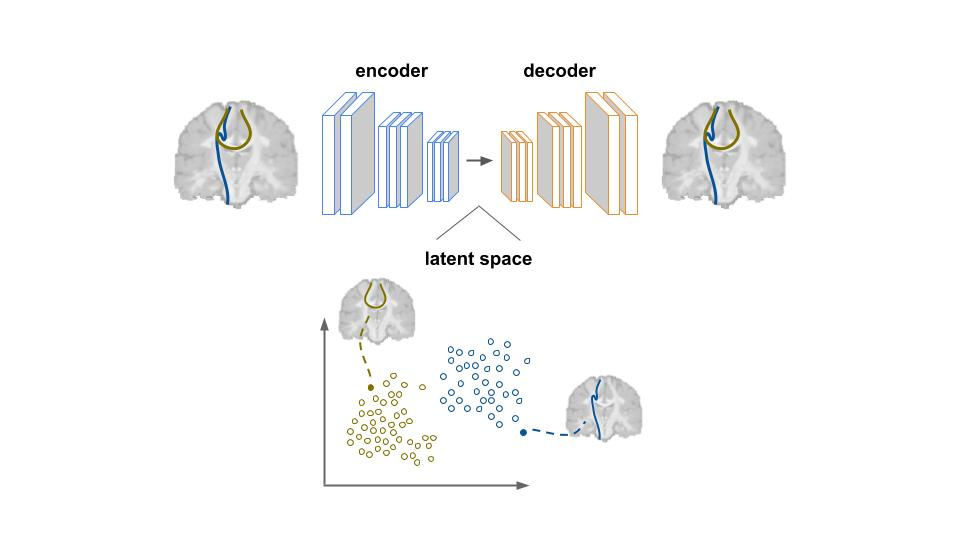}} \\
\textbf{(b)} &
\raisebox{-0.5\height}{\includegraphics[scale=0.95, trim=0.35in 0.3in 0.35in 0.25in, clip=true, width=0.7\linewidth, keepaspectratio=true]{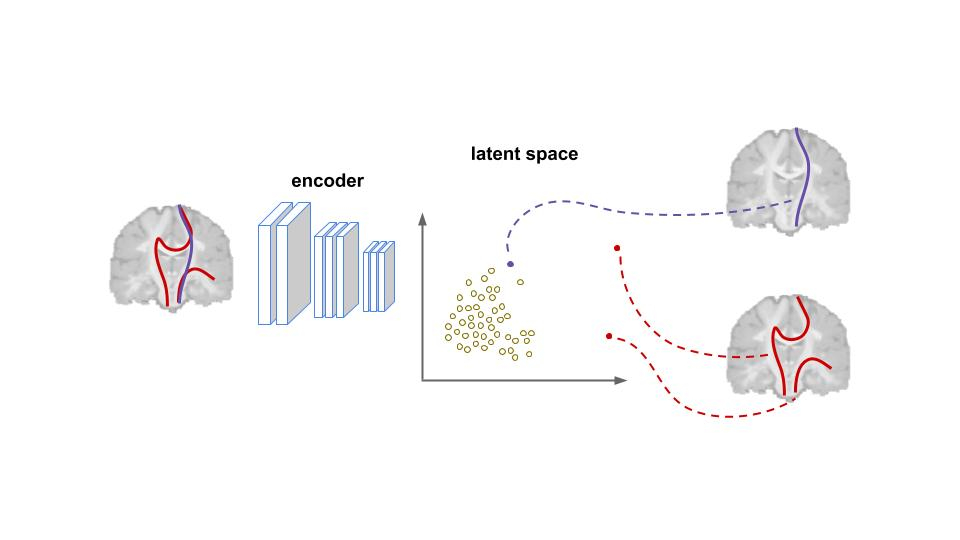}} \\
\end{tabular}
\caption{\label{fig:finta}Schematic representation of tractography filtering with FINTA: (a) training time; (b) test time. Although 1D convolutions are used in the proposed framework, convolutional blocks are depicted using a 2D shape for illustrative purposes.}
\end{figure*}

Autoencoders (AEs) are deep neural networks that have the ability to learn an efficient representation of the data in an unsupervised fashion using a dimensionality reduction approach~\citep{Hinton:Science:2006}. AEs are trained to reconstruct an input signal through a series of compressing operations (known as the {\em encoder}) followed by a series of decompression operations (referred to as the {\em decoder}). As shown in figure~\ref{fig:finta}, between the encoder and the decoder there is the so-called {\em latent space}, where each point is an encoded representation of an input data sample. In this work, the input data is a \textit{streamline}, which is a sequence of three-dimensional points. A streamline is an ordered sequence of points $s = \{c_{1}, c_{2}, \dots, c_{n_{s}}\}, \; c_{i} \in \mathbb{R}^{3}$ that represents a package of similarly oriented fibers describing a neural pathway within the brain white matter. A tractogram representing a set of $M$ streamlines is mathematically expressed as $T = \{s_{1}, s_{2}, \dots, s_{M}\}$.

AEs have two major advantages over other types of neural networks. First, they are trained on raw unlabeled data, which is a major asset in the context of brain tractography. Second, since the latent representation is obtained by a series of compressing operations, two neighboring points in the latent space correspond to input data instances that share similar characteristics~\citep{Painchaud:MICCAI:2019}. Thus, the encoded representation can be used to redefine the notion of inter-data distance. In the context of tractography, instead of measuring the distance between two streamlines with a geometric distance function as is usually the case~\citep{Siless:PRNI:2013}, one can project the streamlines into the latent space and measure their Euclidean distance.

Our filtering approach, shown in figure \ref{fig:finta}, can be summarized by the following steps:
\begin{enumerate}
\item Train an autoencoder with a large, raw, uncurated tractogram.
\item From the training tractogram, label the streamlines of interest. These streamlines could be examples of anatomically plausible fibers, streamlines representing fibers belonging to predefined sets of bundles, or any sorts of streamlines that are of interest for some application. These streamlines are labeled as {\em positive streamlines}.
\item Project the {\em positive} streamlines into the latent space with the encoder network. The positive latent vectors are illustrated by the golden circles in figure~\ref{fig:finta}. Note that in case of more than two classes (as for a multi-label bundling operation) streamlines of several classes could also be projected into the latent space.
\item Given a new tractogram that ought to be filtered, project its streamlines into the latent space (cf. the three bold circles in figure~\ref{fig:finta}(b)) and label them according to the distance to their nearest neighbor.
\end{enumerate}

Hence, FINTA's filtering process actually takes place in the latent space. The nearest neighbor discrimination employs a latent space distance cut-off value, herein named \textit{filtering threshold}, relative to a (labeled) reference set of streamlines to discriminate the uncurated tractogram's streamlines. The filtering threshold represents, thus, the minimum distance at which a streamline is considered to be implausible, and its value is computed on the separate reference set of streamlines. Section \ref{sec:experiment_design} provides details about the criterion adopted to compute the optimal filtering threshold.

The proposed AE is a fully convolutional neural network whose overall structure is summarized in table~\ref{tab:ae_structure}. Our convolutional AE accepts a streamline at its input. Since raw streamlines have a different number of vertices (long streamlines have more vertices than shorter ones), we re-sample the streamlines so that they all have an equal number of $256$ vertices. The latent space length was fixed at a value of $32$ for all of our experiments. The latent space dimensionality choice was made after finding an appropriate dimensionality on the ``Fiber Cup'' dataset (see section \ref{subsubsec:synthetic_data}) through a Bayesian search procedure. The value was found to be $16$ for the dataset; the next power of two ($32$) that fit into the dimensionality reduction scheme was then used as a safety margin to account for the more complex, human brain tractography configurations.

\begin{table}[!htbp]
\caption{\label{tab:ae_structure}FINTA autoencoder structure. The encoder uses strides of size $2$, and the decoder uses strides of size $1$. The decoder's upsampling stages use an upsampling factor of $2$. The autoencoder uses ReLU activations throughout its convolutional layers.}
\centering
\begin{tabular}{cccc}
\hline
\textbf{Part} & \textbf{Type} & \textbf{Features} & \textbf{Size} \\
\hline
\textbf{input} & - & - & $3; 256$ \\
\hline
\multirow{6}{*}{\textbf{encoder}} & 1D conv & $32$ & $256$ \\
& 1D conv & $64$ & $128$ \\
& 1D conv & $128$ & $64$ \\
& 1D conv & $256$ & $32$ \\
& 1D conv & $512$ & $16$ \\
& 1D conv & $1024$ & $8$ \\
\hline
\textbf{latent space} & fully connected & $32$ \\
\hline
\multirow{6}{*}{\textbf{decoder}} & upsampling + 1D conv & $1024$ & $8$ \\
& upsampling + 1D conv & $512$ & $16$ \\
& upsampling + 1D conv & $256$ & $32$ \\
& upsampling + 1D conv & $128$ & $64$ \\
& upsampling + 1D conv & $64$ & $128$ \\
& upsampling + 1D conv & $32$ & $256$ \\
\hline
\textbf{output} & - & - & $3; 256$ \\
\end{tabular}
\end{table}

The autoencoder was trained with an Adam optimizer \citep{Kingma:ICLR:2015} with a mean squared-error loss. The hyper-parameters were adjusted using a Bayesian search method. The learning rate of the optimizer was fixed to a value of $6.68 \times 10^{-4}$, and weight decaying regularization with a $0.13$ valued parameter.

\newpage
\newpage
\newpage

\subsection{Data}
\label{subsec:data}
Experiments were carried out using the ``Fiber Cup'' synthetic data \citep{Fillard:Neuroimage:2011, Cote:MIA:2013}, the ISMRM 2015 Tractography Challenge human-based synthetic dataset \citep{Maier-Hein:Zenodo:2015}, as well as the BIL\&GIN \citep{Mazoyer:Neuroimage:2016, Chenot:BrainStructFunc:2019} and Human Connectome Project (HCP) \citep{Glasser:NatureNeurosc:2016} in vivo human subject brain datasets.

\subsubsection{``Fiber Cup'' synthetic data}
\label{subsubsec:synthetic_data}
A synthetic dataset re-generated (using Fiberfox \citep{Neher:MRM:2014}) from the ``Fiber Cup'' phantom \citep{Fillard:Neuroimage:2011, Cote:MIA:2013} was used as a simplified test set mimicking the human brain white matter bundle anatomy. The ``Fiber Cup'' phantom contains $7$ bundles, and the ground truth consists of a total of $7833$ streamlines. The raw diffusion data were generated using $30$ gradient directions, a diffusion gradient strength of $1000 s/mm^2$, and a $3 mm$ isotropic spatial resolution for a $64 \times 64 \times 3$ volume. Figure \ref{fig:datasets}(a) shows the ground truth tractogram for the synthetic ``Fiber Cup'' phantom.

\subsubsection{ISMRM 2015 Tractography Challenge human-based synthetic data}
\label{subsubsec:human_based_synthetic_data}
The ISMRM 2015 Tractography Challenge dataset of \citet{Maier-Hein:Zenodo:2015} was used in a further step towards a more complex and realistic scenario. The dataset consists of a clinical-style, realistic single subject tractogram (approximately $200\,000$ ground truth streamlines) with $25$ ground truth fiber bundles, raw diffusion-weighted data and a structural T1-weighted MRI volume  generated using Fiberfox \citep{Neher:MRM:2014}. The raw dMRI data were generated at a $2 mm$ isotropic spatial resolution, with $32$ gradient directions, and a \textit{b}-value of $1000 s/mm^2$. Figure \ref{fig:datasets}(b) shows the sagittal left view of the ground truth tractogram for the ISMRM 2015 Tractography Challenge dataset.

\subsubsection{BIL\&GIN human data}
\label{subsubsec:bil_gin_human_data}
\begin{figure*}[!t]
\centering
\begin{tabular}{ccc}
\includegraphics[scale=0.95, trim=0.5in 0.45in 0.5in 0.55in, clip=true, width=0.3\linewidth, keepaspectratio=true]{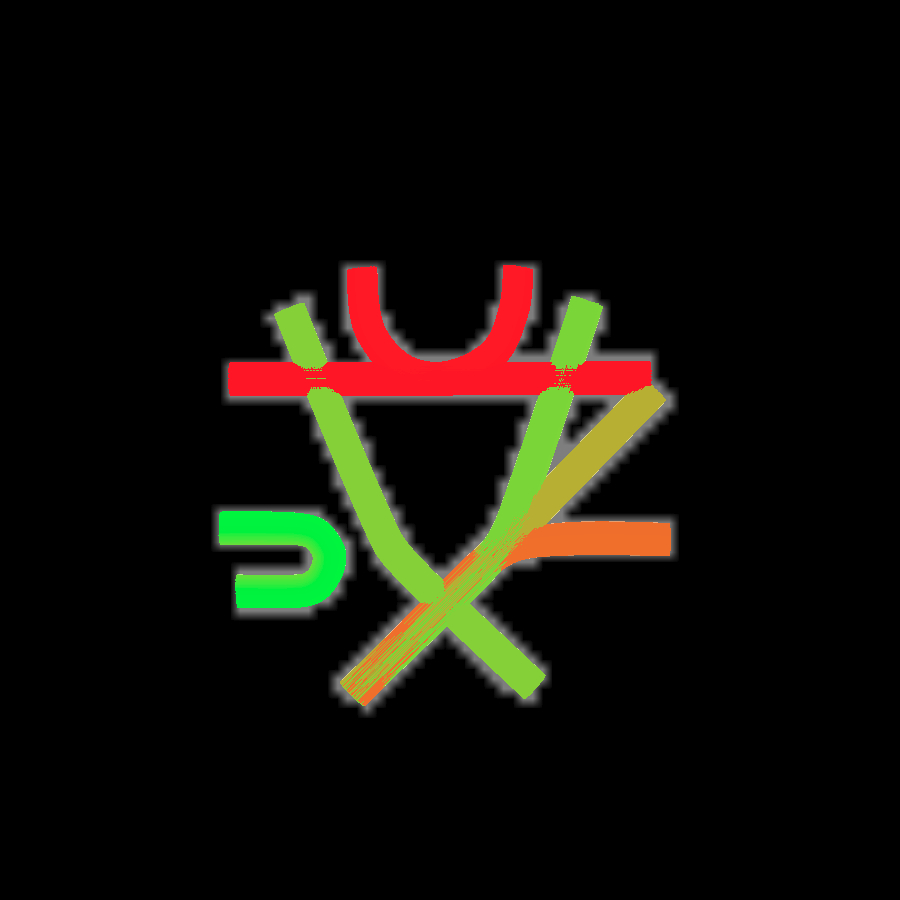} &
\includegraphics[scale=0.95, trim=0.5in 0.45in 0.5in 0.55in, clip=true, width=0.3\linewidth, keepaspectratio=true]{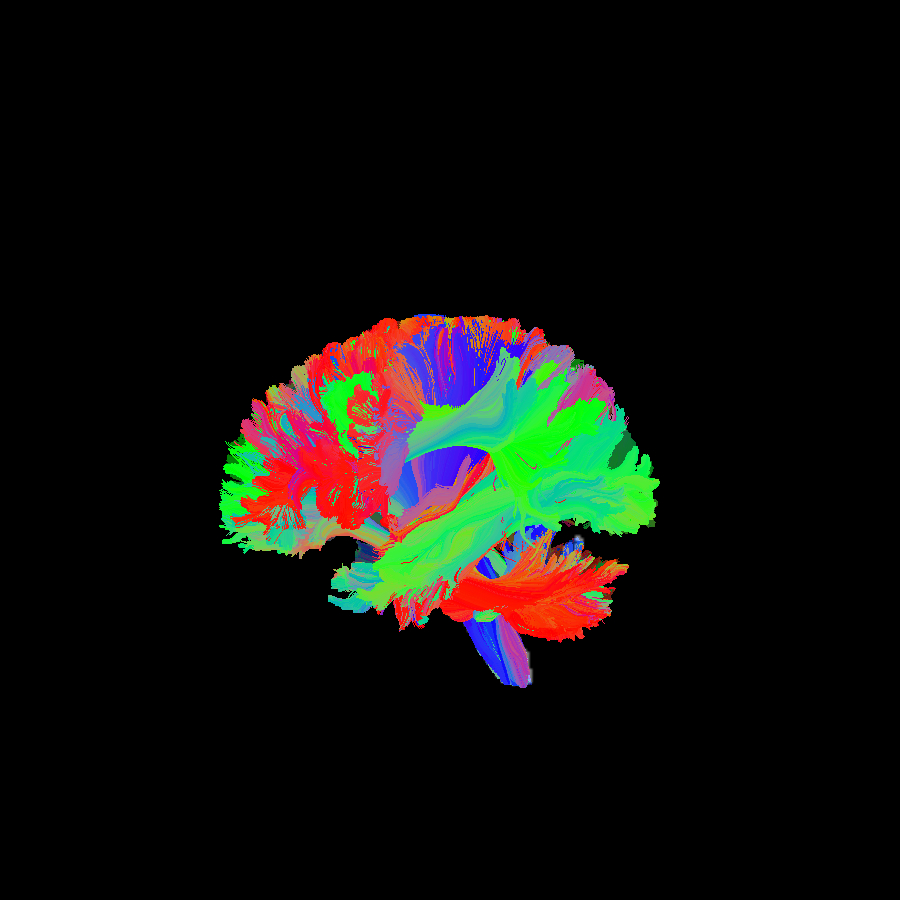} &
\includegraphics[scale=0.95, trim=0.5in 0.45in 0.5in 0.55in, clip=true, width=0.3\linewidth, keepaspectratio=true]{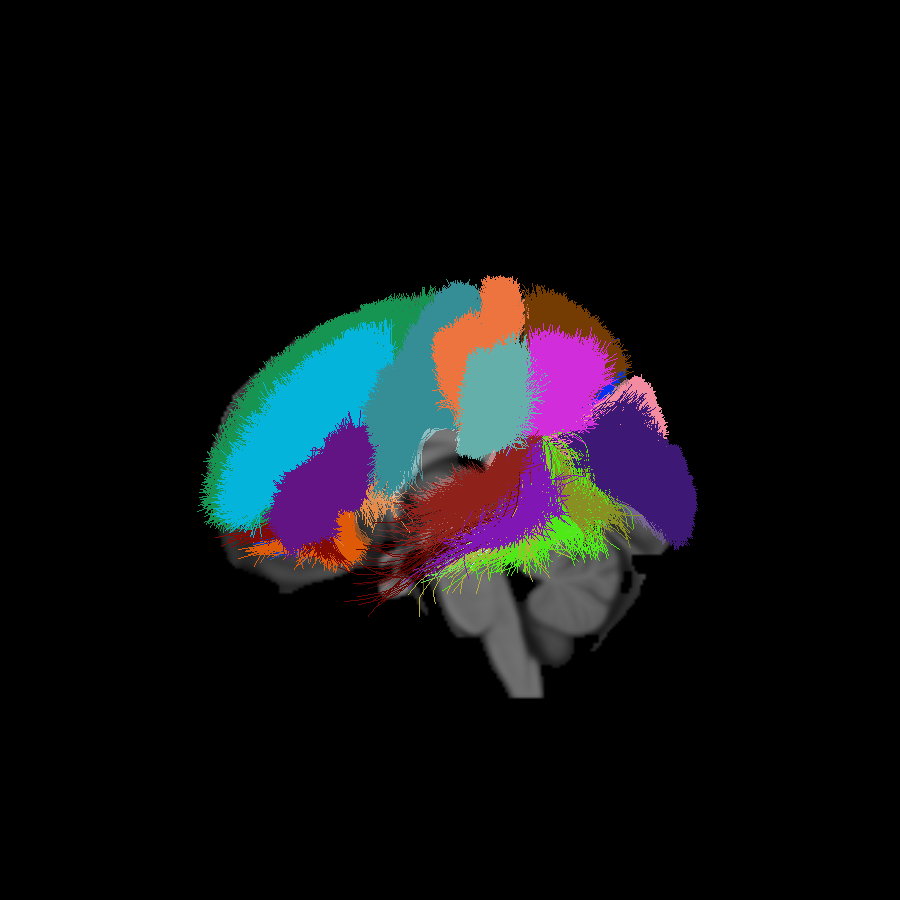} \\
\textbf{(a)} & \textbf{(b)} & \textbf{(c)} \\
\includegraphics[scale=0.95, trim=0.5in 0.45in 0.5in 0.55in, clip=true, width=0.3\linewidth, keepaspectratio=true]{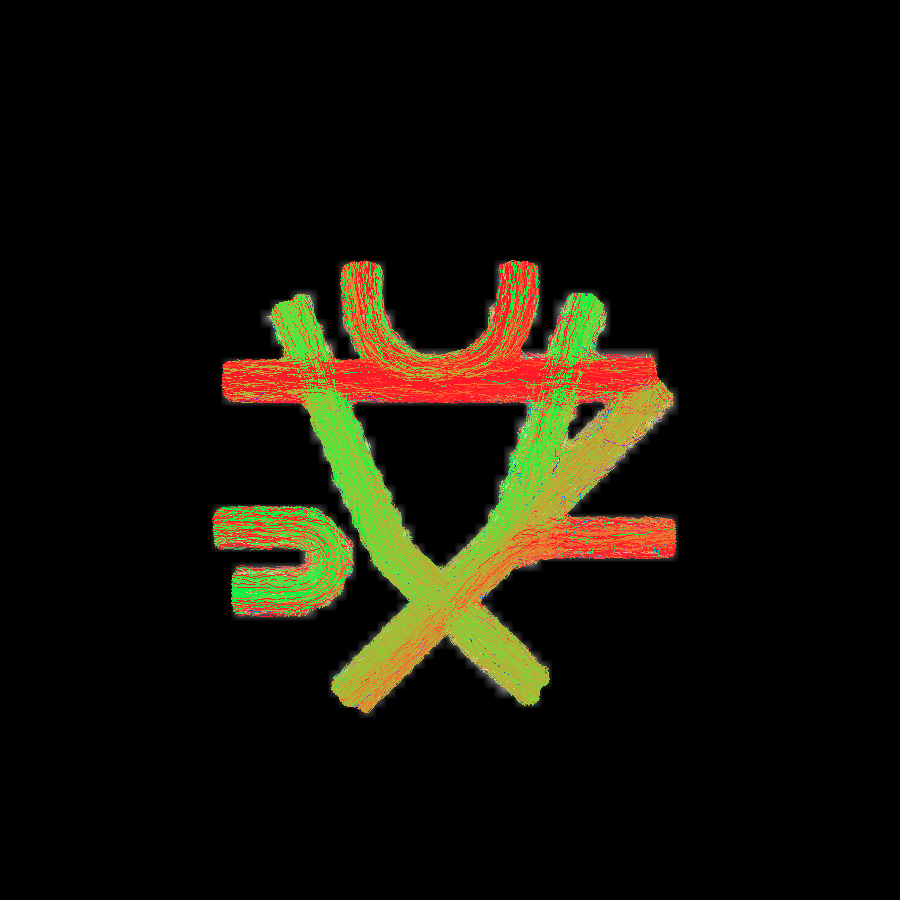} &
\includegraphics[scale=0.95, trim=0.5in 0.45in 0.5in 0.55in, clip=true, width=0.3\linewidth, keepaspectratio=true]{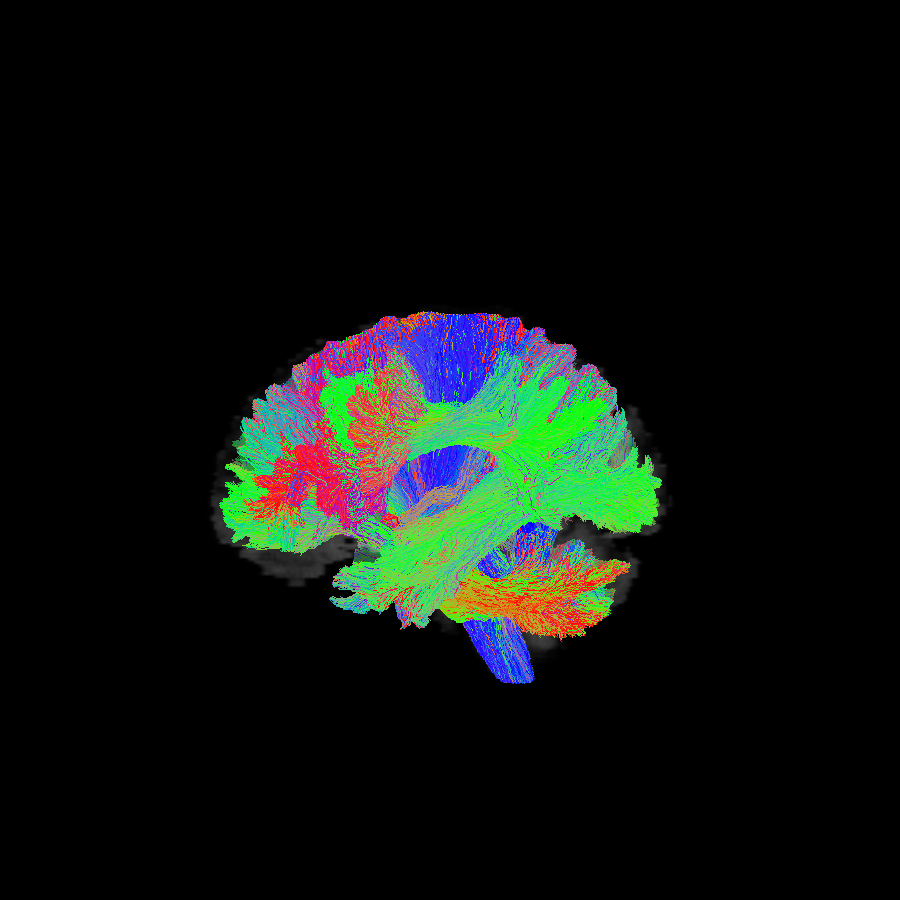} &
\includegraphics[scale=0.95, trim=0.5in 0.45in 0.5in 0.55in, clip=true, width=0.3\linewidth, keepaspectratio=true]{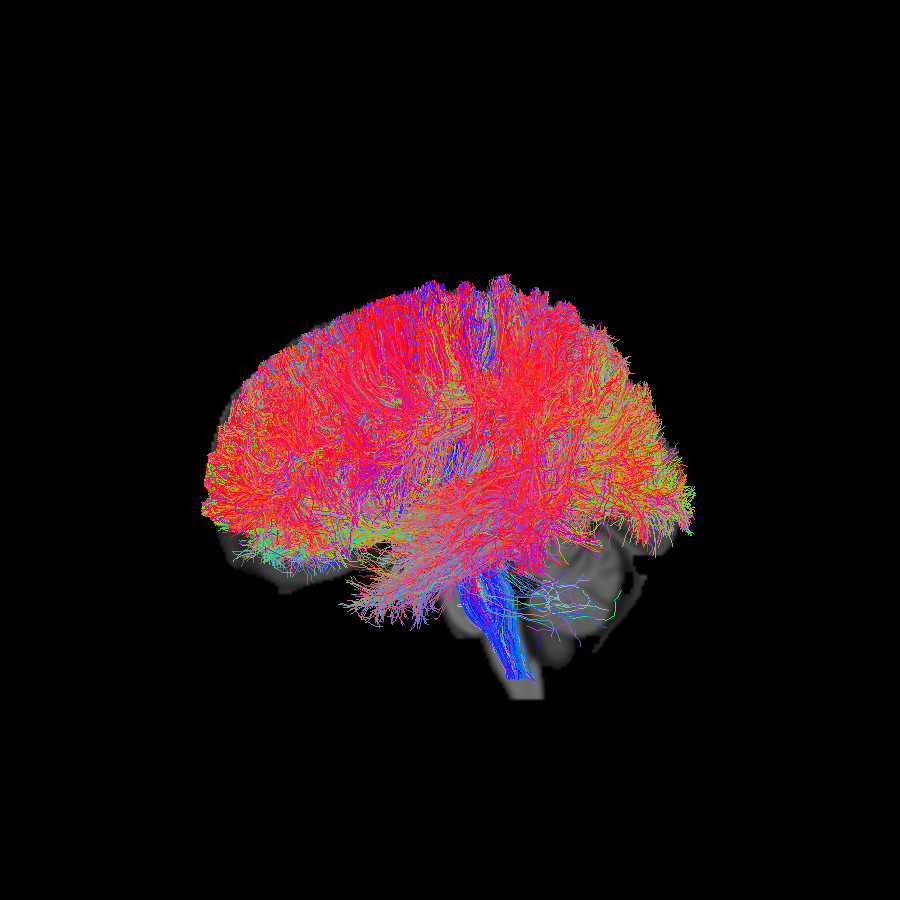} \\
\textbf{(d)} & \textbf{(e)} & \textbf{(f)} \\
\end{tabular}
\caption{\label{fig:datasets}Tractograms corresponding to the considered datasets. Upper row: (a) ground truth tractogram from the ``Fiber Cup'' dataset; (b) ground truth tractogram from the ISMRM 2015 Tractography Challenge dataset; (c) sagittal left view of the gyrus-wise homotopic callosal streamline models corresponding to the BIL\&GIN dataset. Lower row: test sets corresponding to the tracking performed on each of the three datasets: (d) view of the ``Fiber Cup'' test set containing $4963$ streamlines; (e) sagittal left view of ISMRM 2015 Tractography Challenge test set containing $63615$ streamlines; (f) sagittal left view of a randomly picked test subject corresponding to the BIL\&GIN dataset and containing $55160$ streamlines.}
\end{figure*}

A sample of $39$ subjects of the BIL\&GIN ({\em Brain Imaging of Lateralization} by the {\em Groupe d'Imagerie Neurofonctionnelle}) human brain dataset \citep{Mazoyer:Neuroimage:2016, Chenot:BrainStructFunc:2019} was used to gauge the performance of FINTA. Acquisitions were done with a Philips Achieva $3$ Tesla MR scanner using $21$ non-colinear diffusion gradient directions and a diffusion gradient encoding strength of $1000 s/mm^2$ with four averages and an isotropic spatial resolution of $2 mm$. In this work, only the corpus callosum was considered, and within the bundle, homotopic streamlines lying within any of $26$ pairs of gyral-based segments were considered as plausible streamlines. Additional details about the gyral-based callosal segments used for the extraction of the streamlines, shown in figure \ref{fig:datasets}(c), are provided in section \ref{subsec:considered_bil_gin_cc_regions}.

\subsubsection{HCP human data}
\label{subsubsec:hcp_human_data}
A subject from the Human Connectome Project (HCP) dataset \citep{Glasser:NatureNeurosc:2016} was used to perform generalization experiments. The data had been acquired using an HCP scanner equipped with high-end hardware enabling diffusion encoding gradient strengths of $100 \, mT/m$ at $1.25 mm$ isotropic spatial resolution along $270$ gradient directions. The available data had been pre-processed for head motion, eddy currents and susceptibility distortions correction.

\section{Experimental}
\label{sec:experiment_design}
The same experiment design was implemented for the datasets. Noise-free diffusion data were used for the ``Fiber Cup'' and ISMRM 2015 Tractography Challenge datasets, and no denoising procedure was required for the BIL\&GIN dataset. A Constrained Spherical Deconvolution (CSD) \citep{Tournier:Neuroimage:2007} method using a spherical harmonics order of $6$ was used to extract the fiber ODFs from the raw dMRI data. Local probabilistic whole brain tractography was performed using white matter seeding (single seed per voxel; $0.5 mm$ step size; $30^{\circ}$ maximum aperture angle; default values for other parameters) for the ``Fiber Cup'' and the ISMRM 2015 Tractography Challenge datasets to generate the sets for the autoencoder. The extracted streamlines were processed and scored to obtain the (anatomically) implausible streamline and (anatomically) plausible streamline sets. The scoring procedure for the ``Fiber Cup'' labelled as plausible streamline instances obeying constrained length ($20-200 mm$) and shape features (bends $< 335^{\circ}$), as well as mask (streamlines contained within the known bundle masks) and connectivity (streamlines connect the known regions) criteria. The implausible set, thus, comprised streamlines exceeding the given length threshold; streamlines whose trajectories described sharp bends; streamlines that did not reach the corresponding terminal regions; and streamlines connecting regions that were not actually connected in the ground truth (see examples of such trajectories in figure \ref{fig:plausible_vs_implausible}). The ISMRM 2015 Tractography Challenge dataset was scored according to the procedure in \citet{Maier-Hein:NatureComm:2017}, based on the streamline clustering strategy proposed in \citet{Garyfallidis:FrontiersNeurosc:2012}. The terms \textit{Tractometer\_basic} and \textit{Tractometer\_adv} are used, respectively, to refer to the described scoring methods in section \ref{subsec:connectivity_analysis}. For the ISMRM 2015 Tractography Challenge dataset, the anterior commissure (CA) and the posterior commissure (CP) bundles were left out of the experiments since the tracking algorithm was unable to extract them or the extracted bundles did not meet the scoring system criteria. Hence, $23$ bundles were used in the experiments.

Probabilistic tractography was performed on the BIL\&GIN human brain dataset using the Particle Filtering Tracking (PFT) method \citep{Girard:Neuroimage:2014} with white matter seeding ($10$ seeds per voxel; $0.5 mm$ step size; $20^{\circ}$ maximum aperture angle; default values for other parameters). The resulting tractograms were registered \citep{Avants:Neuroimage:2011} to the MNI template common space using the MNI152 2009 standard-space T1-weighted average structural $1 mm$ isotropic resolution template image \citep{Fonov:Neuroimage:2011}. For the experiments in this work, only the callosal subset of the generated tractograms was used. The plausible streamlines within this subset consisted of the streamlines within $26$ gyral-based homotopic callosal segment pairs (see section \ref{subsec:considered_bil_gin_cc_regions} for further details about their definition). The implausible set comprised, as for the ``Fiber Cup'' and ISMRM 2015 Tractography Challenge datasets, streamlines that did not reach the cortical surface, describing sharp bends, having a single endpoint in the callosal region, having endpoints in non-white matter tissues, or connecting both hemispheres through non-callosal segments or tissues (e.g. the brainstem). The procedure was supervised by a neuroanatomist. The procedure yielded, on average, $41838$ streamlines per-subject (standard deviation: $5648$) for the considered callosal dataset. The lower row in figure \ref{fig:datasets} shows the test set tractograms, including both plausible and implausible streamlines, corresponding to all three datasets.

The ratio of implausible streamlines over the resulting tractograms was in-line with the findings in previous studies: $79\%$ for the ``Fiber Cup'' dataset (similar to the figures reported by \citet{Cote:MIA:2013} for probabilistic methods); $46\%$ for the ISMRM 2015 Tractography Challenge dataset (see the valid connection ratios across submissions reported in \citet{Maier-Hein:NatureComm:2017}); and $89\%$ on average (standard deviation: $2\%$) for the callosal BIL\&GIN dataset. All streamlines were resampled to contain an equal number of points ($256$) and their head-to-tail orientations rearranged to make the head vertices closest with respect to the origin for each dataset. The obtained streamlines were split randomly into training/validation and test sets using an $80\%/20\%$ ratio. The proposed convolutional autoencoder was trained in an unsupervised manner using the training split of both the plausible and implausible streamlines issued by the probabilistic tracking methods. A separate training was performed for each dataset. The filtering threshold was set by finding the optimal point in the receiver operating characteristic (ROC) curves evenly rewarding true positives and penalizing false positives.

\subsection{Performance measure definitions}
\label{subsec:performance_measure_definitions}
FINTA's performance was graded according to the following measures:
\begin{itemize}
\item \textit{Accuracy}: proportion of correct predictions (true positives and true negatives) over the total number of streamlines.
\item \textit{Sensitivity}: proportion of relevant instances that are predicted as positives (true positives) among all positive streamlines in the data.
\item \textit{Precision}: proportion of relevant instances that are predicted as true positives among all retrieved (predicted) positive streamlines.
\item \textit{F1-score}: harmonic mean of precision and sensitivity.
\end{itemize}

\subsection{Comparison to other filtering methods}
\label{subsec:benchmarking}
The quantitative comparison was performed on the callosal BIL\&GIN dataset. FINTA was compared to both anatomy-based tractography filtering methods and two variants (single-bundle and multi-bundle) of RecoBundles \citep{Garyfallidis:Neuroimage:2018, Rheault:PhD:2020}, a state-of-the-art method applicable to partial data. The anatomy-based filtering methods were applied successively, each one filtering the previous step's output, as used in several works such as \citet{Maier-Hein:NatureComm:2017, Sarubbo:BrainStructFunc:2019}. The purpose of including the results of each step of this filtering pipeline (baseline filtering methods \#1 to \#4) was to show the successive improvement of the classification performance scores by using anatomical heuristics. The models of plausible streamlines used for the RecoBundles experiments were built according to \citet{Rheault:PhD:2020}, and based on the $26$ gyral-based homotopic callosal segment pairs (see section \ref{subsec:considered_bil_gin_cc_regions}). The multi-bundle RecoBundles experiment considered each gyrus-wise homotopic callosal streamline model as a separate, parameterizable entity, whereas the single-bundle RecoBundles method considered their concatenation as a single target group. This yielded a total of $6$ tractography filtering methods as baselines: 
\begin{enumerate}
\item filtering based on streamline length (named \textit{length} within this text); 
\item filtering with loop removal (\textit{no\_loops}); 
\item filtering with early stops in cerebrospinal fluid (\textit{no\_end\_in\_csf});
\item filtering with early stops in the white matter (\textit{end\_in\_atlas}); 
\item single-bundle RecoBundles (\textit{recobundles\_single}); 
\item multi-bundle RecoBundles (\textit{recobundles\_multi}). 
\end{enumerate}

In addition to the measures mentioned in section \ref{subsec:performance_measure_definitions}, the balanced accuracy was computed on the predictions on the callosal BIL\&GIN dataset as the mean of the sensitivity on the plausible and implausible classes. Similarly, the subscripts \textit{m} and \textit{w} are used (see \ref{subsec:human_data_benchmarking_supplementary_results}) to distinguish the unweighted (or macro) and weighted measures. The last terms account for the class imbalance between the plausible and implausible streamline classes.

The goodness of the retrieved white matter anatomy was measured in terms of the ability to preserve homotopic streamlines across the gyral segments of interest. A filtered tractogram was considered to preserve successfully the underlying white matter anatomy if at least one homotopic streamline was present in each gyral-based homotopic callosal segment pair. Thus, the anatomical goodness measure, named \textit{valid gyrus-wise rate} (VGR), is defined for a given subject as the ratio between the number of segments ($n_{VG}$) containing at least one homotopic streamline after filtering and the total number of segments containing homotopic streamlines ($N_{VG}$) in the underlying ground truth, i.e. $n_{VG}/N_{VG}$. Note that $N_{VG}$ might be different across subjects. The measure is bounded in the $[0, 1]$ range. Note also that the preserved streamlines were recovered from the positives predicted by each method using the anatomical criteria defined for each segment (see section \ref{subsec:considered_bil_gin_cc_regions}).

To estimate the task performance uniquely, a \textit{success rate} (SR) measure was computed as the unweighted average of the classification measures and the mentioned valid gyrus-wise rate.

\subsection{Generalization}
\label{subsec:generalization_experimental}
In order to test the generalization ability of FINTA, we performed an additional local deterministic tracking using an ensemble tractography protocol \citep{Takemura:PlosComputBiol:2016} on the ISMRM 2015 Tractography Challenge dataset and a global tracking \citep{Reisert:Neuroimage:2011} on the Human Connectome Project subject. The ensemble local deterministic tractography was performed using white matter seeding (single seed per voxel), $\{0.1, 0.5, 1.0\} mm$ step sizes, and $\{10, 20, 30, 40\}^{\circ}$ maximum aperture angles. The tractograms resulting from all permutations of parameters were merged into a single tractogram file. The global tracking used $900\,000\,000$ iterations, a particle length of $1 mm$, a particle width of $0.1 mm$, and a particle weight of $0.002$. Diffusion data was tracked six (6) times and the resulting whole-brain tractograms were combined into a single set. In each case $500\,000$ streamlines were randomly sampled and scored using the same method as described in the experimental setting earlier. The scoring yielded a plausible-to-implausible ratio of $57/43\%$ on the ensemble local deterministic tracking data and $6/94\%$ on the global tracking data. For each set the performance measures described in section \ref{subsec:performance_measure_definitions} were computed using two different filtering threshold values (i.e. the latent space minimum distance at which a streamline is considered to be implausible):
\begin{itemize}
\item Computed on the ISMRM 2015 Tractography Challenge dataset for the local probabilistic setting (\textit{Generalized thr}): i.e. the reference streamline set and filtering threshold remained unaltered relative to the ISMRM 2015 Tractography Challenge local probabilistic experiment.
\item Computed on a subset of each new dataset (\textit{Specific thr}): i.e. the reference streamline sets were re-defined as new dataset-specific subsets to provide a filtering threshold value.
\end{itemize}

In either case, the autoencoder remained unaltered (i.e. was not retrained) with respect to the training done on the ISMRM 2015 Tractography Challenge local probabilistic streamlines.

\subsection{Time requirements}
\label{subsec:time_requirements_experimental}
The filtering clock time of FINTA was measured on each of the datasets. For quantitative comparison purposes, tractograms of different sizes were generated from the ISMRM 2015 Tractography Challenge data using the same procedure described earlier. In total, six ($6$) different tractograms (generated using a local probabilistic method) were generated containing $20\,000$, $40\,000$, $100\,000$, $200\,000$, $600\,000$, and $1\,000\,000$ streamlines, respectively. The time required to filter each of these tractograms was measured for both FINTA and RecoBundles three ($3$) times, and the mean and standard deviation values were computed for each tractogram and method. In all cases, solely the time required for filtering was measured, excluding I/O operation time. All time tests were performed on a conventional desktop machine (Intel(R) Xeon(R) W-$2133$ CPU @ $3.60$GHz $6$ core processor; $16$ GB RAM; NVIDIA GeForce GTX $1080$ Ti $12$ GB graphics card).

\section{Results}
\label{sec:results}

\subsection{Tractogram latent space}
\label{subsec:latent_space}
\begin{figure}[!htb]
\centering
\begin{tabular}{ccc}
\includegraphics[scale=0.95, width=0.32\linewidth, keepaspectratio=true]{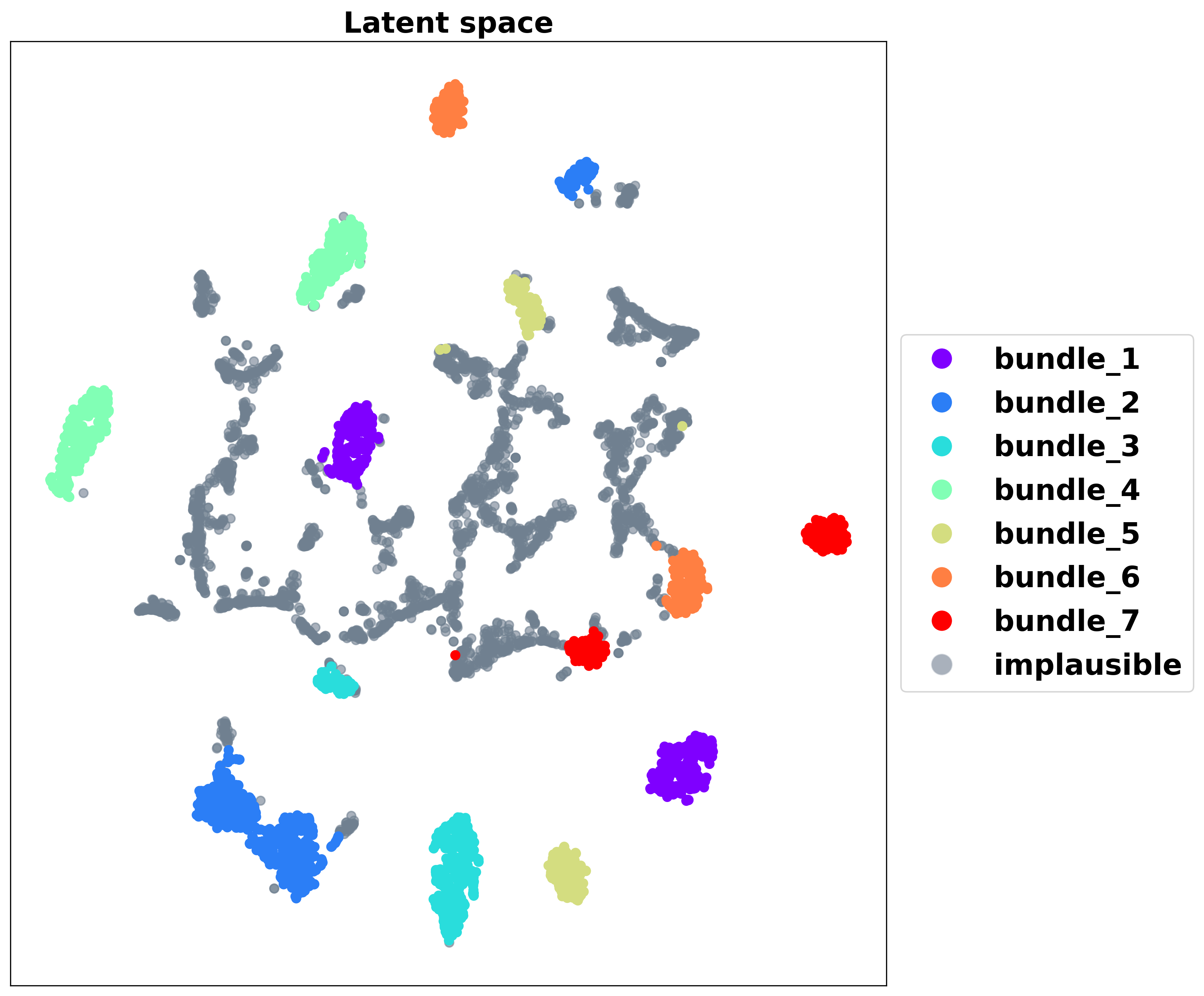} &
\includegraphics[scale=0.95, clip=true, width=0.32\linewidth, keepaspectratio=true]{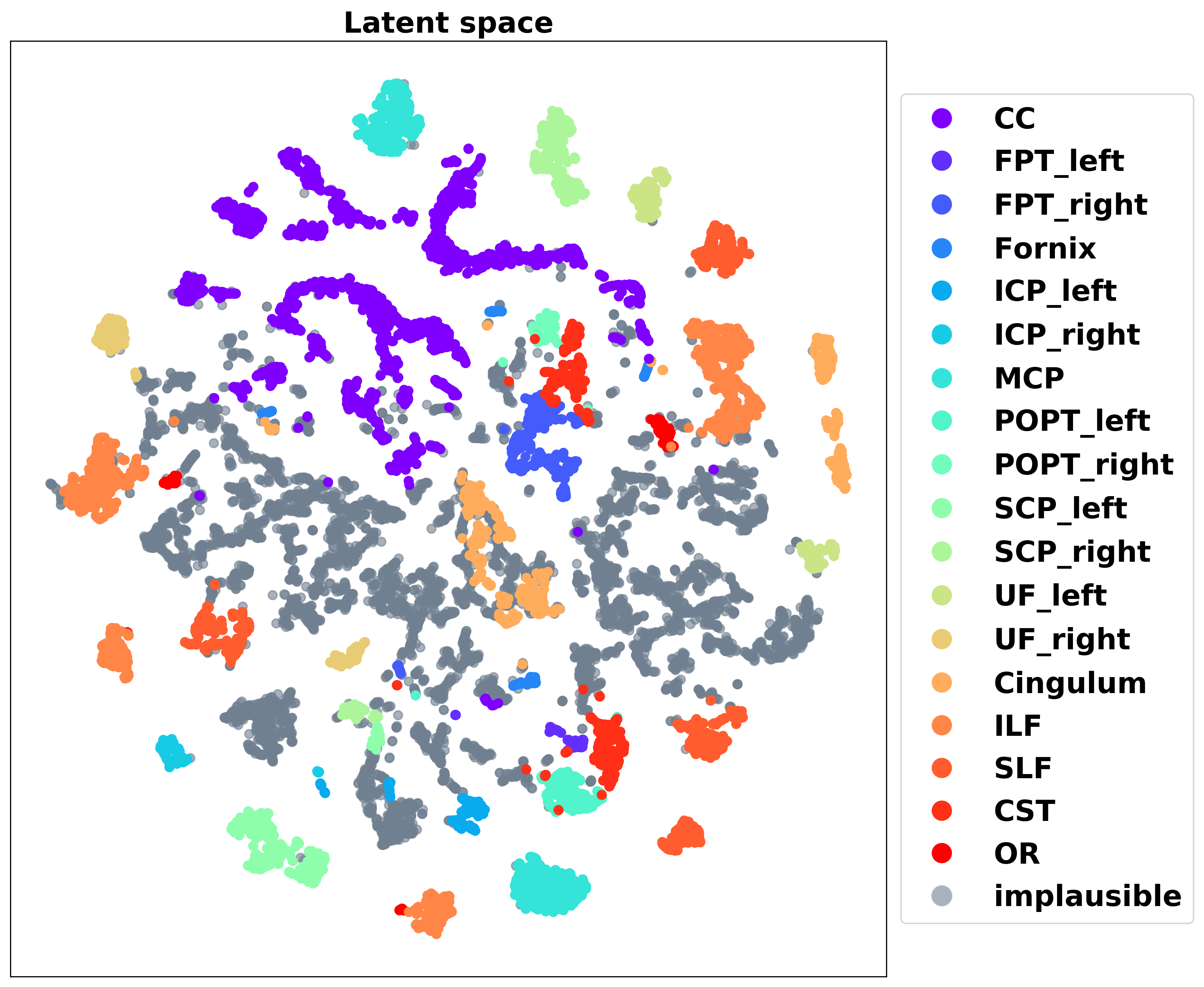} &
\includegraphics[scale=0.95, clip=true, width=0.32\linewidth, keepaspectratio=true]{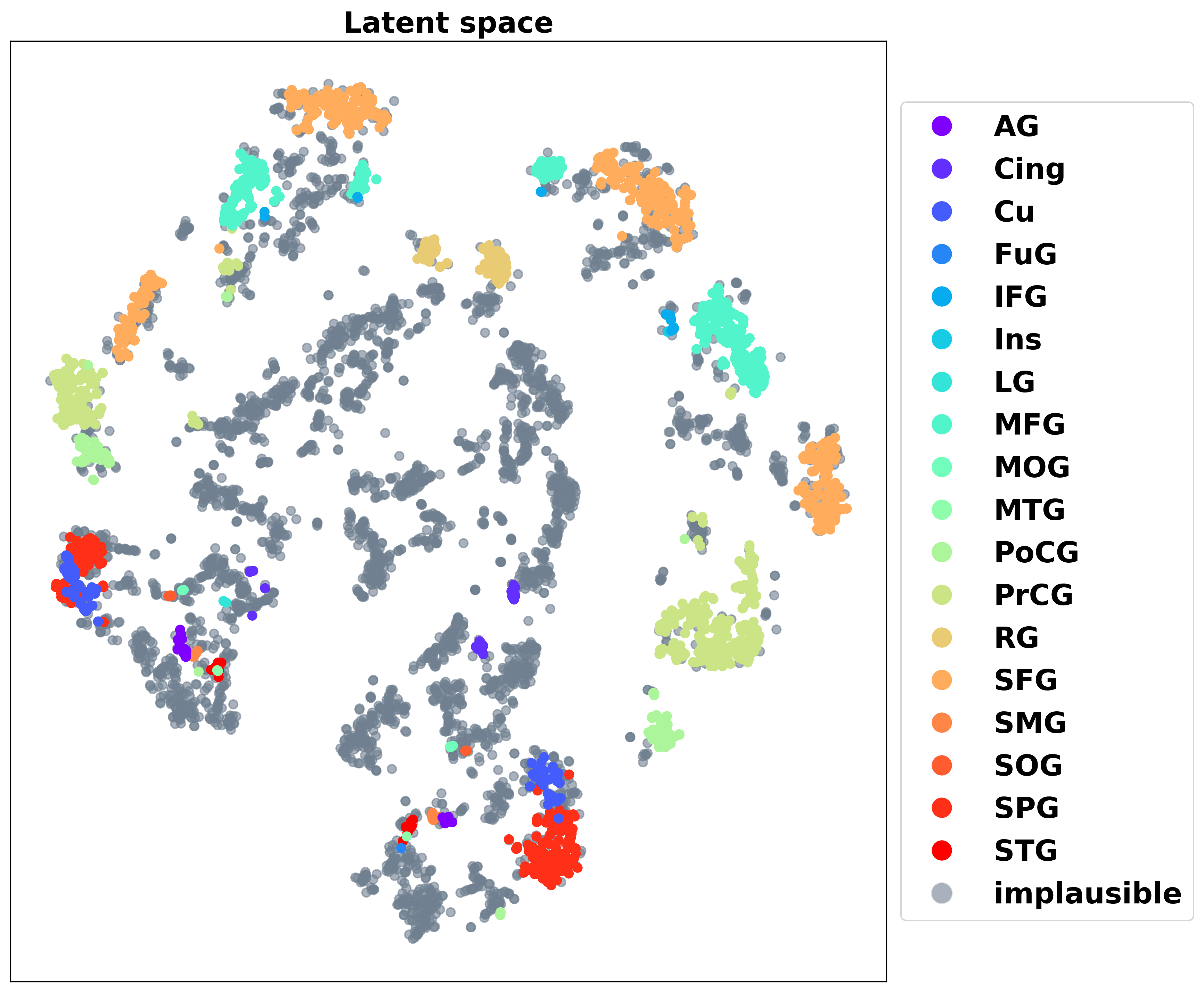} \\
\textbf{(a)} & \textbf{(b)} & \textbf{(c)} \\
\end{tabular}
\caption{\label{fig:latent_space_tsne}Latent space views using a t-SNE dimensionality reduction. (a) ``Fiber Cup'' data; (b) ISMRM 2015 Tractography Challenge data; (c) Callosal BIL\&GIN human brain data. The number of projected streamlines was limited in each case for visualization purposes. The callosal BIL\&GIN data corresponds to a randomly picked subject which was missing all streamlines in $8$ of the gyral-based segments.}
\end{figure}

The autoencoder's latent space allows to re-interpret the streamline pair-wise distance. For illustrative purposes, a number of randomly picked plausible streamlines were assigned to their corresponding bundle (or gyrus-wise homotopic callosal streamline group for BIL\&GIN), and together with some implausible streamlines, they were projected into the latent space of a trained autoencoder. Figure \ref{fig:latent_space_tsne} shows a view of the latent spaces corresponding to each of the three datasets. The views were generated by applying a t-SNE dimensionality reduction \citep{vanderMaaten:JMLR:2008} to the autoencoder's latent space vectors. The figure suggests that streamlines belonging to the same bundle (or streamline group) lie at a short distance from each other whereas implausible streamlines are uniformly distributed. Similarly, it provides evidence supporting that the autoencoder setting can be generalized and give way to multi-label streamline classification tasks such as bundling. 

\begin{figure}[!htb]
\centering
\begin{tabular}{cc}
\multicolumn{2}{c}{\includegraphics[scale=0.95, trim=0.05in 0.4in 0.05in 0.4in, clip=true, width=\linewidth, keepaspectratio=true]{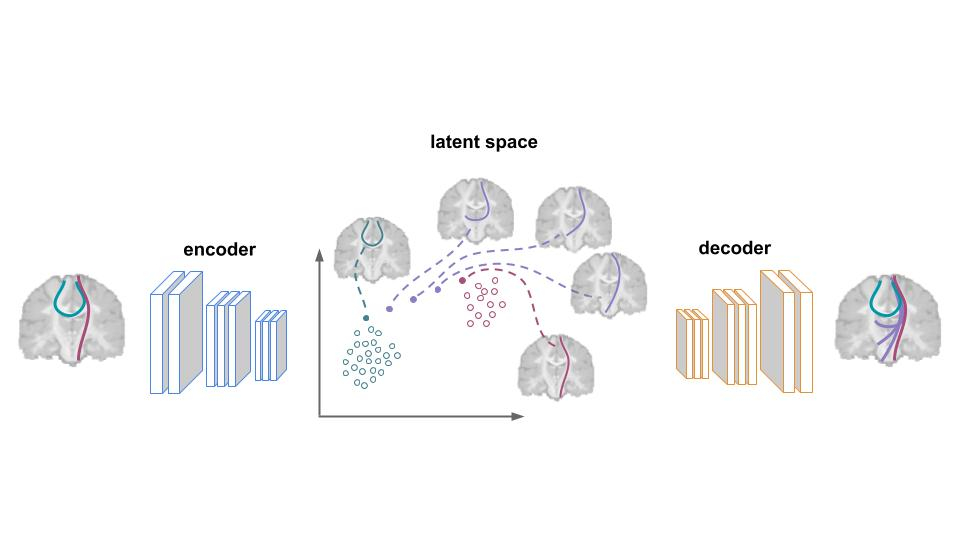}} \\
\multicolumn{2}{c}{\textbf{(a)}} \\
\includegraphics[scale=0.95, trim=2.175in 0.55in 2.025in 0.865in, clip=true, width=0.45\linewidth, keepaspectratio=true]{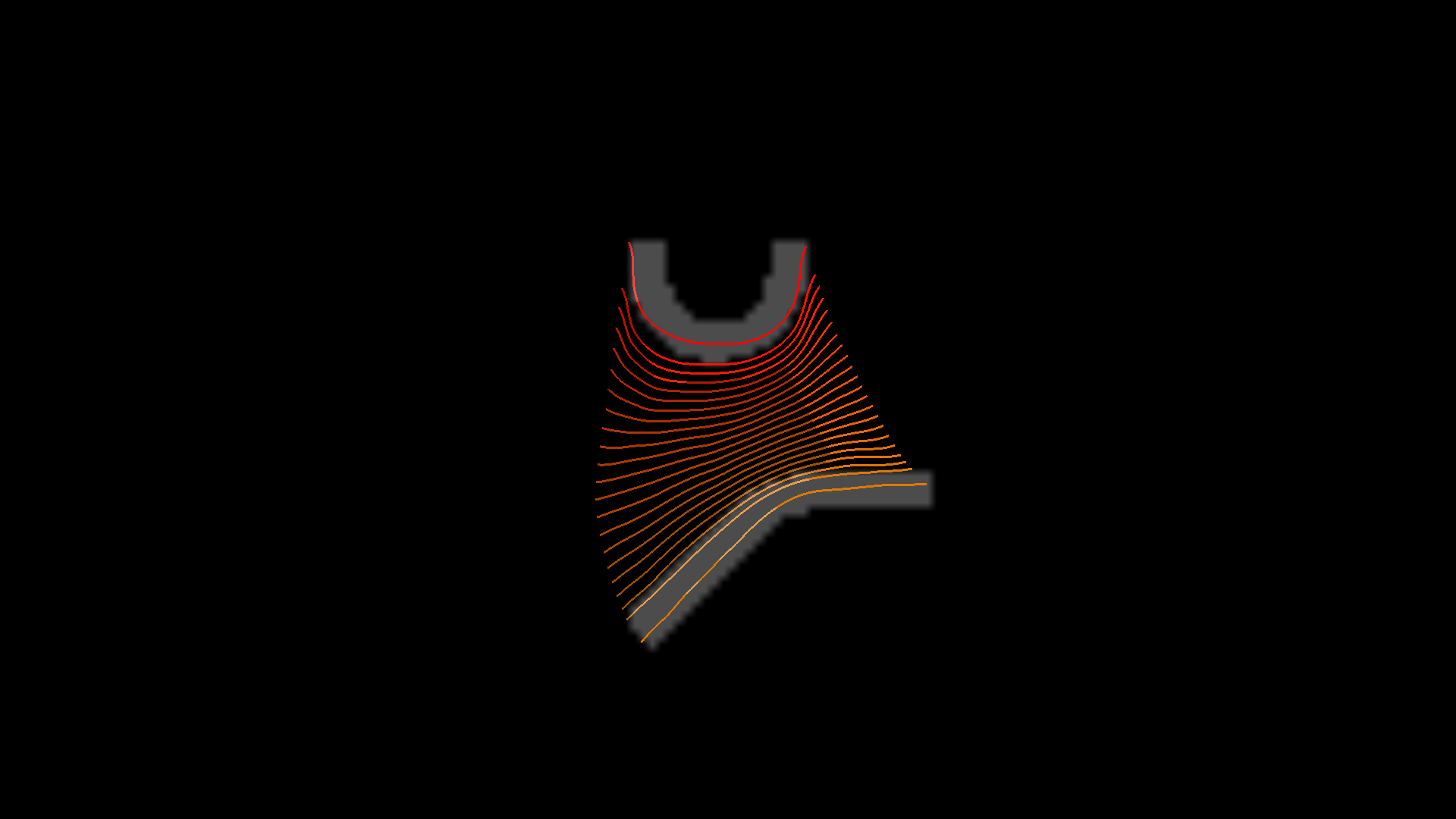} &
\includegraphics[scale=0.95, trim=2.025in 0.55in 2.175in 0.865in, clip=true, width=0.45\linewidth, keepaspectratio=true]{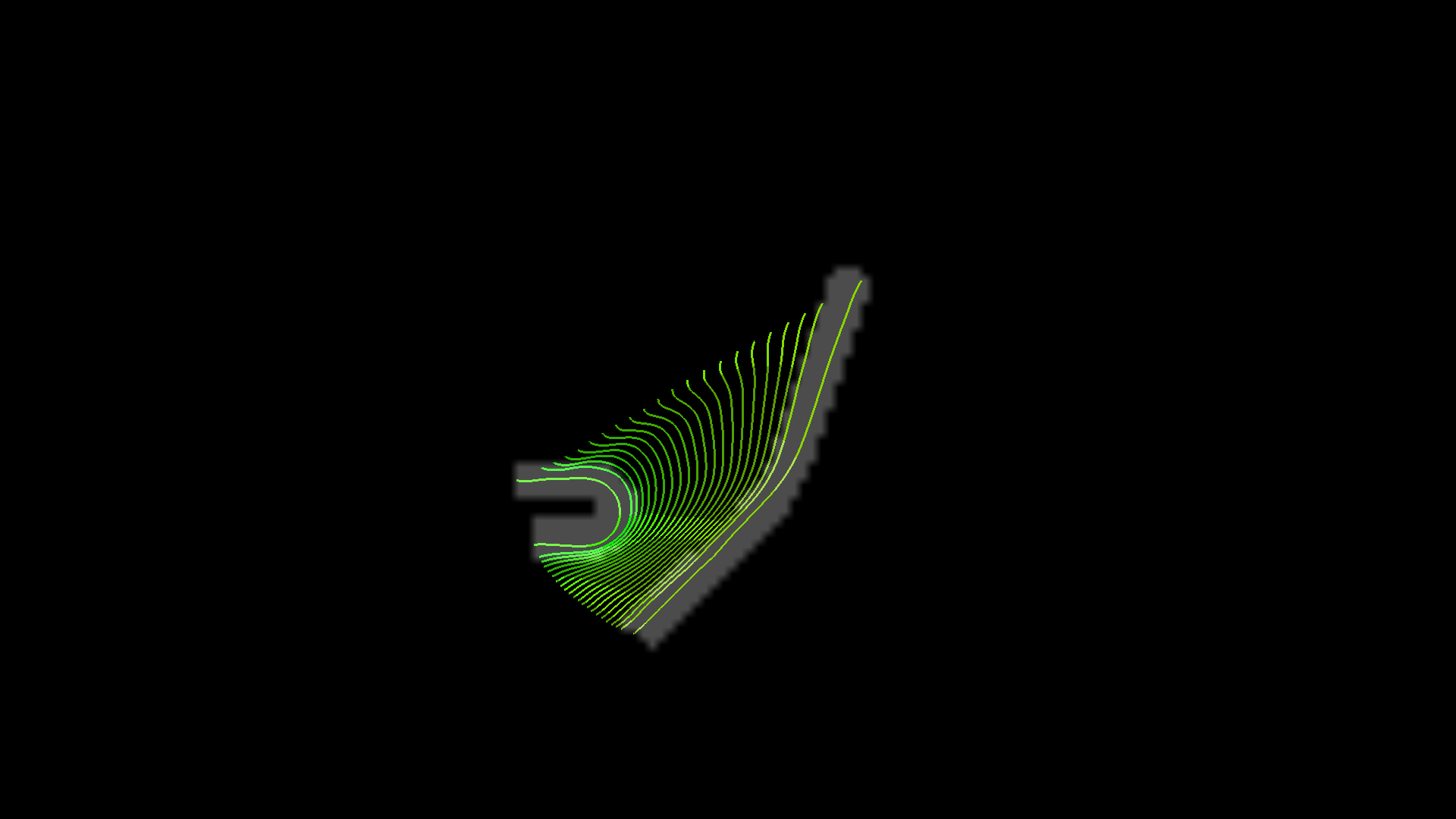} \\
\textbf{(b)} & \textbf{(c)} \\
\end{tabular}
\caption{\label{fig:fibercup_latent_space_interpolation}Latent space interpolation. (a) Schematic representation of latent space streamlines interpolation. The streamline representation in the latent space allows to choose a pair of encoded streamlines and fill the space between them through a series of interpolated streamlines. (b) and (c) ``Fiber Cup'' latent space streamline interpolation: streamlines interpolated between the instances at the boundaries of each series: (b) bundles 1 and 5; (c) bundles 4 and 7 (see the bundle numbering in \citet{Cote:MIA:2013}). Only the structural shapes of the bundles used for the interpolation are shown for the sake of clearness. The streamlines are shown with a larger diameter with respect to the one used for the rest of the figures.}
\end{figure}

Furthermore, the application of mathematical tools in the latent space is straightforward. Figure \ref{fig:fibercup_latent_space_interpolation}(a) shows the interpolation principle applied to the latent space. Interpolating in the latent space allows to explore the space through a series of intermediate representations between any number of input streamlines of choice. Figure \ref{fig:fibercup_latent_space_interpolation}(b) and (c) show streamlines interpolated in the latent space between streamline pairs corresponding to two distinct sets of bundles of the ``Fiber Cup'' dataset. This example stresses the underlying significance of the latent space streamline pair-wise distance. Our proposed algorithm relies on the underlying hypothesis that the Euclidean distance between the latent representation of two streamlines is a good proxy to measure their structural similarity. Thus, the fact that a linear interpolation between two latent vectors leads to a smooth transition between their reconstructed streamlines underlines that the learned manifolds are smooth and that the latent Euclidean distance is a good similarity metric.

The reconstruction of the streamlines at the output of the autoencoder involves decoding the samples in the latent space. Section \ref{subsec:reconstruction} shows the reconstructed tractograms corresponding to the ``Fiber Cup'' test set streamlines.

\newpage
\subsection{``Fiber Cup'' synthetic data}
\label{subsec:synthetic_data_results}
\begin{figure}[!ht]
\centering
\begin{tabular}{cc}
\includegraphics[width=0.45\linewidth]{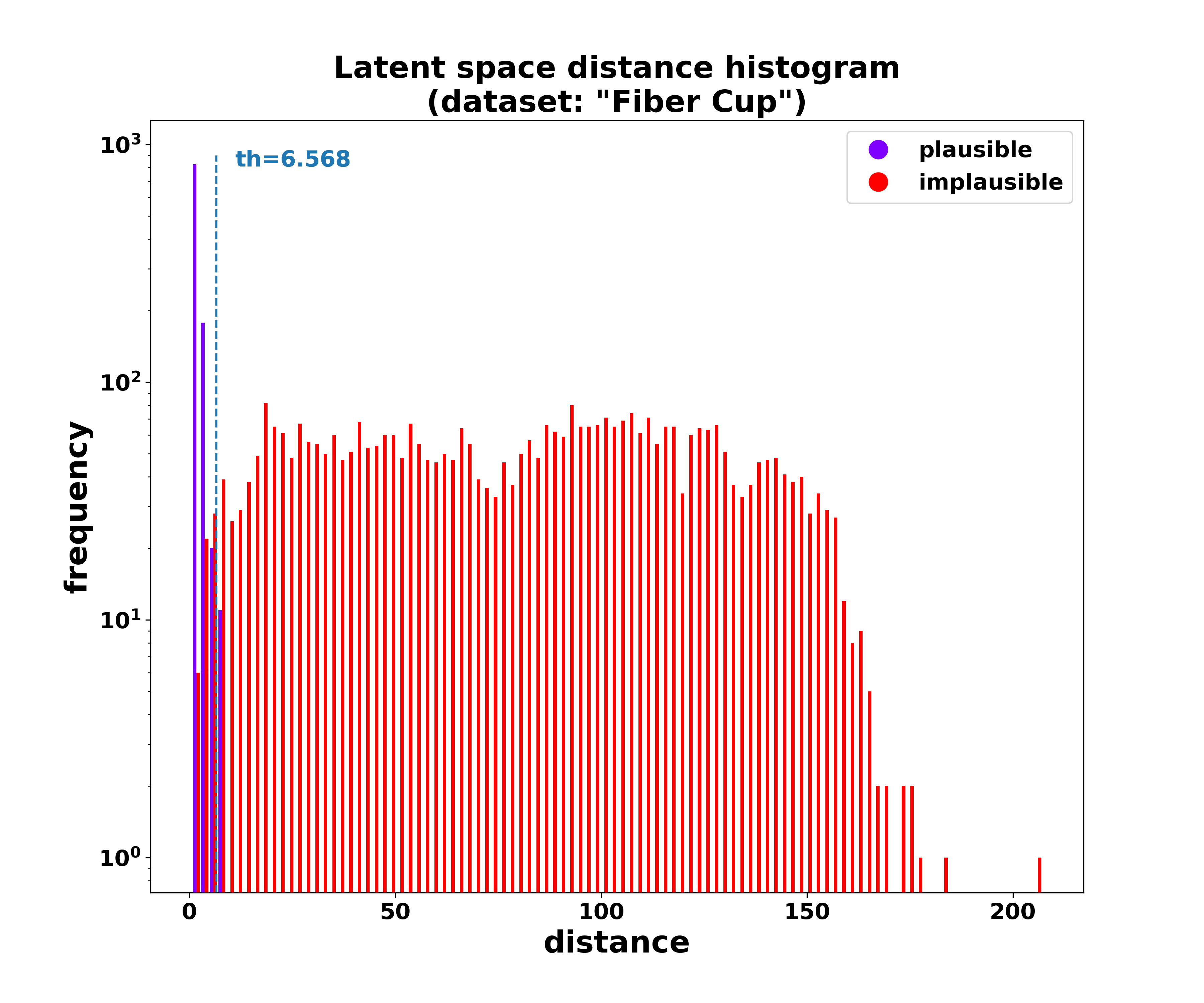} &
\includegraphics[width=0.45\linewidth]{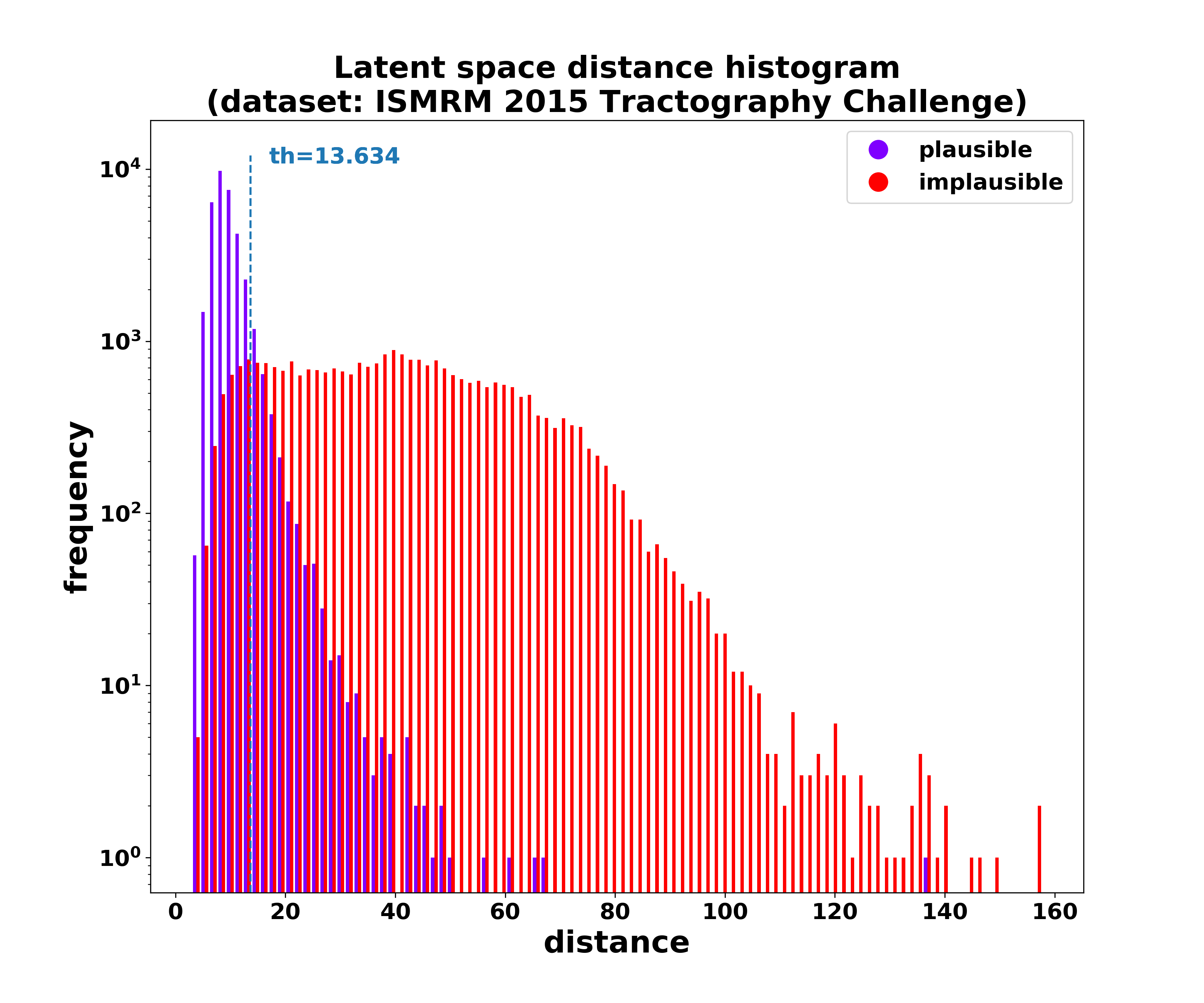} \\
\textbf{(a)} & \textbf{(b)} \\
\includegraphics[width=0.45\linewidth]{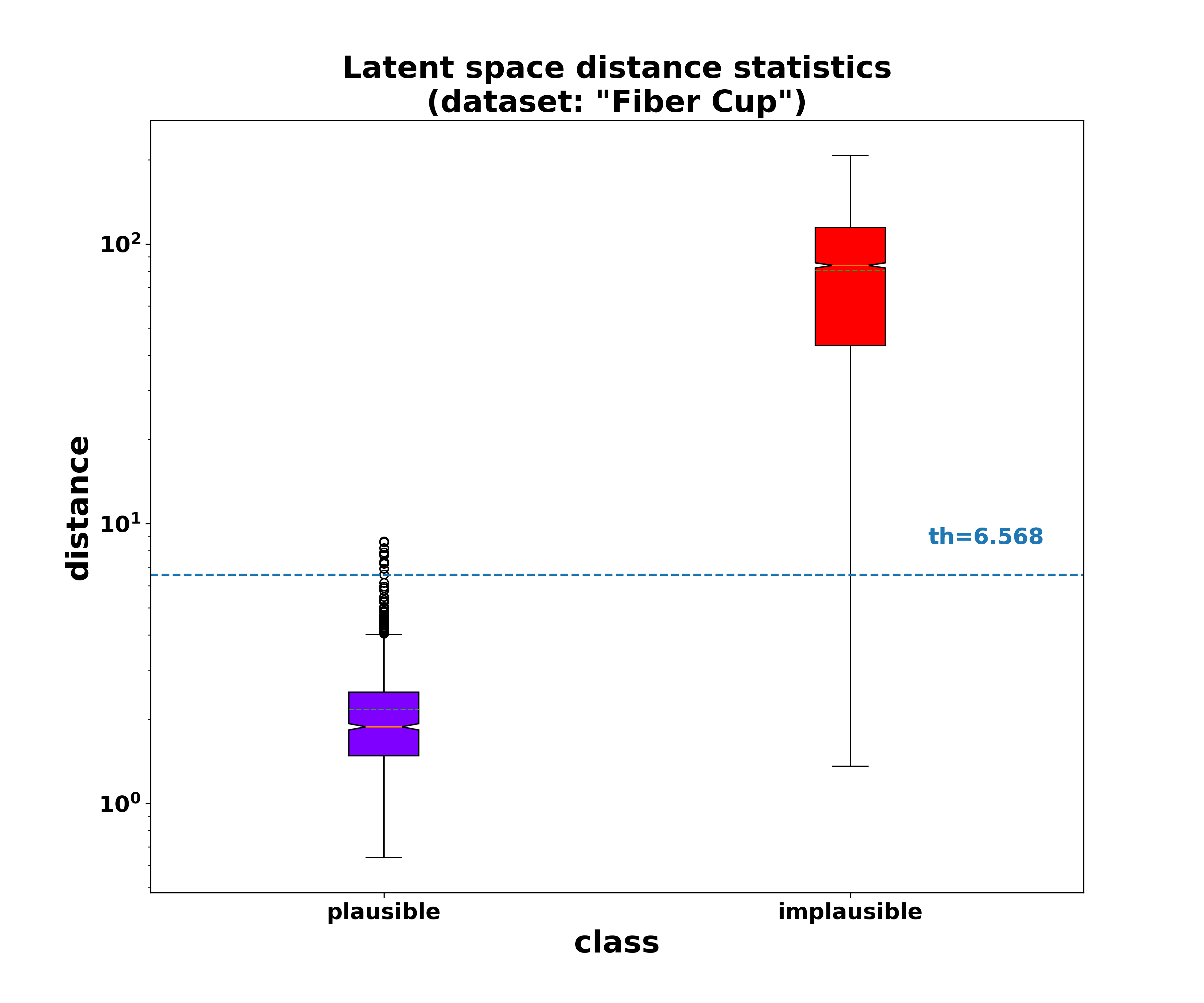} &
\includegraphics[width=0.45\linewidth]{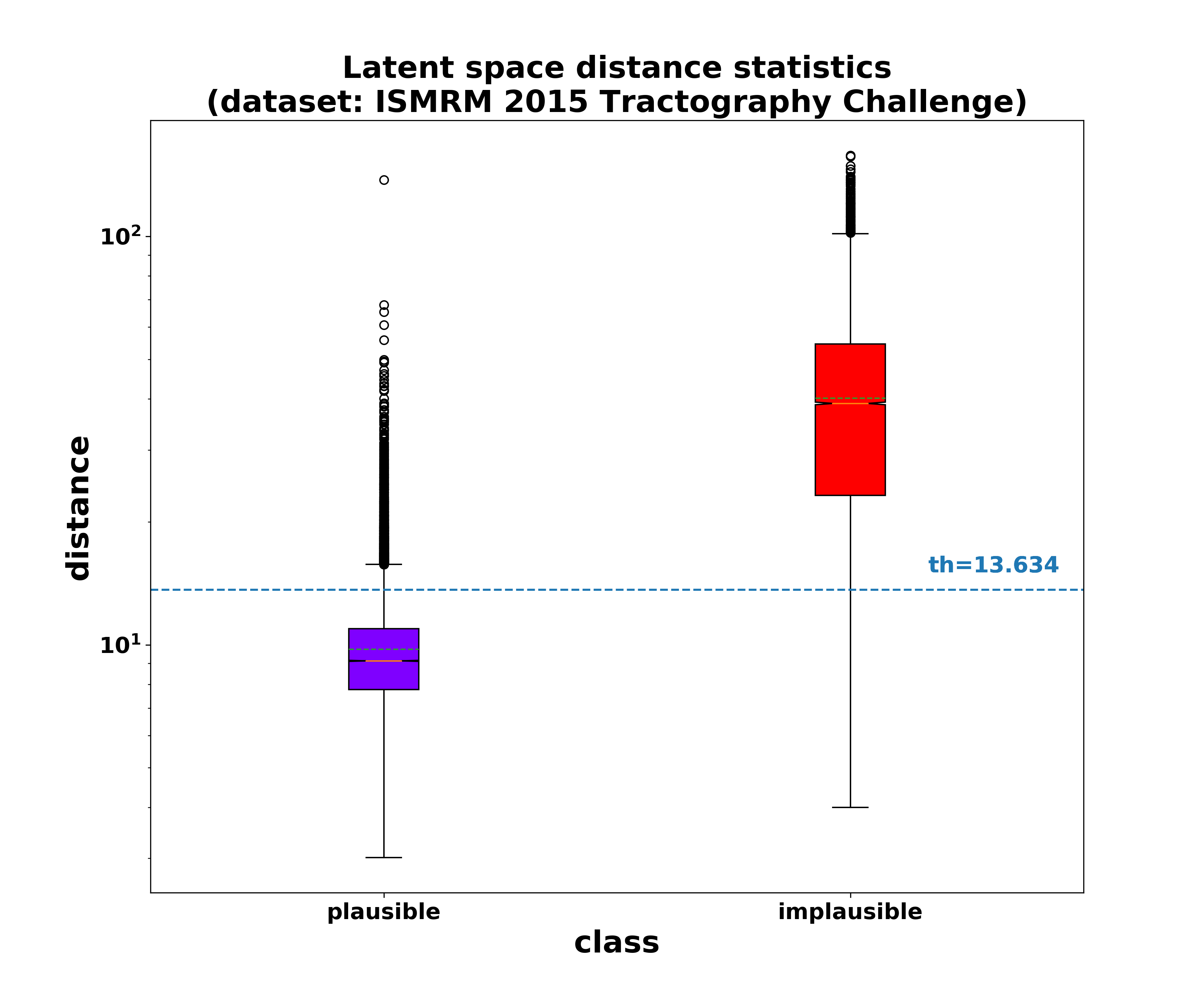} \\
\textbf{(c)} & \textbf{(d)} \\
\end{tabular}
\caption{\label{fig:latent_space_distance_distr}Nearest neighbor latent space distance distribution: histogram (a) and within-class mean and variance (c) corresponding to the ``Fiber Cup'' dataset test set streamlines; histogram (b) and within-class mean and variance (d) corresponding to the ISMRM 2015 Tractography Challenge. The dashed lines indicate the filtering threshold distance in each case. Note that the vertical axes are in logarithmic scale.}
\end{figure}

Figure \ref{fig:latent_space_distance_distr} (a) and (c) show the nearest neighbor latent space distance distribution histograms and within-class mean and variance for the ``Fiber Cup'' dataset. The figure illustrates the filtering principle of FINTA based on the boundary between the inliers and outliers: the plausible streamline within-class nearest neighbor distance is low (close to $0$ for the ``Fiber Cup'') and shows a reduced variance. As a contrast, the implausible streamlines show a much higher mean distance to their nearest neighbor streamline, as well as a higher variance. Note that the distance measure is computed considering the plausible streamlines from the train set as the reference.

\clearpage

As quantitatively shown in table \ref{tab:dataset_classification_performance}, the filtering distance threshold obtained by means of the autoencoder-based strategy allows to filter out the undesired implausible streamlines with a high degree of success.

\begin{table}[!hb]
\caption{\label{tab:dataset_classification_performance}Filtering performance for the ``Fiber Cup'' and the ISMRM 2015 Tractography Challenge datasets.}
\centering
\begin{tabular}{ccc}
\hline
\textbf{Measure} & \multicolumn{2}{c}{\textbf{Dataset}}\\
& \textbf{``Fiber Cup''} & \textbf{ISMRM 2015 Tractography Challenge} \\
\hline
Accuracy & 0.99 & 0.91 \\
Sensitivity & 0.99 & 0.91 \\
Precision & 0.97 & 0.91 \\
F1-score & 0.98 & 0.91
\end{tabular}
\end{table}

\begin{figure}[!ht]
\centering
\includegraphics[scale=0.35, trim=0.5in 0.45in 0.5in 0.65in, clip=true, width=0.45\linewidth, keepaspectratio=true]{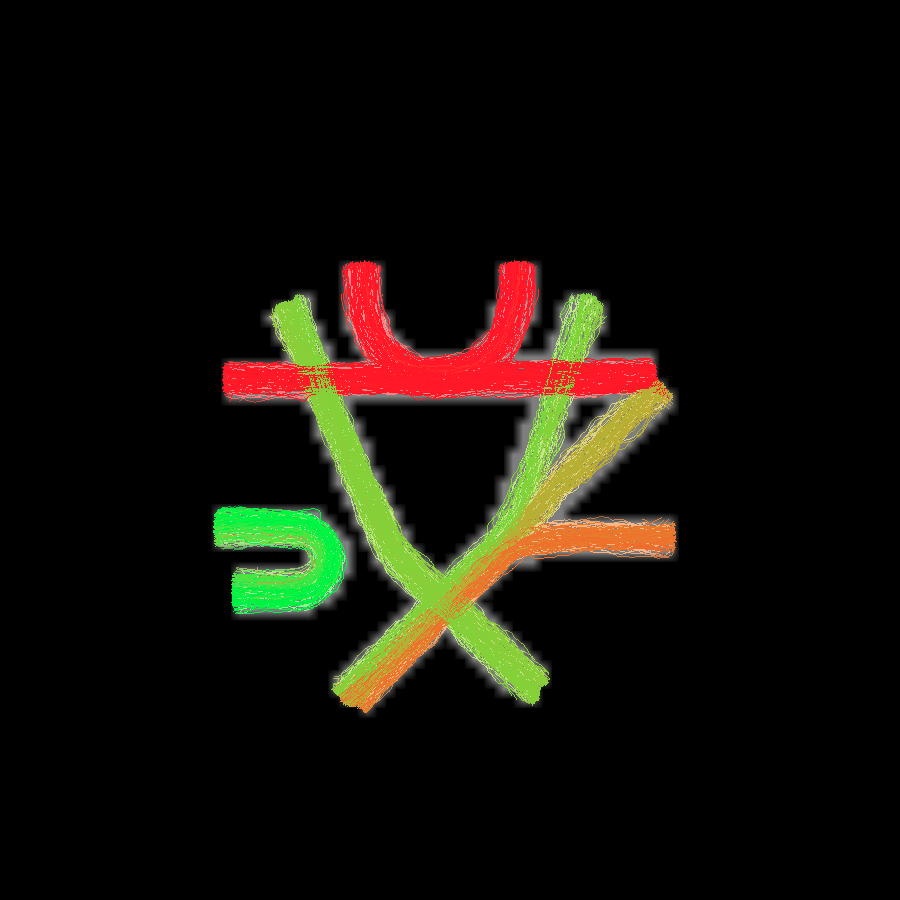}
\caption{\label{fig:fibercup_predicted_positives}Predicted positives on the test set corresponding to the ``Fiber Cup'' dataset. Note that, in contrast to the input dataset shown in figure \ref{fig:datasets}(d), the filtered tractogram shows a bundle color code information that matches better the ``Fiber Cup'' ground truth shown in figure \ref{fig:datasets}(a), implying that implausible streamlines have been eliminated to a large extent.}
\end{figure}

The predicted positive streamlines for the ``Fiber Cup'' are shown in figure \ref{fig:fibercup_predicted_positives}. These are the streamlines considered as plausibles by FINTA from the test set, and hence include both the true positives and the false positives. Qualitatively, looking at the tractogram bundle color code information, and in agreement with the quantitative results, the implausibles seem to have been substantially filtered out, implying a decreased influence of the false positives.

\begin{figure}[H]
\centering
\begin{tabular}{ccc}
\includegraphics[scale=0.95, trim=1.975in 0.55in 2in 0.89in, clip=true, width=0.445\linewidth, keepaspectratio=true]{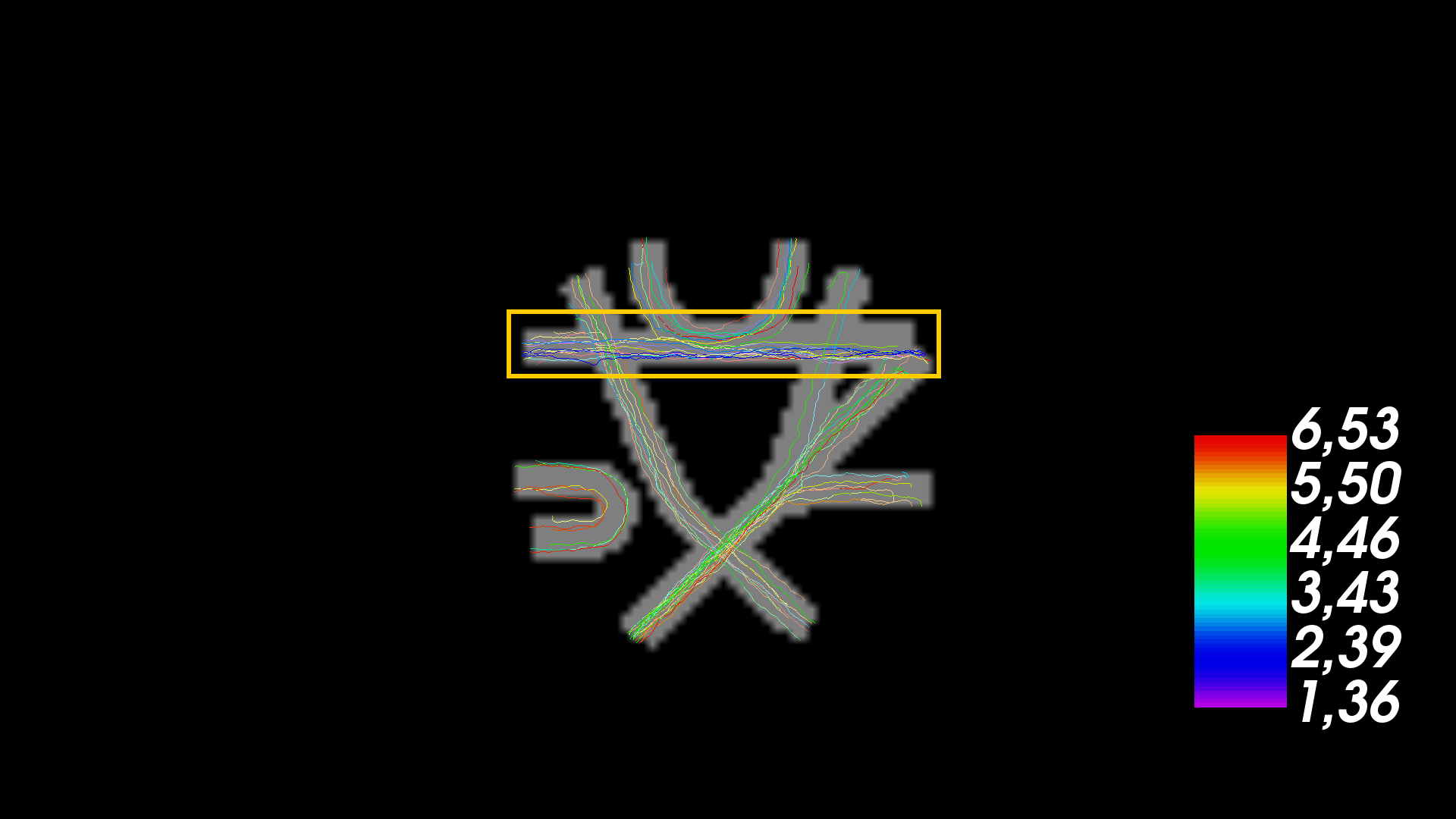} &
\includegraphics[scale=0.95, trim=1.975in 0.45in 1.95in 0.75in, clip=true, width=0.3025\linewidth, keepaspectratio=true]{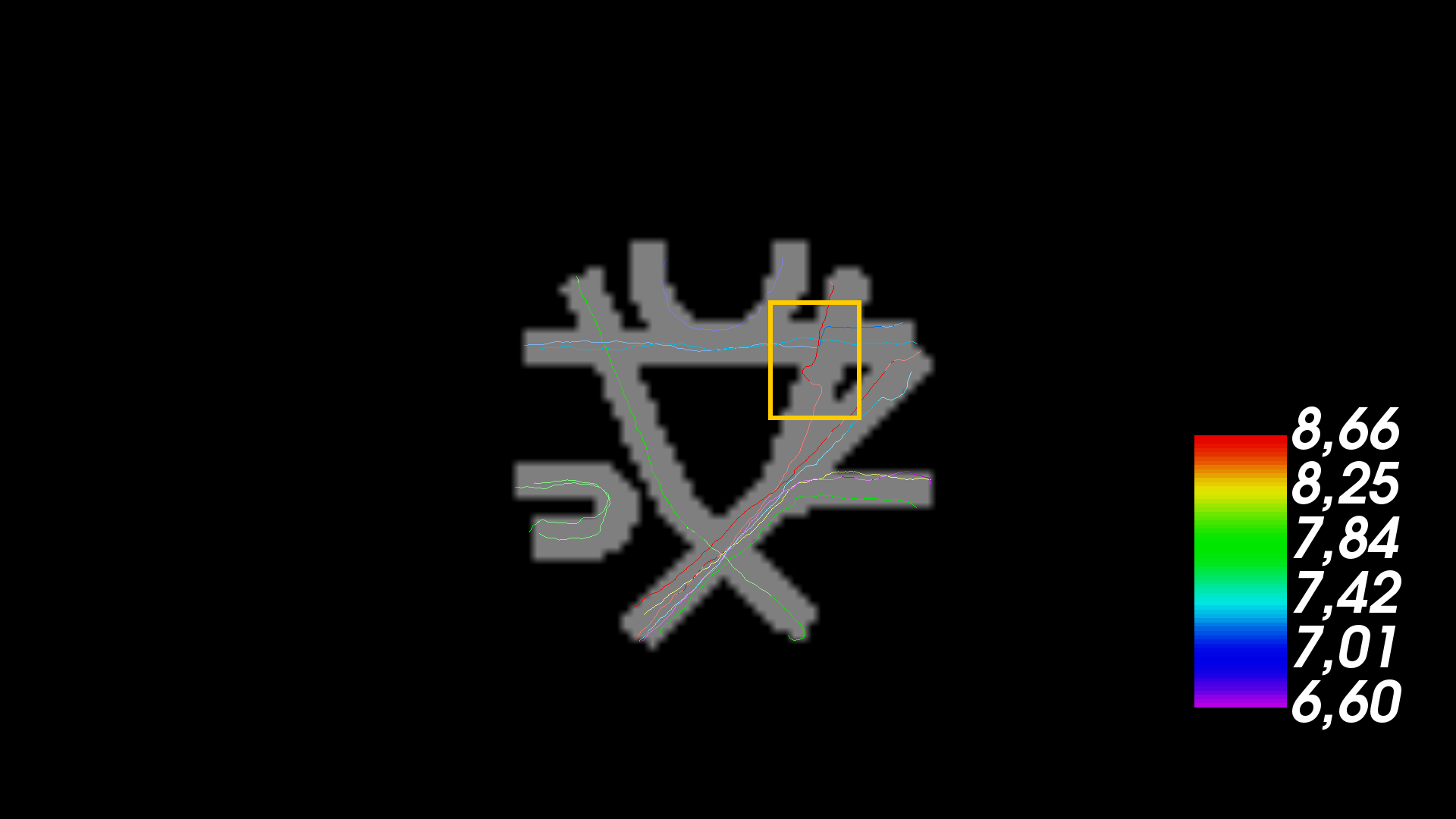} &
\includegraphics[scale=0.95, width=0.185\linewidth, keepaspectratio=true]{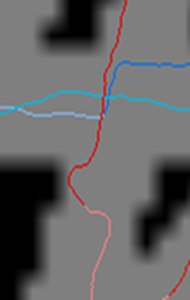} \\
\includegraphics[scale=0.95, width=0.445\linewidth, keepaspectratio=true]{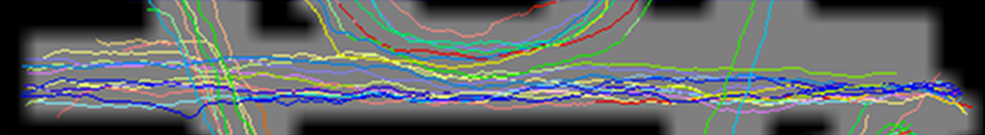} \\
\textbf{(a)} & \multicolumn{2}{c}{\textbf{(b)}} \\
\end{tabular}
\caption{\label{fig:fibercup_dataset_streamline_classification_falses} False positive and false negative streamlines issued by FINTA on the ``Fiber Cup'' dataset test set. (a) false positives: streamlines labeled as plausibles by FINTA but considered implausibles by the scoring method; (b) false negatives: streamlines labeled as being implausibles by FINTA but considered plausibles by the scoring method.}
\end{figure}

Figures \ref{fig:fibercup_dataset_streamline_classification_falses}(a) and (b) show, respectively, the false positive and false negative streamlines as detected by FINTA for the ``Fiber Cup'' dataset, and color-coded according to the latent space nearest neighbor distance. From the figures and the corresponding highlighted areas it can be drawn that, at least for part of the streamlines, FINTA has identified that these streamlines were misclassified in the reference set: in figure \ref{fig:fibercup_dataset_streamline_classification_falses}(a) a large part of the highlighted streamlines seem to be close to an anatomically plausible streamline within the bundle at issue. Many of these false positives are labeled so because they end $1$ or $2$ voxels away from the terminal regions (which would correspond to the cortical surface or gray matter border on human subject brain data). Conversely, part of the highlighted streamlines in figure \ref{fig:fibercup_dataset_streamline_classification_falses}(b) describe trajectories that could have been identified as anatomically implausible due to erratic bends. This illustrates how close from the ground truth our results are.

Training on the ``Fiber Cup'' dataset took approximately $25$ min. FINTA takes around $0.2$s to filter the ``Fiber Cup'' test tractogram.

\subsection{ISMRM 2015 Tractography Challenge human-based synthetic data}
\label{subsec:human_based_synthetic_data_results}
The nearest neighbor latent space distance distribution histograms and within-class mean and variance for the ISMRM 2015 Tractography Challenge dataset are shown in figure \ref{fig:latent_space_distance_distr}(b) and (d). As it is the case with the ``Fiber Cup'' dataset, although a larger degree of overlap between classes is observed, our autoencoder-based filtering strategy allows to obtain a threshold to filter out the undesired implausible streamlines with a degree of success of approximately $91\%$ over all measures (see table \ref{tab:dataset_classification_performance}). Here again, the latent space similarity metric shows the same behavior in both experiments: implausible streamlines have a larger distance to the reference nearest neighbor.

\begin{figure*}[!t]
\centering
\setlength{\tabcolsep}{0pt}
\begin{tabular}{cccc}
\includegraphics[scale=0.95, trim=0.55in 0.55in 0.55in 0.55in, clip=true, width=0.295\linewidth, keepaspectratio=true]{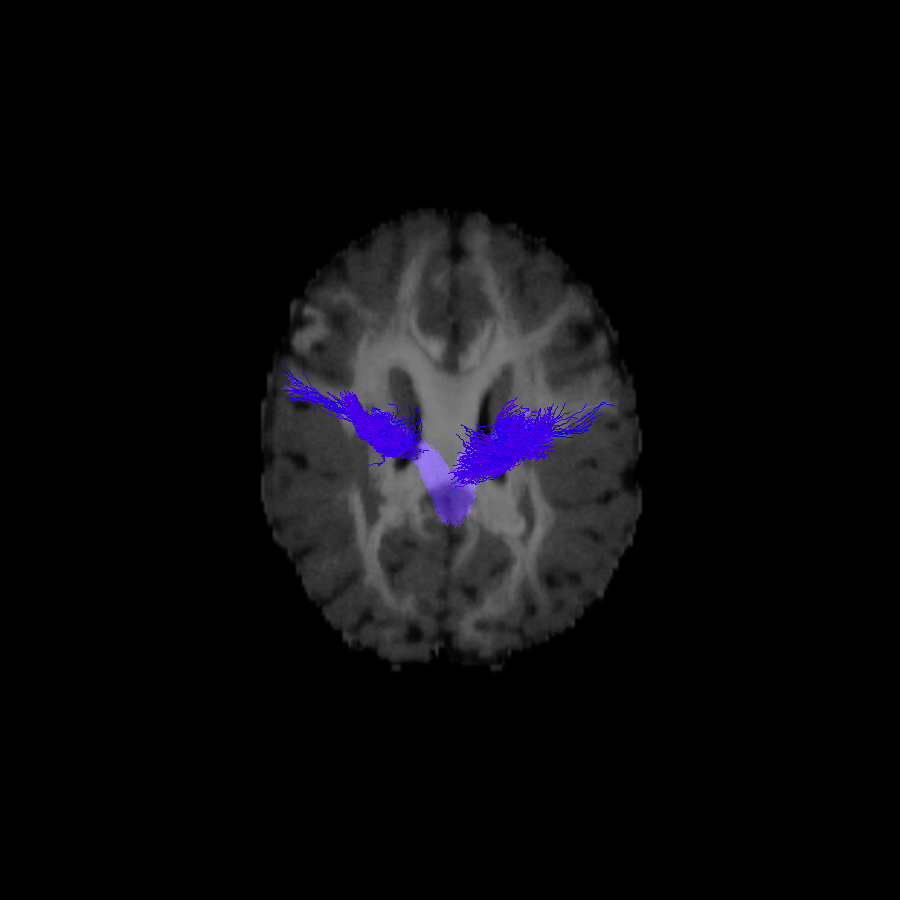} &
\includegraphics[scale=0.95, trim=0.5in 0.5in 0.5in 0.5in, clip=true, width=0.295\linewidth, keepaspectratio=true]{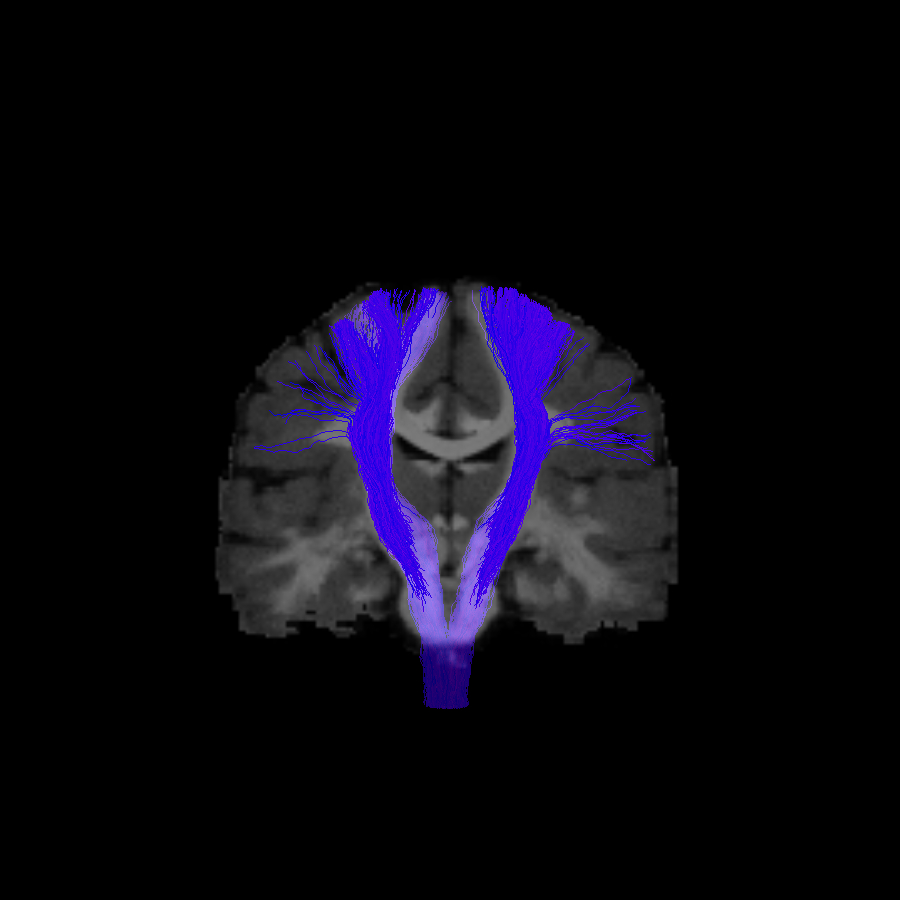} &
\includegraphics[scale=0.95, trim=0.5in 0.5in 0.5in 0.5in, clip=true, width=0.295\linewidth, keepaspectratio=true]{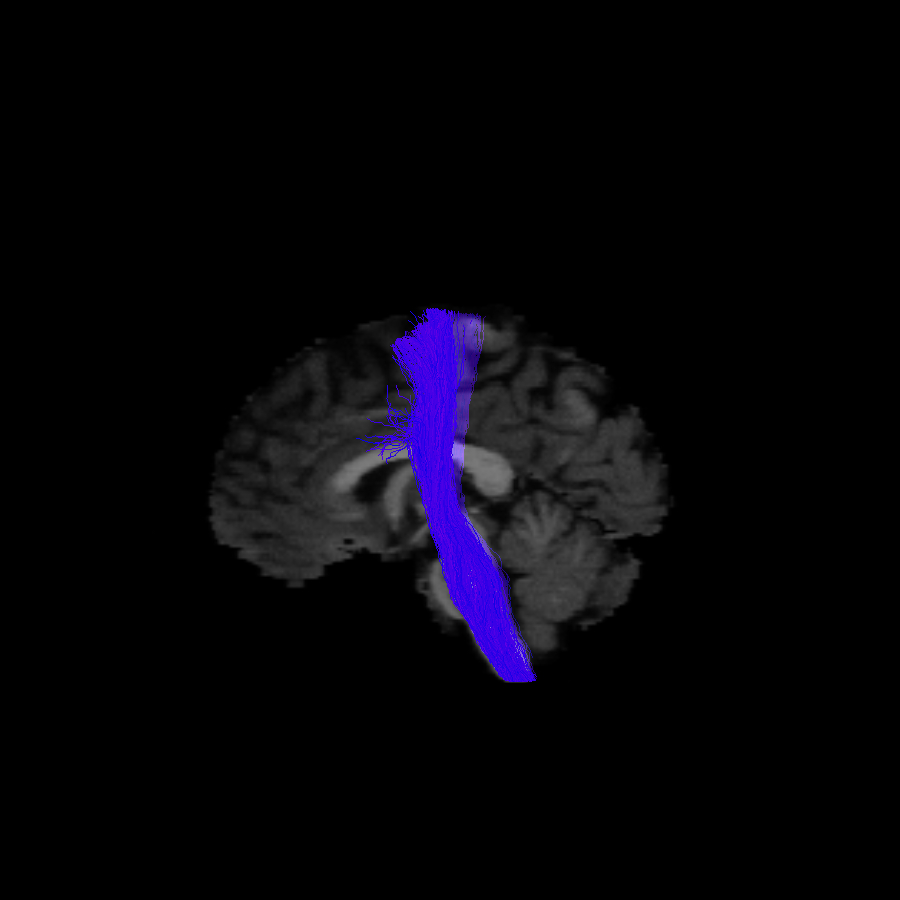} &
\includegraphics[scale=0.95, trim=2.15in 0.155in 0.015in 0.155in, clip=true, width=0.09155\linewidth, keepaspectratio=true]{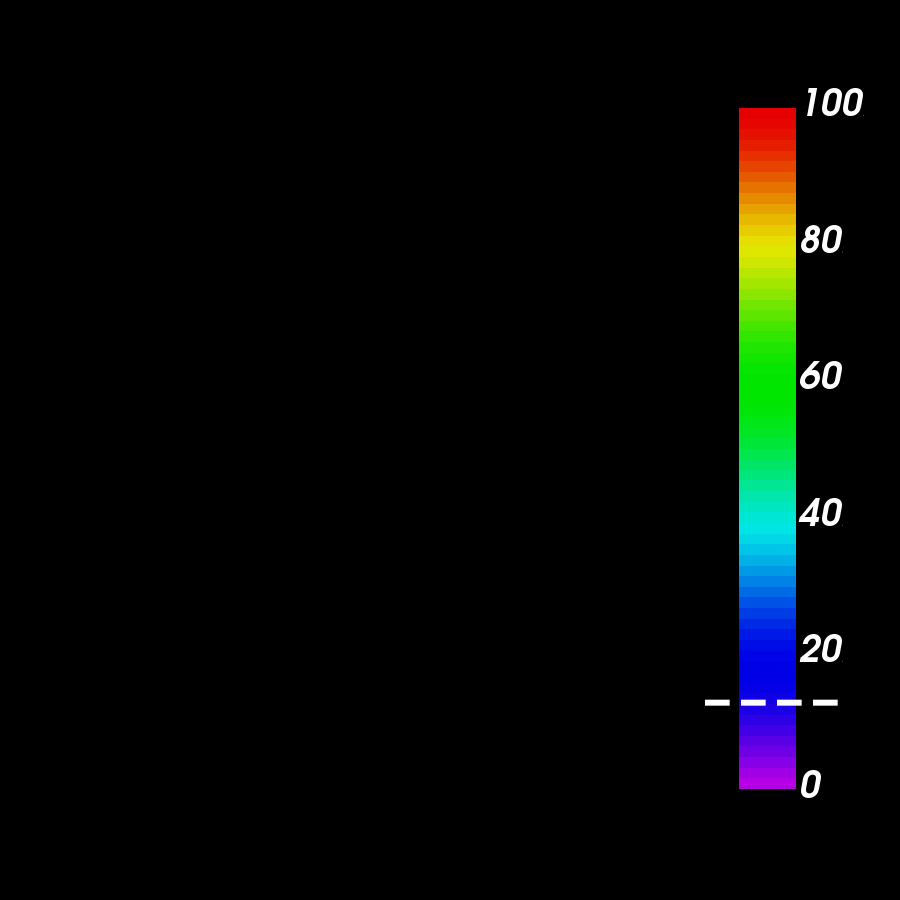} \\
\textbf{(a)} & \textbf{(b)} & \textbf{(c)} & \\
\includegraphics[scale=0.95, trim=0.55in 0.55in 0.55in 0.55in, clip=true, width=0.295\linewidth, keepaspectratio=true]{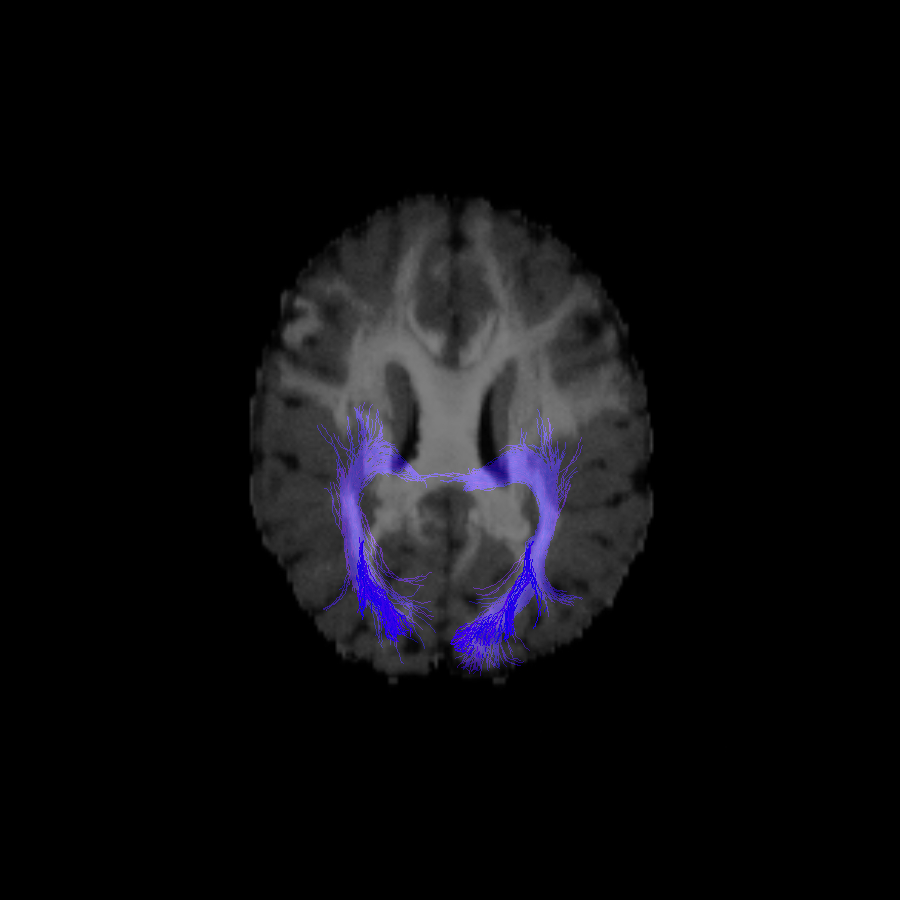} &
\includegraphics[scale=0.95, trim=0.5in 0.45in 0.5in 0.55in, clip=true, width=0.295\linewidth, keepaspectratio=true]{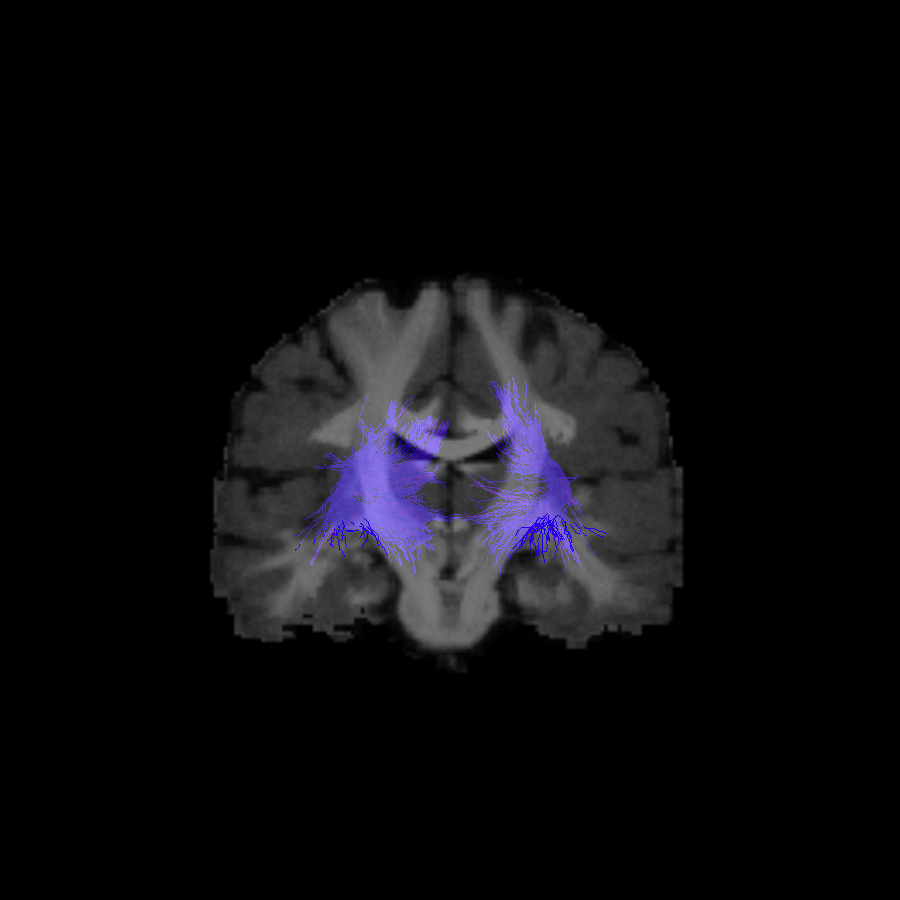} &
\includegraphics[scale=0.95, trim=0.5in 0.5in 0.5in 0.5in, clip=true, width=0.295\linewidth, keepaspectratio=true]{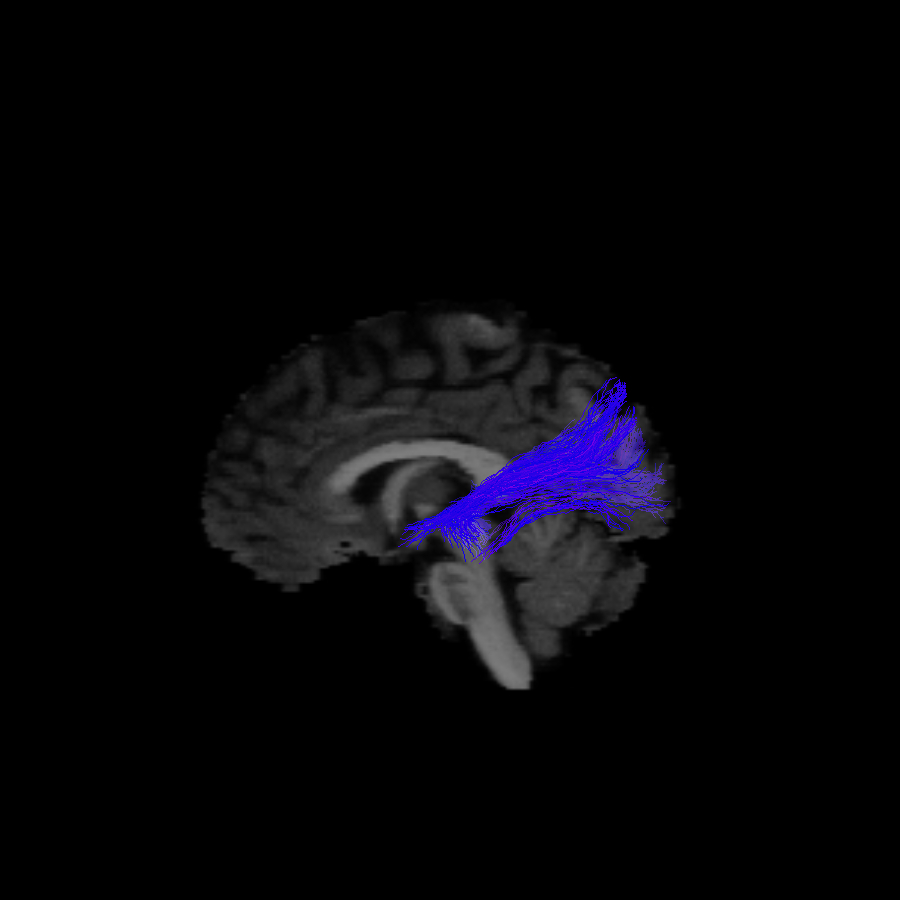} &
\includegraphics[scale=0.95, trim=2.15in 0.155in 0.015in 0.155in, clip=true, width=0.09155\linewidth, keepaspectratio=true]{./figures/results/ismrm_2015_tractography_challenge/ismrm_2015_tractography_challenge_test_data_colorbar_threshold.jpg} \\
\textbf{(d)} & \textbf{(e)} & \textbf{(f)} & \\
\includegraphics[scale=0.95, trim=0.55in 0.55in 0.55in 0.55in, clip=true, width=0.295\linewidth, keepaspectratio=true]{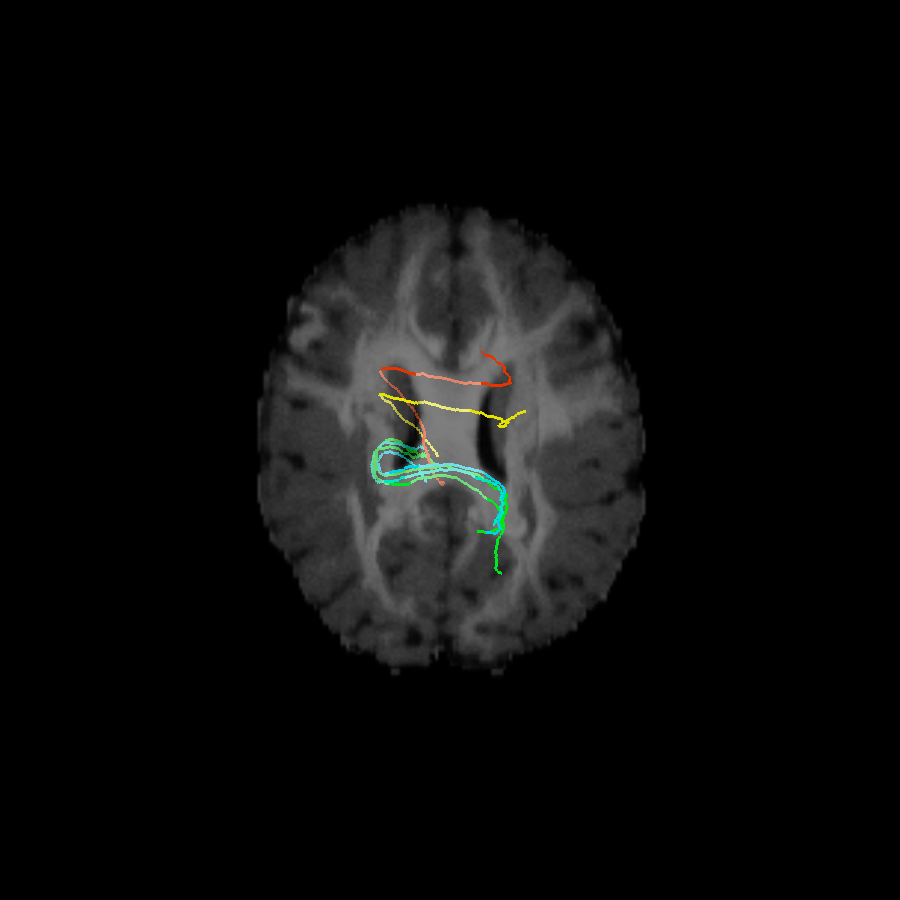} &
\includegraphics[scale=0.95, trim=0.5in 0.5in 0.5in 0.5in, clip=true, width=0.295\linewidth, keepaspectratio=true]{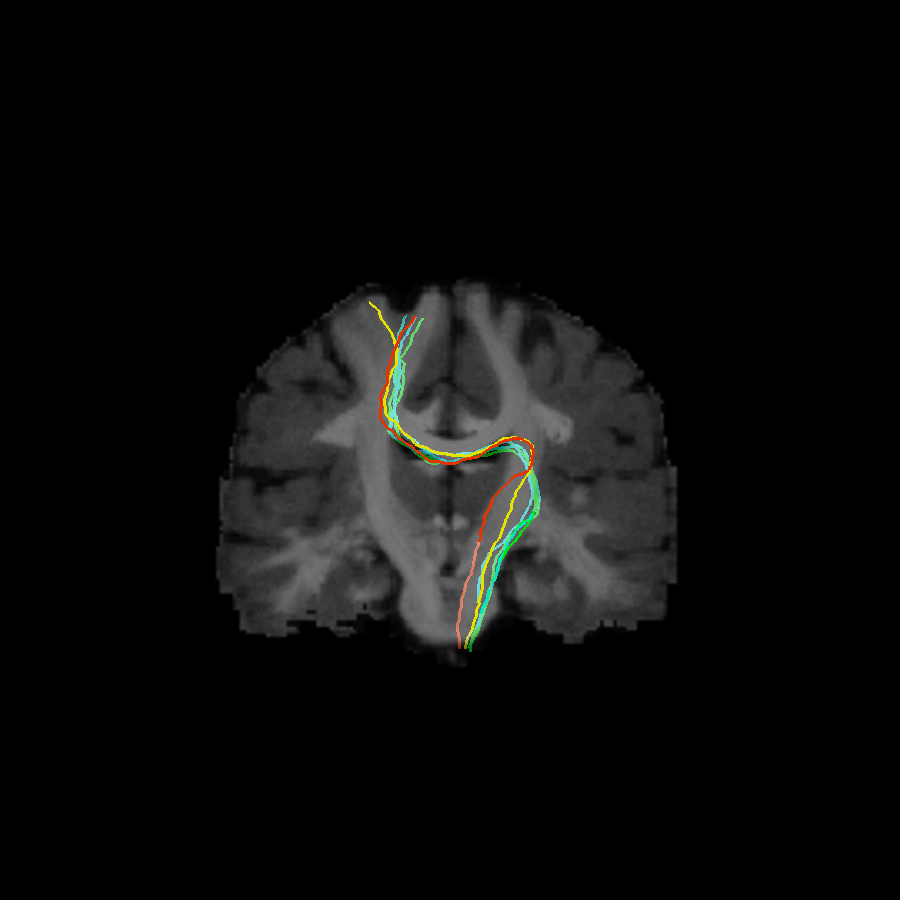} &
\includegraphics[scale=0.95, trim=0.5in 0.5in 0.5in 0.5in, clip=true, width=0.295\linewidth, keepaspectratio=true]{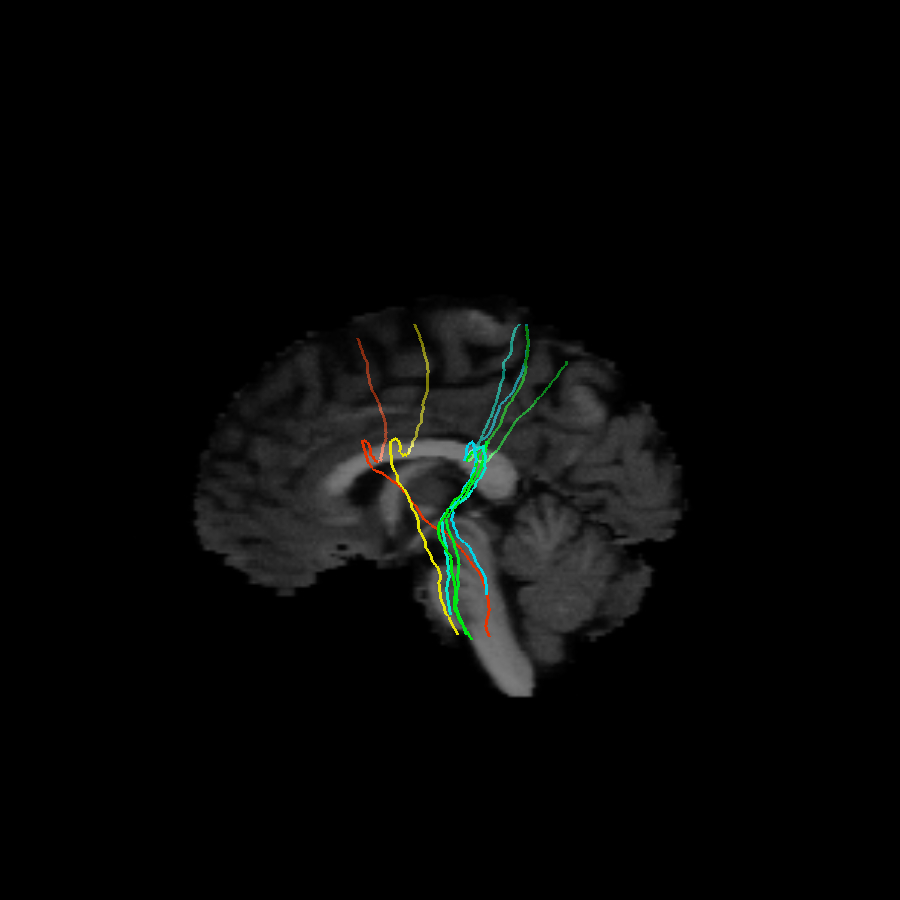} &
\includegraphics[scale=0.95, trim=2.15in 0.155in 0.015in 0.155in, clip=true, width=0.09155\linewidth, keepaspectratio=true]{./figures/results/ismrm_2015_tractography_challenge/ismrm_2015_tractography_challenge_test_data_colorbar_threshold.jpg} \\
\textbf{(g)} & \textbf{(h)} & \textbf{(i)} & \\
\end{tabular}
\caption{\label{fig:ismrm_2015_tractography_challenge_dataset_streamline_classification}Streamline classification predictions corresponding to some cases of the ISMRM 2015 Tractography Challenge dataset test set: (a), (b), (c) corticospinal tract (CST); (d), (e), (f) optic radiation (OR); (g), (h), (i) example corresponding to implausible streamlines. All axial superior, coronal anterior and sagittal left views. Streamlines are color-coded according to their nearest neighbor distance in the latent space. The dashed white lines on the scale bars mark the filtering threshold. The streamlines corresponding to the implausible class are shown with a larger diameter with respect to the one used for the rest of the figures.}
\end{figure*}

For qualitative visualization purposes, figure \ref{fig:ismrm_2015_tractography_challenge_dataset_streamline_classification} shows the streamline-wise classification tractograms for some selected cases corresponding to the ISMRM 2015 Tractography Challenge dataset test set, color-coded according to their nearest neighbor distance in the latent space. In particular, plausible predictions corresponding to the the corticospinal tract (CST) and the optic radiation (OR) bundles are shown, together with an example corresponding to implausible streamlines. As it can be drawn from the scale bar, the latent space dissimilarity distances for the true positive samples (a) to (f) are below the distance threshold determined shown in figure \ref{fig:latent_space_distance_distr}(b, d) ($13.634$) while the distances for the implausible predictions shown are above that threshold. The filtered whole brain tractogram is shown in \ref{subsec:predicted_positives_human_based_synthetic_data_results}.

Training on the ISMRM 2015 Tractography Challenge dataset lasted around 5 h. FINTA takes around $12$s to filter the corresponding test set tractogram.

\clearpage
\subsection{BIL\&GIN human data}
\label{subsec:bil_gin_human_data_results}
\begin{figure*}[!b]
\centering
\includegraphics[width=\linewidth]{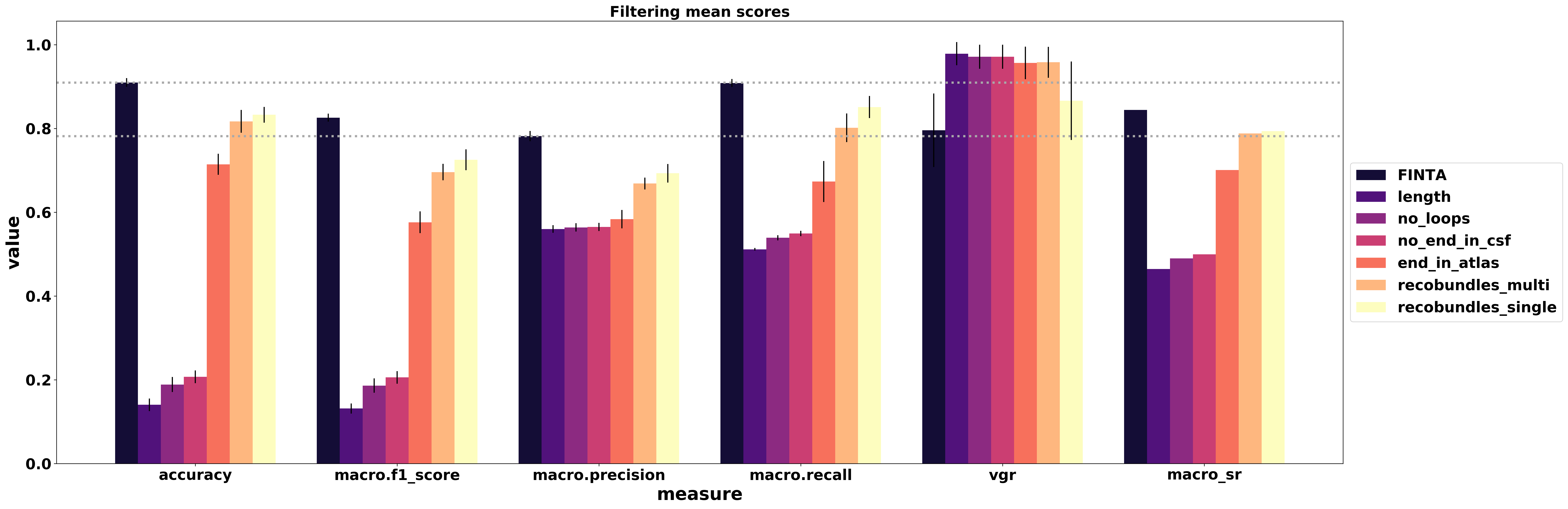}
\caption{\label{fig:bil_gin_dataset_classification_performance}Classification performance measures on the callosal BIL\&GIN dataset. For each of the compared filtering methods, the mean and standard deviation of the measures are presented. The horizontal dotted lines indicate FINTA's minimum and maximum scores across the measures.}
\end{figure*}

FINTA's filtering ability on in vivo human data is consistent with the results on synthetic datasets. The quantitative results for the comparison to other filtering methods are shown in table \ref{tab:bil_gin_dataset_results_macro}, and plot in figure \ref{fig:bil_gin_dataset_classification_performance}\footnote{The reader might notice that methods \#1 to \#4 in table \ref{tab:bil_gin_dataset_results_macro} do not have a perfect score of $1.0$ for the VGR measure. This is because such pipeline: (i) involves the use of steps and thresholding values that might differ from those used by the annotation tool; (ii) is affected by the effects of the registration; and (iii) uses an atlas that is slightly different from that of the annotator.}. As it can be drawn from these results, FINTA achieves a better classification performance than its competitors. Similarly, compared to any of the baseline methods, it is observed that FINTA shows a reduced variability in all measures consistently, with the sole exception of the valid gyrus-wise rate (VGR). Supplementary results of the comparison to the baseline filtering methods on the callosal BIL\&GIN dataset are reported in the appendix sections \ref{subsec:human_data_filtering_supplementary_results} and \ref{subsec:human_data_benchmarking_supplementary_results}.

The recovery of the existing anatomy, measured by the valid gyrus-wise rate, requires to be interpreted with care. Although results show that the anatomy-based baseline methods show a superior ability in this aspect, this comes at the expense of significantly lower scores for the rest of the measures (see section \ref{subsec:human_data_discussion} for a discussion). Yet, FINTA still achieves an average valid gyrus-wise rate of $80\%$, performing coherently across all measures.

\begin{table*}[!ht]
\caption{\label{tab:bil_gin_dataset_results_macro}Callosal BIL\&GIN dataset macro results. Macro mean (standard deviation) values over test subjects. The highest mean score is marked in bold face.}
\centering
\begin{tabular}{cccccccc}
\hline
& \textbf{Method} & \textbf{Accuracy\textsubscript{m}} & \textbf{Sensitivity\textsubscript{m}} & \textbf{Precision\textsubscript{m}} & \textbf{F1-score\textsubscript{m}} & \textbf{VGR} & \textbf{SR\textsubscript{m}}\\
\hline
\#1 & \textbf{length} & 0.14 (0.02) & 0.51 (0.0) & 0.56 (0.01) & 0.13 (0.01) & \textbf{0.98} (0.03) & 0.46 \\
\#2 & \textbf{no\_loops} & 0.19 (0.02) & 0.54 (0.01) & 0.56 (0.01) & 0.19 (0.02) & 0.97 (0.03) & 0.49 \\
\#3 & \textbf{no\_end\_in\_csf} & 0.21 (0.02) & 0.55 (0.01) & 0.57 (0.01) & 0.21 (0.01) & 0.97 (0.03) & 0.5 \\
\#4 & \textbf{end\_in\_atlas} & 0.71 (0.02) & 0.67 (0.05) & 0.58 (0.02) & 0.58 (0.03) & 0.96 (0.04) & 0.70 \\
\#5 & \textbf{recobundles\_single} & 0.83 (0.02) & 0.85 (0.03) & 0.69 (0.02) & 0.73 (0.02) & 0.87 (0.1) & 0.79 \\
\#6 & \textbf{recobundles\_multi} & 0.82 (0.03) & 0.8 (0.03) & 0.67 (0.01) & 0.7 (0.02) & 0.96 (0.04) & 0.79  \\
\#7 & \textbf{FINTA} & \textbf{0.91} (0.01) & \textbf{0.91} (0.01) & \textbf{0.78} (0.01) & \textbf{0.83} (0.01) & 0.8 (0.09) & \textbf{0.84}  \\
\end{tabular}
\end{table*}

\begin{figure*}[!t]
\centering
\begin{tabular}{ccc}
\includegraphics[scale=0.95, trim=0.5in 0.45in 0.5in 0.55in, clip=true, width=0.3\linewidth, keepaspectratio=true]{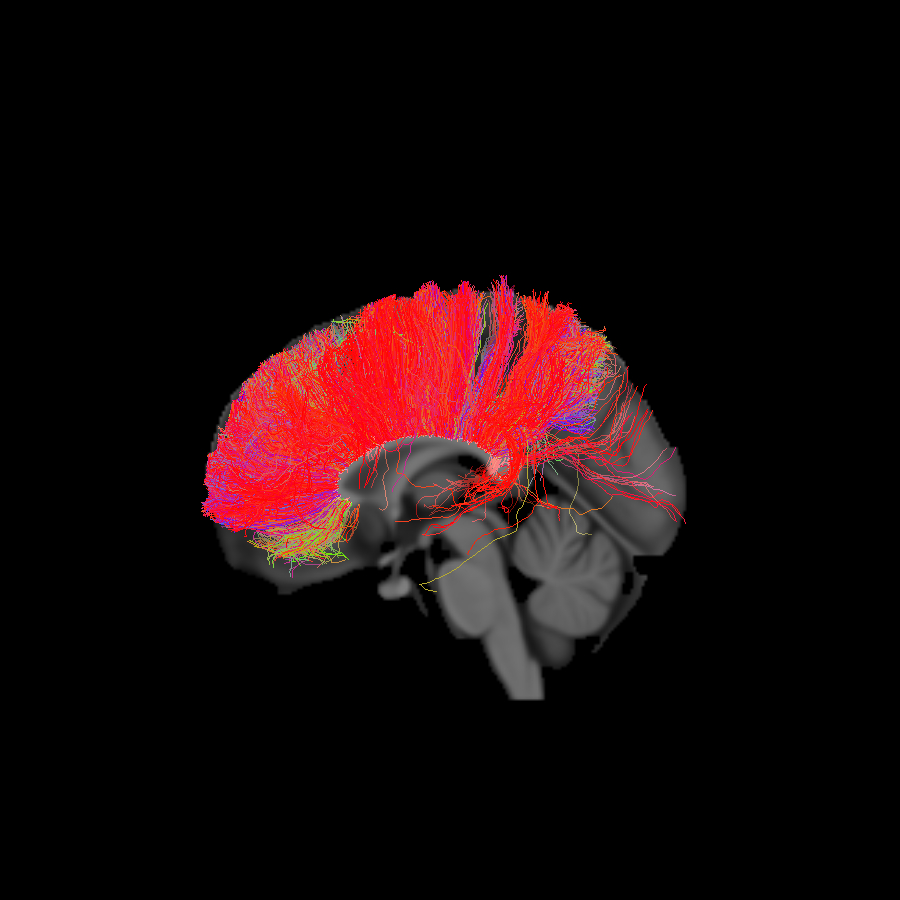} &
\includegraphics[scale=0.95, trim=0.5in 0.45in 0.5in 0.55in, clip=true, width=0.3\linewidth, keepaspectratio=true]{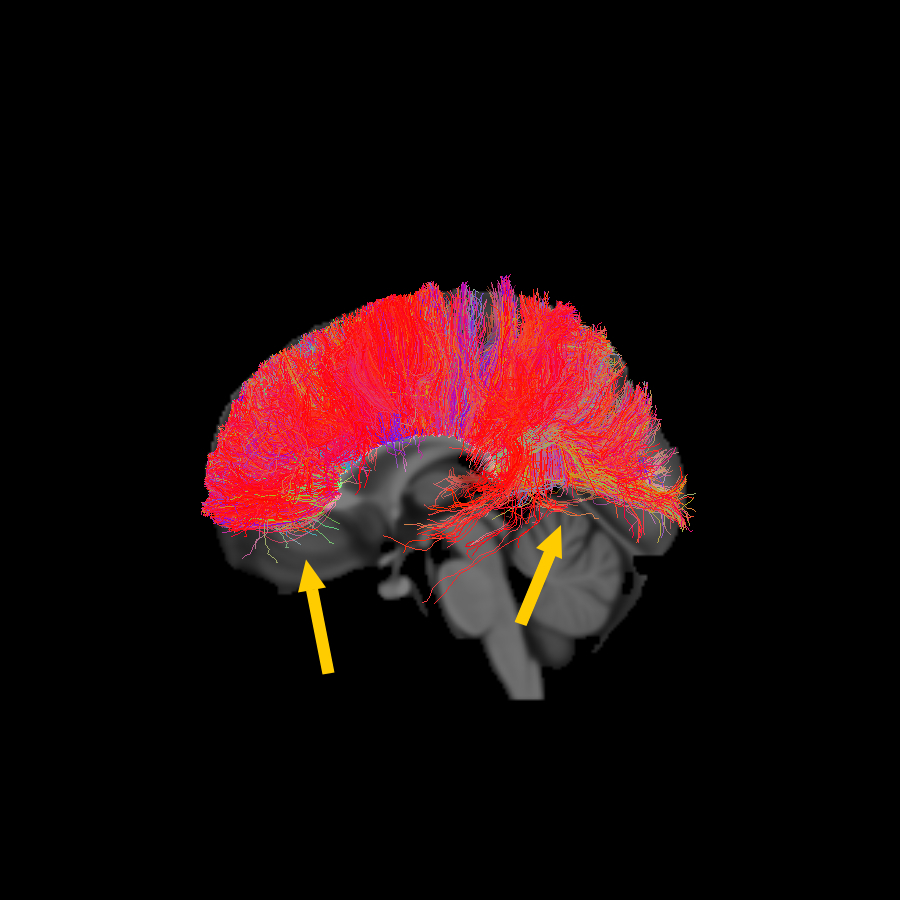} &
\includegraphics[scale=0.95, trim=0.5in 0.45in 0.5in 0.55in, clip=true, width=0.3\linewidth, keepaspectratio=true]{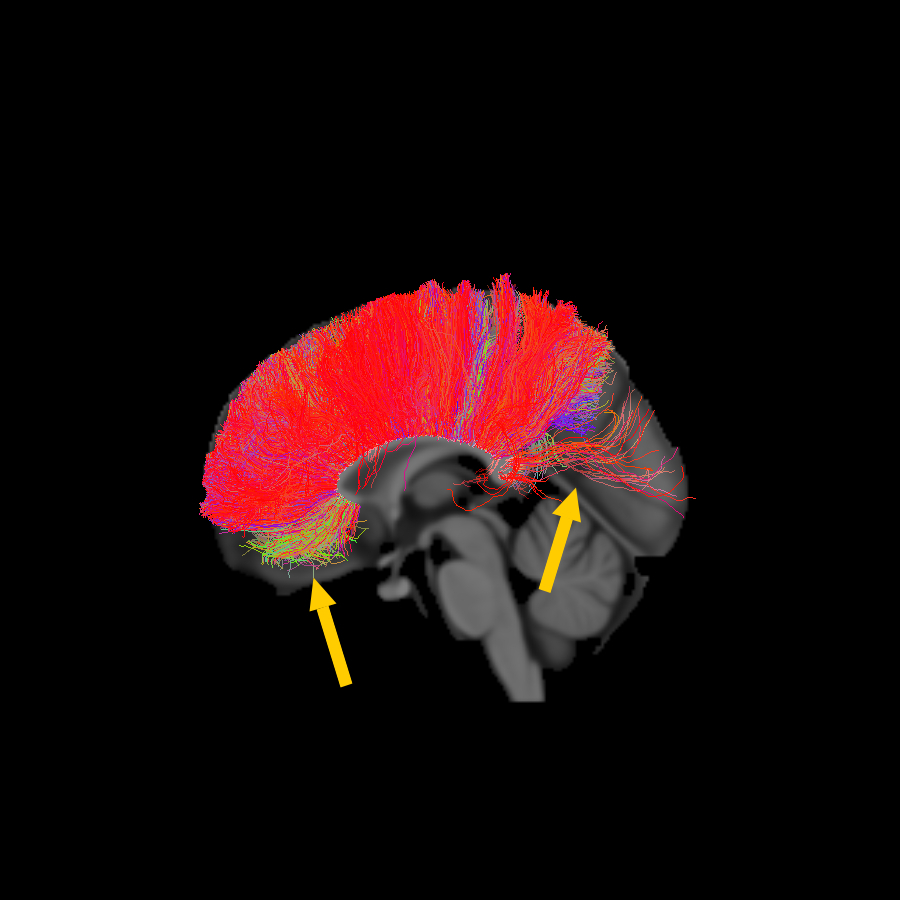} \\
\textbf{(a)} & \textbf{(b)} & \textbf{(c)} \\
\end{tabular}
\caption{\label{fig:bil_gin_dataset_reference_predicted_positives}Predicted positives on a randomly picked test subject corresponding to the callosal BIL\&GIN dataset. Predicted positives represent the passband of the filtering process. (a) reference streamlines (homotopic connections); (b) single-bundle RecoBundles; (c) FINTA. All sagittal left views. The pointed regions show areas where FINTA's filtered tractogram is qualitatively closer to the reference tractogram compared to the single-bundle RecoBundles baseline method.}
\end{figure*}

Qualitatively, figure \ref{fig:bil_gin_dataset_reference_predicted_positives} shows that, compared to FINTA, the single-bundle RecoBundles method misses streamlines in the anterior, ventral part of the frontal lobe, and includes false positives in the occipital lobe (pointed regions).

Table \ref{tab:bil_gin_dataset_classification_performance_regionwise} reports a gyrus-wise analysis on the callosal BIL\&GIN dataset of the sensitivity measure on the predicted positives for FINTA. The reported data correspond to the averaged values over the test subjects. Given that the callosal BIL\&GIN dataset gyral-based segments allow solely to extract homotopic streamlines, only the sensitivity is reported (see the appendix section \ref{subsec:bil_gin_regionwise_analysis} for further details). The gyral-based Hippo and TPole segments are not reported since none of the test subjects contain plausible streamlines in such segments. Similarly, the mean streamline count being rounded to the nearest integer, some segments contain $0$ streamlines on average (IOG, Ins, LFOG, MFOG, and PHG) and hence are not reported. Also, note that some others have a very low (as low as $1$) streamline count, and thus missing a few streamlines bears a significant impact on the results. That is the case, for example, for the FuG segment, where $2$ of the test subjects have both $2$ streamlines, and another $2$ have both $1$ streamline, the remaining $4$ lacking streamlines in that segment. As seen in the table, the sensitivity values for these are noticeably lower than for the segments containing a high streamline density, such as the MFG, PrCu, or SFG, all of the latter being above $86\%$.

\begin{table}[!htbp]
\caption{\label{tab:bil_gin_dataset_classification_performance_regionwise}Gyral segment-wise callosal BIL\&GIN dataset sensitivity of FINTA. Mean (standard deviation; [min, max]) values over test subjects for FINTA. The \textit{Count} column contains the mean streamline count for each segment across the test subjects.}
\centering
\begin{tabular}{rcl||rcl}
\hline
\textbf{Segment} & \textbf{Count} & \textbf{Sensitivity} &  \textbf{Segment} & \textbf{Count} & \textbf{Sensitivity} \\
\hline
\textbf{AG} & 14 & 0.14 (0.14) [0, 0.4] & \textbf{PoCG} & 122 & 0.98 (0.03) [0.91, 1] \\
\textbf{Cing} & 23 & 0.81 (0.12) [0.63, 0.96] & \textbf{PrCG} & 177 & 0.98 (0.01) [0.95, 1] \\
\textbf{Cu} & 6 & 0.08 (0.11) [0, 0.3] & \textbf{PrCu} & 978 & 0.88 (0.05) [0.77, 0.94] \\
\textbf{FuG} & 1 & 0 (0) [0, 0] & \textbf{RG} & 87 & 0.99 (0.01) [0.98, 1] \\
\textbf{IFG} & 12 & 0.5 (0.23) [0, 0.74] & \textbf{SFG} & 3428 & 0.93 (0.01) [0.9, 0.95] \\
\textbf{ITG} & 1 & 0 (0) [0, 0] & \textbf{SMG} & 10 & 0.64 (0.18) [0.45, 1]\\
\textbf{LG} & 4 & 0.2 (0.28) [0, 0.75] & \textbf{SOG} & 23 & 0.52 (0.14) [0.25, 0.71]\\
\textbf{MFG} & 274 & 0.87 (0.28) [0.77, 0.93] & \textbf{SPG} & 142 & 0.67 (0.22) [0.23, 0.91] \\
\textbf{MOG} & 9 & 0.21 (0.17) [0, 0.41] & \textbf{STG} & 5 & 0.1 (0.17) [0, 0.5] \\
\textbf{MTG} & 4 & 0 (0) [0, 0] & \\
\end{tabular}
\end{table}

FINTA was trained for 24 h on the callosal BIL\&GIN dataset. FINTA takes around $21$s, on average, to filter a callosal tractogram corresponding to a BIL\&GIN test subject. The single-bundle RecoBundles baseline method takes, on average, $66$s, in our experiments.

\subsection{Generalization}
\label{subsec:generalization_results}
Table \ref{tab:generalization_classification_performance} shows the generalization ability of FINTA according to the experimental setting described in section \ref{subsec:generalization_experimental}. Results reveal that the filtering performance decreases when using the generalized threshold value with respect to the results reported in table \ref{tab:dataset_classification_performance}. Computing a dataset-specific threshold notably improves the filtering ability of the method.

\begin{table}[!htb]
\caption{\label{tab:generalization_classification_performance}Filtering performance ability on generalization.}
\centering
\begin{tabular}{c|cc||cc}
\hline
& \multicolumn{2}{c}{\textbf{ISMRM 2015 Tractography Challenge}} & \multicolumn{2}{c}{\textbf{Human Connectome Project}} \\
& \multicolumn{2}{c}{\textbf{Deterministic tracking}} & \multicolumn{2}{c}{\textbf{Global tracking}} \\
\textbf{Measure} & \textbf{Generalized thr} & \textbf{Specific thr} & \textbf{Generalized thr} & \textbf{Specific thr} \\
\hline
Accuracy & 0.8 & 0.94 & 0.95 & 0.93 \\
Sensitivity & 0.81 & 0.94 & 0.67 & 0.92 \\
Precision & 0.82 & 0.94 & 0.85 & 0.72 \\
F1-score & 0.8 & 0.94 & 0.73 & 0.78 
\end{tabular}
\end{table}

\subsection{Time requirements}
\label{subsec:time_requirements_results}
\begin{figure*}[!htb]
\centering
\includegraphics[scale=0.95, width=0.6\linewidth]{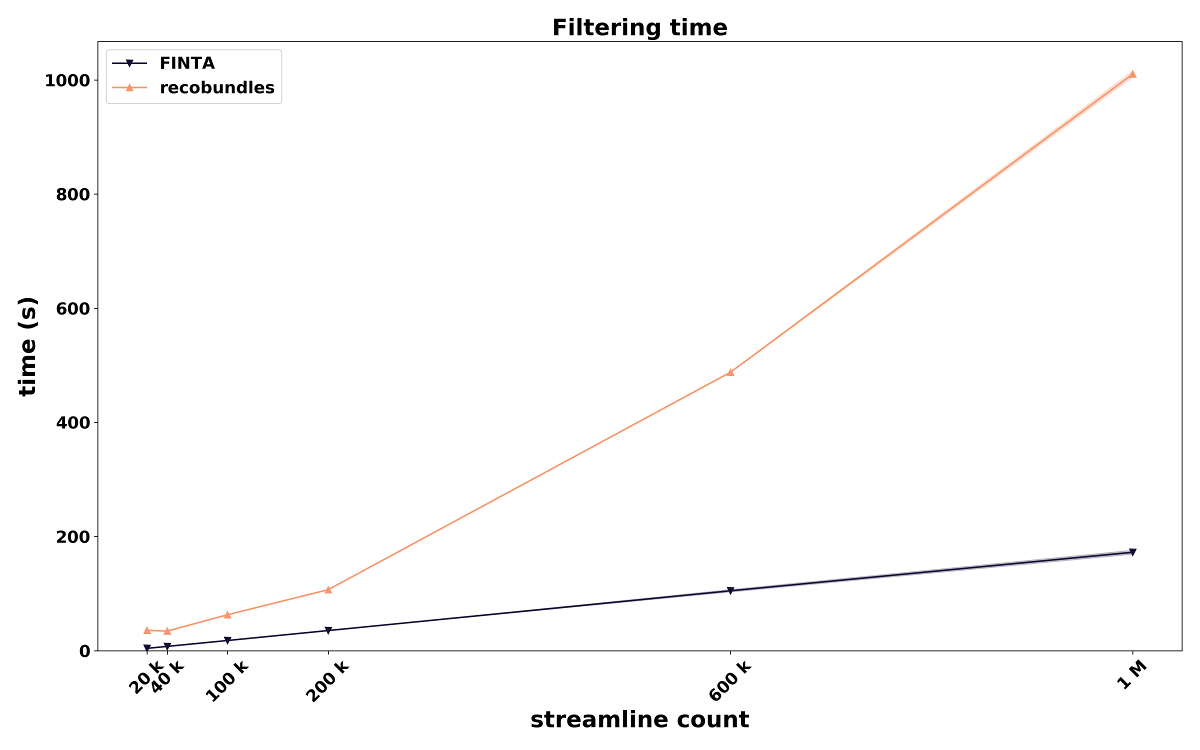}
\caption{\label{fig:ismrm_2015_tractography_challenge_filtering_time_stats}Filtering time requirements for FINTA and RecoBundles. The curves demonstrate that FINTA is fast, having linear time complexity with the streamline count, while RecoBundles requires a significantly longer time to filter a given tractogram. Note that due to the vertical scale and reduced standard deviation values, the latter are hardly noticeable around the mean value. Similarly, note that the streamline counts are expressed with SI prefixes and engineering notation.}
\end{figure*}

Figure \ref{fig:ismrm_2015_tractography_challenge_filtering_time_stats} compares the time required by FINTA and RecoBundles to filter tractograms of different sizes corresponding to the ISMRM 2015 Tractography Challenge data. For each method, the curves account for the mean filtering time (standard deviation values are below $10$s in all cases). FINTA requires less time than RecoBundles to filter a tractogram of the same size. Also, it follows from the curves that FINTA is linear in terms of the streamline count in a tractogram, whereas RecoBundles requires a time that tends to be several order of magnitudes longer than FINTA as the streamline count increases.

\section{Discussion}
\label{sec:discussion}
The results across the three different datasets show that FINTA is a state-of-the-art alternative for tractography filtering. The latent space views suggest that streamlines belonging to the same class lie close to each other in the latent space, supporting the hypothesis that the autoencoder successfully clusters such streamlines in an unsupervised manner. The proposed filtering method achieves good performance statistics on both synthetic and in vivo human brain data, and consistently ranks higher in the considered measures compared to the baseline methods.

\subsection{Performance on synthetic datasets}
\label{subsec:synthetic_data_discussion}
Although less convoluted than a tractogram derived from a human brain dataset, the ``Fiber Cup'' dataset still presents some challenges to tractography filtering methods, and hence it offers a good test bed to show the behavior of a given method. Our results show that FINTA provides almost perfect performances on it. Additionally, the dataset allows to gain further insight into the behavior of a given filtering method. Our findings illustrated in figure \ref{fig:fibercup_dataset_streamline_classification_falses} suggest that: (1) false positives and false negatives generated by FINTA are very close to true positives and true negatives; and (2) in some cases, FINTA can reveal examples of streamlines that were misclassified by the scoring method. Such inconsistencies with respect to the reference highlight the ability our proposed method has to exploit the learned features from the streamlines. At the same time, it is a demonstration of some potential limitations of conventional filtering approaches, which employ voxel-based or anatomical criteria, and pair-wise streamline dissimilarity measures that parallel those employed by existing scoring methods.

The slight decrease in performance registered by FINTA on the ISMRM 2015 Tractography Challenge (also seen in the human in vivo callosal BIL\&GIN dataset) compared to the ``Fiber Cup'' is explained by the complexity of the former, where a larger variety of bundle and streamline configurations are present. Nonetheless, with a mean accuracy of $91\%$, our method still gets an excellent score. Together, the results show that there is a good agreement between the performance on the ISMRM 2015 Tractography Challenge and callosal BIL\&GIN datasets, suggesting that these results are fully descriptive of the power of the method on human in vivo tractogram settings.

\subsection{Performance on human data}
\label{subsec:human_data_discussion}
Although the anatomy-based tractography filtering methods (methods \#1 to \#4) were expected to record lower scores than their competitors, they are still an important part of routine tractography filtering pipelines prior to downstream tasks, such as connectivity studies \citep{Jeurissen:NMRBiomed:2019, Yeh:JMRI:2020}. No additional steps, such as obtaining the gyral-based homotopic streamlines, were added to avoid reproducing the scoring method providing the reference.

Classical filtering methods obtain a better valid gyrus-wise rate than FINTA at the cost of including an excessive number of false positives among the kept streamlines. Penalizing more severely missing groups would have had some impact on the reported success rates without harm to the rest of the filtering performance measures. Similarly, the superior ability of the RecoBundles bundle recognition methods to recover the existing anatomy may be explained by the use of gyrus-wise homotopic callosal streamline models specifically built for that purpose. The single-bundle RecoBundles obtained slightly better scores than the multi-bundle version except for the valid gyrus-wise rate measure. This might be explained by a greed of the multi-bundle version to assimilate streamlines to every model of the corpus callosum gyral-based segment pairs, leading it to an excessive number of false positives. Compared to FINTA, the improved valid gyrus-wise rate of the multi-bundle RecoBundles baseline comes at the cost of potentially recognizing bundles that are not contained in the data (i.e. invalid bundles).

Note that the valid gyrus-wise rate does not account for the potential invalid (i.e. not existing in the underlying anatomy) groups that some of the methods may recover. It must be noted that the $26$ pairs of gyral-based callosal segments (see the appendix section \ref{subsec:considered_bil_gin_cc_regions}) are representative of the entire BIL\&GIN population, but subjects may be missing streamlines on an entire segment. Furthermore, the streamline count on some segments may be as low as a single streamline for some subjects, which may impact considerably the valid gyrus-wise rate measure.

In this work, true positives and false positives were evenly weighted to find the optimal filtering threshold. We found that maximizing only the accuracy penalized excessively the least numerous class in a class imbalance situation. Likewise, the results in figure \ref{fig:bil_gin_dataset_classification_performance} show that this strategy leads to a high specificity, and allows to keep false positives at low rates in FINTA. Meanwhile, the rest of methods suffer from an excessive number of false positives. Note that this effect is observed consistently in the measures where false positives are involved (accuracy, F1-score, and precision). This finding is relevant for connectome studies, where keeping a good specificity is essential \citep{Zalesky:Neuroimage:2016}.

The reduced variability in the measures shown by FINTA suggests that the method is less prone to be influenced by the local, potentially noisy streamline trajectories. Conventional methods' ability to filter streamlines is determined by local decisions based on either local (e.g. curvature, regions of interest, endpoint location) or non-local (e.g. length) features, and thus a single streamline coordinate not complying with the given criterion (e.g. falling outside a structural mask) is enough to filter out the streamline. This reaffirms the ability of FINTA to encode those features of the input data which are most relevant for an accurate reconstruction.

The corpus callosum spans a large part of the brain volume in the anterior-posterior axis across both hemispheres; it is one of the largest fiber systems in the white matter, both in terms of the absolute number of streamlines and the occupied brain volume. Compared to some other white matter systems, it presents a less compact configuration. As a consequence, it is challenging to precisely delineate the corpus callosum homotopies in the reference set \citep{DeBenedictis:HBM:2016}. Similarly, as previously noted, the $26$ gyral segment pairs used in this work differ greatly in the streamline density they hold, and thus bear an impact on the gyrus-wise sensitivity scores shown in table \ref{tab:bil_gin_dataset_classification_performance_regionwise}. Yet, FINTA records high classification performance scores, and shows a reduced variability in the valid gyrus-wise rate measure.

The performance assessment on human data was restricted by the nature of the callosal BIL\&GIN dataset and its corresponding ground truth (see section \ref{subsec:considered_bil_gin_cc_regions} for the relevant aspects). This constrained the baseline filtering methods, excluding some state-of-the-art techniques such as COMMIT2 or SIFT2. However, the baselines used are commonly found in current white matter analysis practice.

\subsection{General considerations}
\label{subsec:general_considerations}
As opposed to the conventional filtering methods that assume that the estimated streamline population at every voxel should be proportionally supported by the diffusion data (which is the case of COMMIT/COMMIT2, LiFE, or SIFT/SIFT2), our autoencoder works on tracking data only. This makes it immune against well-known domain adaptation issues~\citep{Perone:Neuroimage:2019}. Furthermore, compared to a quadratic or super-quadratic complexity of some methods, FINTA is linear in terms of the streamline count at test time as demonstrated by the results in figure \ref{fig:ismrm_2015_tractography_challenge_filtering_time_stats}.

Compared to other deep learning approaches that may be specifically oriented to classification tasks (e.g. regular classification convolutional neural networks), FINTA offers the benefit of using an unsupervised learning approach. Thus, it does not depend on the number of classes in the input data as the output of the network does not look for maximizing the probability of a given class among the possible ones. As such, the method is better suited to the reality of tractography, where there is a limited knowledge about the ground truth \citep{Yeh:Neuroimage:2016, Sotiropoulos:NMRBiomed:2019}, and where the analysis lends itself to different organizational levels, and thus different classification degrees. As such, our method naturally adapts to new sets of classes without the need of having to retrain the network.

A potential side-benefit of the proposed autoencoder approach is that the reconstructed streamlines describe a locally smooth trajectory. Although not explicitly used or exploited in the present work, succeeding pipeline or visualization processes may benefit from it, providing a less complex or more realistic long-range apparent fiber trajectory representation.

Downstream tractography tasks benefit from filtering strategies applied at earlier stages when done in a coherent manner. One of such tasks is the structural connectivity analysis (or connectomics). Any proposed filtering strategy to reject implausible streamlines is expected to preserve the existing connectivity. Estimating the accuracy of a connectome yielded by a tractography and filtering pipeline can only be done faithfully on synthetic data, and yet many confounding factors (e.g. parcellation and discretization effects), including those from upstream steps (such as the seeding method), are involved. Conceding such limitations to an analysis, we provide evidence in section \ref{subsec:connectivity_analysis} that FINTA is able to regenerate the connectome to a high degree of agreement with the ground truth on the synthetic datasets used in this work.

\subsection{Generalization, streamline diversity and balance effects}
\label{subsec:streamline_diversity_balance_effects}
FINTA's reported global performances stress the fact that the proposed framework is robust across a varying spectrum of settings. The proposed autoencoder-based method assumes that, provided that the implausible streamlines have overall distinctive features from the plausible ones, they will be cast to different regions in the latent space, and hence the filtering framework will be able to separate them, regardless of their location and arrangement in the native space or within-fascicle balance. Similarly, as highlighted in section \ref{subsec:latent_space}, the autoencoder shows the ability to disentangle a varying number of bundles or streamline groups.

The results in section \ref{subsec:generalization_results} show that FINTA's optimal filtering threshold computation framework obtains best results when tailoring its value to the dataset at issue. Otherwise, a trade-off value can be proposed by finding a threshold on data from a variety of tracking settings (e.g. probabilistic, deterministic, global, etc.) and datasets. It is worthwhile noting that the autoencoder was not retrained across datasets for the generalization experiments. These results show how immune against domain adaptation problems FINTA is.

The datasets used in this work varied greatly in the overall plausible \textit{versus} implausible streamline density: the ``Fiber Cup'' dataset had an overall plausible/implausible ratio of $21/79\%$, whereas the ISMRM 2015 Tractography Challenge settings had $54/46\%$ (probabilistic) and $57/43\%$ (deterministic) respectively. Meanwhile, the callosal BIL\&GIN dataset had an $11/89\%$ ratio, and the Human Connectome Project global tracking data showed a $6/94\%$ ratio.

It can be hypothesized that the unsupervised optimization training phase and the sensitivity \textit{versus} specificity trade-off used to set the global filtering threshold might have a larger impact on the performance on the least numerous class. Particularly, this effect is visible in the global tracking setting of the Human Connectome Project dataset (table \ref{tab:generalization_classification_performance}), where the performance might be influenced by the misclassifications in the plausible streamlines. Nevertheless, even in the unfavorable conditions of the callosal BIL\&GIN dataset, the weighted performance scores reported in tables \ref{tab:bil_gin_dataset_results_weighted} and \ref{tab:bil_gin_dataset_results_weighted_min_max} for the callosal BIL\&GIN dataset show that FINTA successfully filters implausible streamlines when accounting for the class imbalance.

\subsection{How good is the ground truth in tractography?}
\label{subsec:ground_truth_tractography}
Inter-subject variability, and the lack of a ``gold standard'' to reliably determine bundles on in vivo human brain data \citep{Rheault:JNeuralEng:2020} add a degree of uncertainty to tractography and downstream tasks. Our work demonstrates the ability of a 1D convolutional autoencoder to reveal uncertainties inherent to current state-of-the-art tractography filtering methods. Our filtering method is a powerful tool to uncover disparities and biases in current data scoring practices that can otherwise go unnoticed. The insight gained from such a model can be useful to build upon existing knowledge in tractography. Re-defining some of the current tractography filtering approaches may be essential when dealing with human data, where consensus on the best predictors of the underlying white matter anatomy is still hard to achieve. Together, this also highlights the need of a larger set of validated, shared data to converge towards a more quantitative tractography approach.

\subsection{Future work}
\label{subsec:future_work}
The proposed architecture requires all streamlines to have an equal number of points. A limitation of this work is that this number is larger than the minimal number of points required to express a streamline in regular tractography pipelines \citep{Presseau:Neuroimage:2015}. Further work is needed to determine a more efficient characterization of a streamline in terms of a representation learning framework. Following from this, the proposed framework would also benefit from investigating alternatives to avoid the streamline flipping pre-processing step. Note that we did not study the effect of the number of parameters of the network, and only focused on the optimization of the reconstruction accuracy. Additional investigations are required to quantify the impact of the network size on the reconstructed streamlines and filtering performance.

We did not investigate the streamline or tractography features the autoencoder-based approach focuses on. Further work is required to gain more insight on the data aspects that the model is able to capture or tends to discard. This is still an active area of research in deep learning, where many efforts are being put to explain how a model behaves in terms of explainable concepts related to the data and its features. This would allow, among others, to estimate reliably how a given ground truth labeling constraint may affect FINTA's performance.

Additional work includes confirming the robustness of the method against particular anatomical conditions. This involves, for example, measuring the ability of our method to generalize on infant brain tractography data or on brain data presenting a condition, such as a tumor, potentially affecting the tractogram distribution, without having to re-train the system. These experiments so could lead to further classification categories or to the need of a conditional mechanism attached to the autoencoder.

The current findings on the effects of the proposed filtering method in downstream tractography tasks have a limited scope. FINTA records low false positive rates, which is paramount to connectome analyses \citep{Zalesky:Neuroimage:2016}, and successfully reconstructs the connectome on the synthetic datasets. However, misses in sparse streamline fascicles penalize its performance on the callosal BIL\&GIN human brain \textit{in vivo} dataset. Further work would be required to improve FINTA's ability to preserve the white matter fascicle support in complex tractography settings by possibly incorporating additional a priori information into the autoencoder learning framework.

\section{Conclusions}
\label{sec:conclusions}
We have proposed an unsupervised, deep learning-based dimensionality reduction method for dMRI tractography filtering. We have dubbed our method FINTA, {\em Filtering in Tractography using Autoencoders}. We have shown that, working with a simple architecture, FINTA is able to robustly learn the structure of streamlines in tractography, and have applied such a framework to discriminate between an input tractogram's ``positive'' and ``negative'' streamlines. These categories may target plausible and implausible streamlines or a subset of bundles \textit{versus} the rest of the tractogram. For the plausible \textit{versus} implausible setting, once the neural network has been trained on unlabeled tractograms, simply labeling a subset of the streamlines of interest to obtain the partitioning threshold suffices to use this value to filter new tractograms using FINTA. We have demonstrated that the method allows to successfully and reliably filter tractograms in this context with accuracies above the $90\%$ bar, obtaining improved scores compared to state-of-the-art filtering methods. To this end, we have proposed a means for redefining the notion of inter-streamline distance which can be similarly used for any streamline filtering tasks without the need of having to re-train the network. Additionally, we have shown that the proposed convolutional neural network autoencoder generalizes effectively under a number of varying tractography settings, such as partial tractograms or different tracking methods across datasets.

The proposed FINTA framework offers a fast, robust, data-driven approach for tractography filtering. Its unique characteristics unfold the potential of autoencoders to improve the accuracy and reliability of downstream tractometry and connectomics derivatives. Thus, it may assist neuroanatomists in better describing the white matter anatomy using tractography, and hence bring a significant positive impact to research in neuroscience. FINTA is expected to be integrated into a specialized diffusion MRI analysis open source toolkit.

\section*{Acknowledgments}
This work has been partially supported by the Centre d'Imagerie M\'{e}dicale de l’Universit\'{e} de Sherbrooke (CIMUS); the Axe d'Imagerie M\'{e}dicale (AIM) of the Centre de Recherche du CHUS (CRCHUS); the R\'{e}seau de Bio-Imagerie du Qu\'{e}bec (RBIQ)/Quebec Bio-imaging Network (QBIN) (FRSQ - R\'{e}seaux de recherche th\'{e}matiques File: 35450); and the Samuel de Champlain 2019-2020 Program of the Conseil Franco-Qu\'{e}b\'{e}cois de Coop\'{e}ration Universitaire (CFQCU). This research was enabled in part by support provided by Calcul Qu\'{e}bec (\href{https://www.calculquebec.ca/en/}{www.calculquebec.ca/en/}) and Compute Canada (\href{https://www.computecanada.ca/}{www.computecanada.ca}). We also thank the research chair in Neuroinformatics of the Universit\'{e} de Sherbrooke. Data were provided in part by the Human Connectome Project, WU-Minn Consortium (Principal Investigators: David Van Essen and Kamil Ugurbil; 1U54MH091657) funded by the 16 NIH Institutes and Centers that support the NIH Blueprint for Neuroscience Research; and by the McDonnell Center for Systems Neuroscience at Washington University.

\bibliographystyle{abbrvnat}
\setcitestyle{authoryear,open={((},close={))}}
\bibliography{./bibliography/Legarreta20_-_arXiv_-_Tractography_Filtering_Using_Autoencoders}

\newpage
\appendix
\section{Appendix}
\label{sec:appendix}

\subsection{Reconstruction}
\label{subsec:reconstruction}
\begin{figure}[!ht]
\centering
\begin{tabular}{cc}
\includegraphics[scale=0.95, trim=2in 0.55in 2in 0.85in, clip=true, width=0.45\linewidth, keepaspectratio=true]{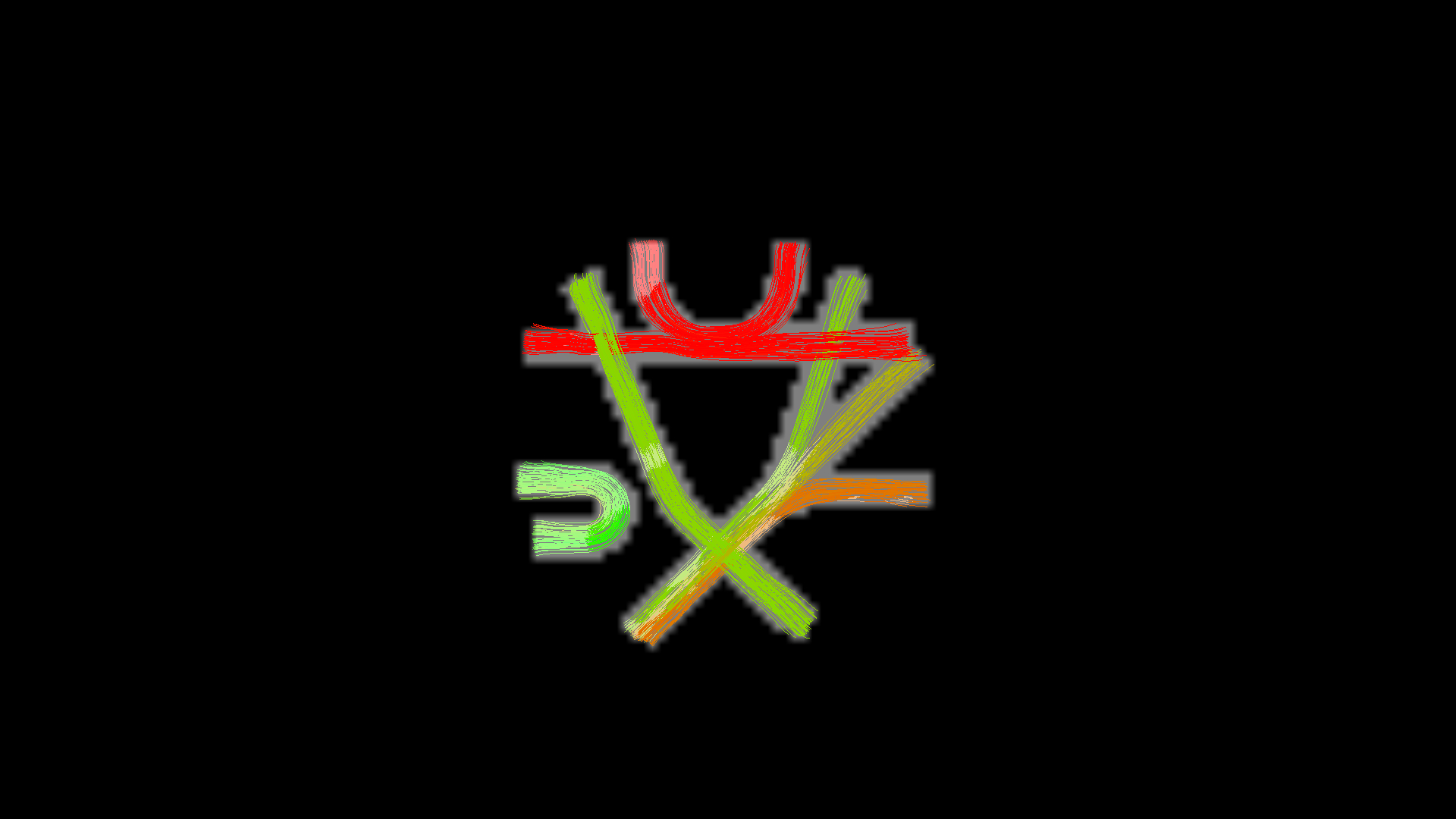} &
\includegraphics[scale=0.95, trim=2in 0.55in 2in 0.85in, clip=true, width=0.45\linewidth, keepaspectratio=true]{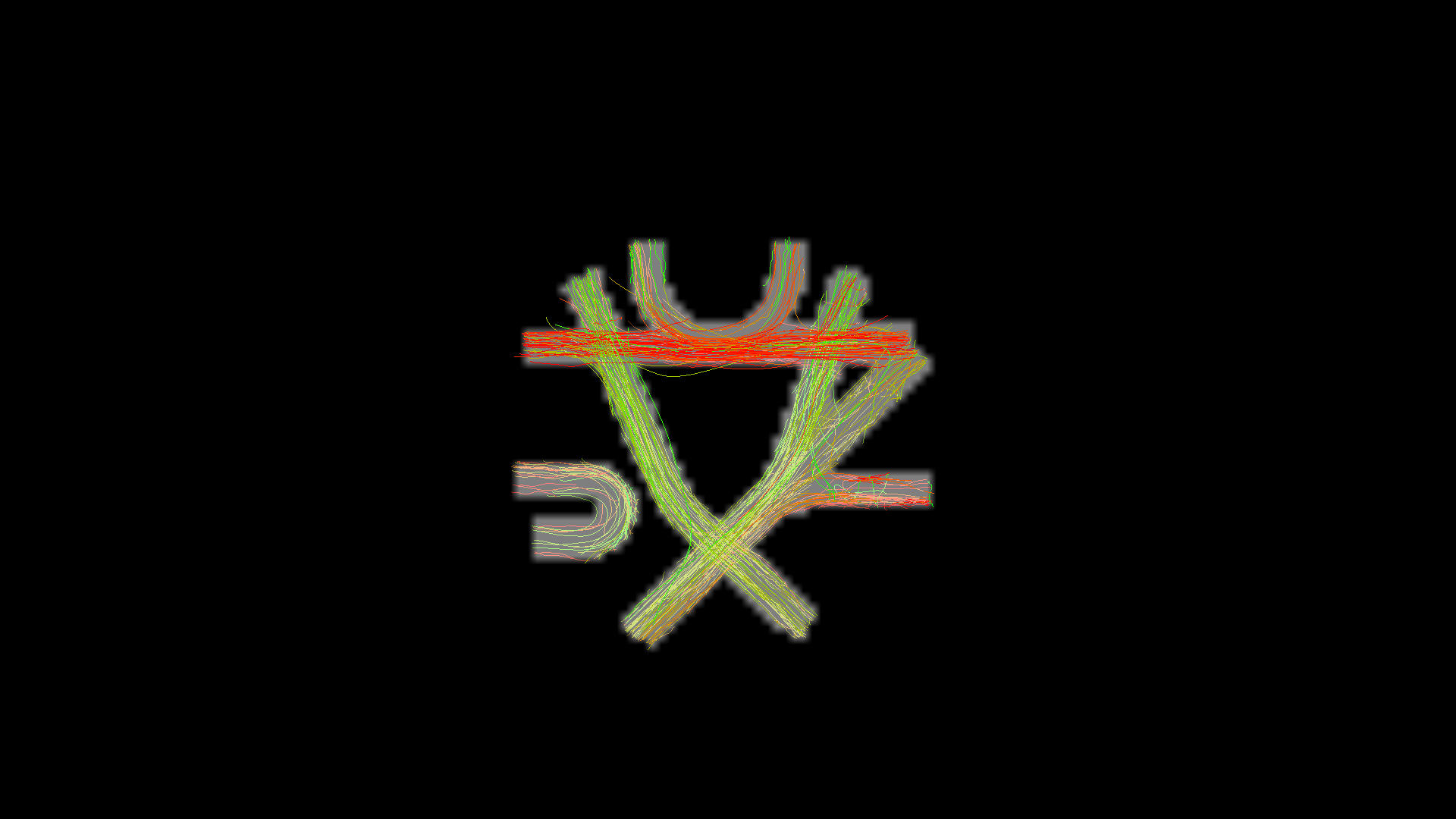} \\
\textbf{(a)} & \textbf{(b)} \\
\end{tabular}
\caption{\label{fig:fibercup_dataset_reconstruction}``Fiber Cup'' test set filtered streamlines as reconstructed by the autoencoder: (a) plausible streamlines; (b) implausible streamlines. The number of streamlines displayed is lower than the actual number in the test set for illustrative purposes.}
\end{figure}

Figure \ref{fig:fibercup_dataset_reconstruction} demonstrates the result of the autoencoder reconstruction corresponding to a subset of the ``Fiber Cup'' dataset test set streamlines. The considered classes (plausibles and implausibles) are shown separately according to the FINTA filtering procedure. The figures show that the plausible streamlines match the expected bundle-wise color code information from the synthetic ground truth streamlines (see figure \ref{fig:datasets}(a)). The implausible set can be seen to contain streamlines whose tracking had prematurely stopped within the tissue of interest, as well as streamlines connecting regions that are not connected in the synthetic ground truth. Additionally, and as anticipated by the autoencoder theory, it can be seen that the reconstructions are a smoothed version of the input tractogram streamlines in figure \ref{fig:datasets}(d) (also visible in figure \ref{fig:fibercup_latent_space_interpolation}).

\subsection{Predicted positives on the ISMRM 2015 Tractography Challenge human-based synthetic data}
\label{subsec:predicted_positives_human_based_synthetic_data_results}
\begin{figure*}[!ht]
\centering
\setlength{\tabcolsep}{0pt}
\begin{tabular}{ccc}
\includegraphics[scale=0.95, trim=0.5in 0.5in 0.5in 0.5in, clip=true, width=0.33\linewidth, keepaspectratio=true]{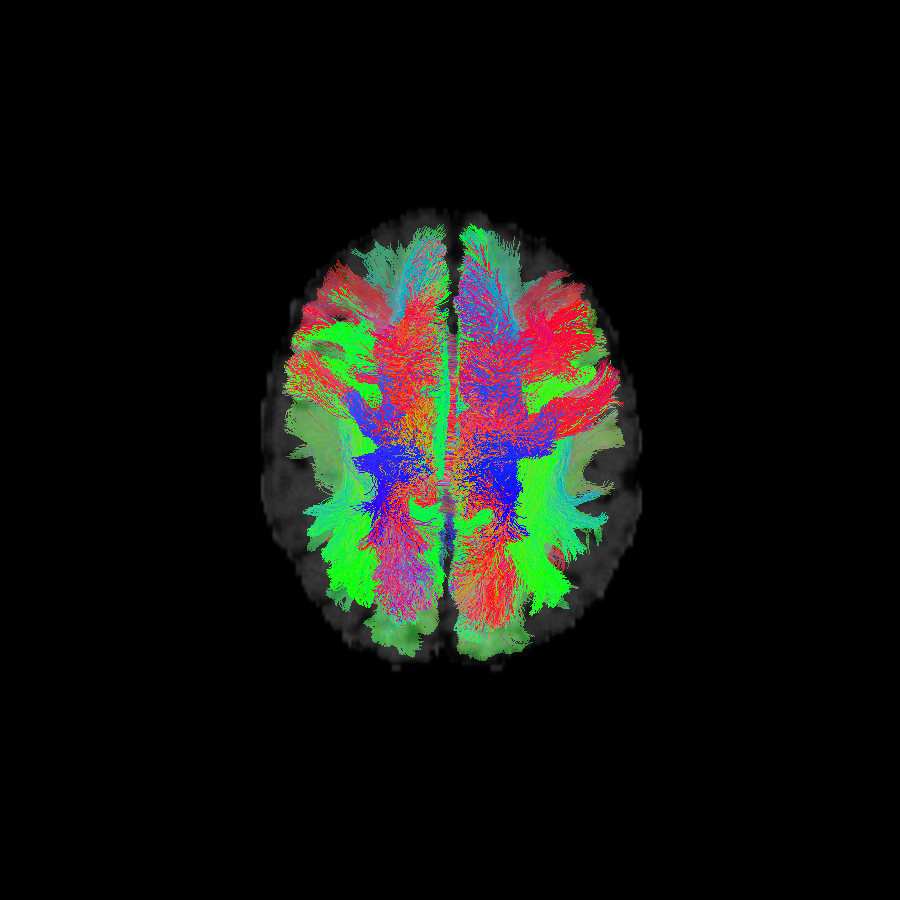} &
\includegraphics[scale=0.95, trim=0.5in 0.45in 0.5in 0.55in, clip=true, width=0.33\linewidth, keepaspectratio=true]{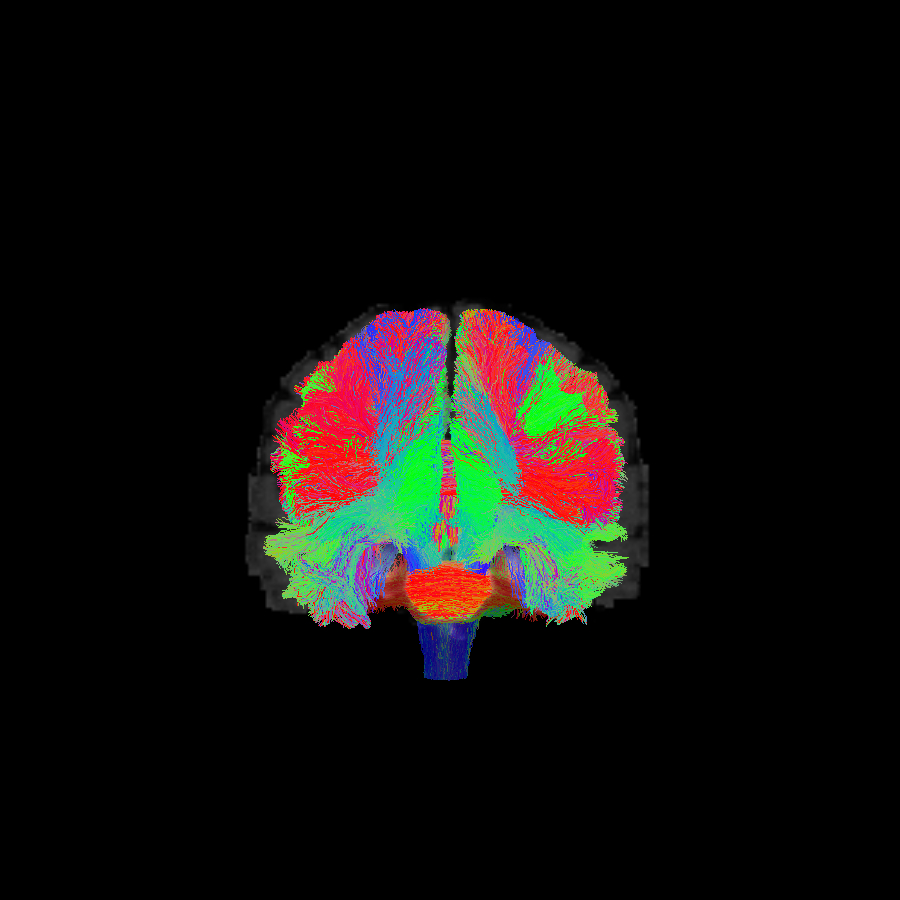} &
\includegraphics[scale=0.95, trim=0.5in 0.45in 0.5in 0.55in, clip=true, width=0.33\linewidth, keepaspectratio=true]{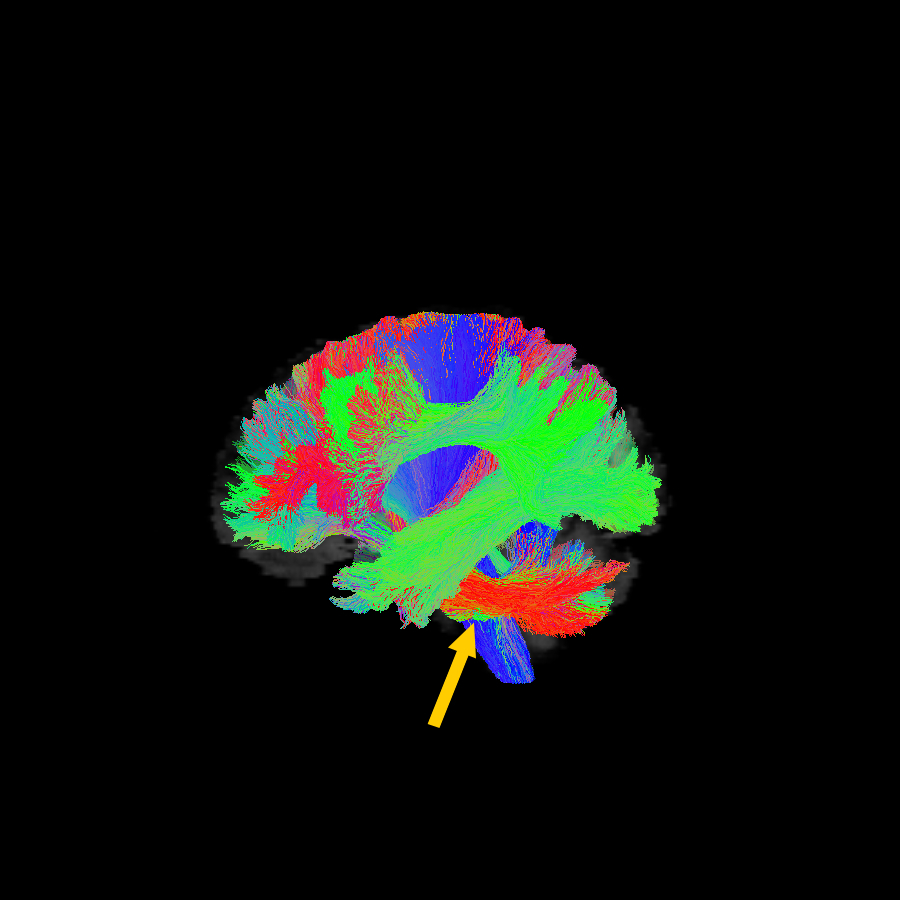}\\
\textbf{(a)} & \textbf{(b)} & \textbf{(c)} \\
\end{tabular}
\caption{\label{fig:ismrm_2015_tractography_challenge_dataset_predicted_positives}Predicted positives on the test set corresponding to the ISMRM 2015 Tractography Challenge dataset. (a) Axial superior view; (b) coronal anterior view; (c) sagittal left view. Although due to the streamline density the visualization can be difficult to interpret, the middle cerebellar peduncle (pointed bundle) clearly shows that the filtered tractogram has an orientation information that matches better the dataset ground truth shown in figure \ref{fig:datasets}(b) in comparison to the raw, unfiltered tractogram in figure \ref{fig:datasets}(e).}
\end{figure*}

Figure \ref{fig:ismrm_2015_tractography_challenge_dataset_predicted_positives} shows the predicted positive streamlines on the ISMRM 2015 Tractography Challenge dataset test tractogram. A visual inspection of the middle cerebellar peduncle (pointed bundle) shows that the filtered tractogram shows an orientation information that matches better the dataset ground truth shown in figure \ref{fig:datasets}(b) in comparison to the raw, unfiltered tractogram in figure \ref{fig:datasets}(e).

\newpage
\subsection{BIL\&GIN human data filtering supplementary results}
\label{subsec:human_data_filtering_supplementary_results}
\begin{figure*}[!hb]
\centering
\includegraphics[width=0.75\linewidth]{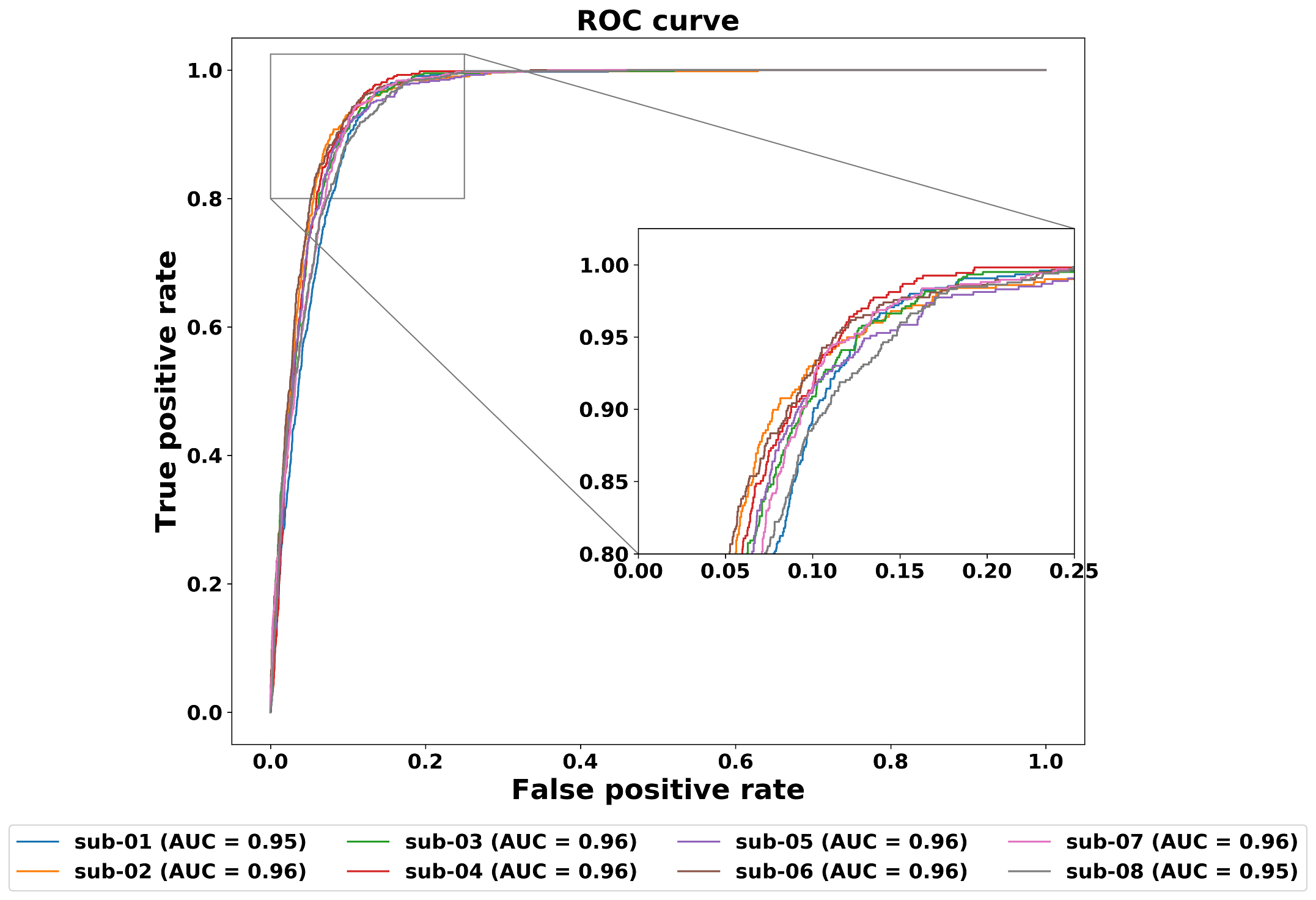}
\caption{\label{fig:bil_gin_dataset_roc_aggregated}Callosal BIL\&GIN dataset receiver operating characteristic (ROC) curves. The ROC curve shows the true positive rate (TPR) against the false positive rate (FPR) at various threshold settings. The figure shows the curves for the subjects used to compute the filtering threshold. The area under the curve (AUC) is indicated for each subject.}
\end{figure*}

Figure \ref{fig:bil_gin_dataset_roc_aggregated} shows the ROC curves corresponding to the callosal BIL\&GIN dataset subjects used to compute the filtering threshold. As it can be concluded from the curves, FINTA is consistent at filtering the tractograms across subjects. At the same time, results are in agreement with the experiments performed on the synthetic datasets.

\subsection{BIL\&GIN human data performance supplementary results}
\label{subsec:human_data_benchmarking_supplementary_results}
Tables \ref{tab:bil_gin_dataset_results_weighted}, \ref{tab:bil_gin_dataset_results_macro_min_max}, and \ref{tab:bil_gin_dataset_results_weighted_min_max} present supplementary results of the comparison to the baseline filtering methods on the callosal BIL\&GIN human data: table \ref{tab:bil_gin_dataset_results_weighted} shows the weighted mean and standard deviation values, and tables \ref{tab:bil_gin_dataset_results_macro_min_max} and \ref{tab:bil_gin_dataset_results_weighted_min_max} show the minimum and maximum values on the macro and weighted mean values. The weighted measures still show that FINTA provides an improved performance compared to the baseline methods. As mentioned, the latter yield an excessively large amount of false positives, and thus their valid gyrus-wise rates are generally higher than the values scored by FINTA. Nevertheless, FINTA still scores values over $0.68$, with a maximum value of $0.94$ across subjects.

\begin{table*}[!htbp]
\caption{\label{tab:bil_gin_dataset_results_weighted}Callosal BIL\&GIN dataset weighted results. Weighted mean (standard deviation) values over test subjects. The highest mean score is marked in bold face.}
\centering
\begin{tabular}{cccccccc}
\hline
& \textbf{Method} & \textbf{Balanced Accuracy} & \textbf{Sensitivity\textsubscript{w}} & \textbf{Precision\textsubscript{w}} & \textbf{F1-score\textsubscript{w}} & \textbf{VGR} & \textbf{SR\textsubscript{w}} \\
\hline
\#1 & \textbf{length} & 0.51 (0.0) & 0.14 (0.02) & 0.89 (0.01) & 0.07 (0.01) & \textbf{0.98} (0.03) & 0.52 \\
\#2 & \textbf{no\_loops} & 0.54 (0.01) & 0.19 (0.02) & 0.9 (0.01) & 0.16 (0.02) & 0.97 (0.03) & 0.55 \\
\#3 & \textbf{no\_end\_in\_csf} & 0.55 (0.01) & 0.21 (0.02) & 0.9 (0.01) & 0.19 (0.02) & 0.97 (0.03) & 0.56 \\
\#4 & \textbf{end\_in\_atlas} & 0.67 (0.05) & 0.71 (0.02) & 0.85 (0.03) & 0.76 (0.02) & 0.96 (0.04) & 0.79 \\
\#5 & \textbf{recobundles\_single} & 0.85 (0.03) & 0.83 (0.02) & 0.91 (0.01) & 0.86 (0.02) & 0.87 (0.1) & 0.86 \\
\#6 & \textbf{recobundles\_multi} & 0.8 (0.03) & 0.82 (0.03) & 0.89 (0.02) & 0.84 (0.03) & 0.96 (0.04) & 0.86 \\
\#7 & \textbf{FINTA} & \textbf{0.91} (0.01) & \textbf{0.91} (0.01) & \textbf{0.94} (0.01) & \textbf{0.92} (0.01) & 0.8 (0.09) & \textbf{0.89} \\
\end{tabular}
\end{table*}

\begin{table*}[!htbp]
\caption{\label{tab:bil_gin_dataset_results_macro_min_max}Callosal BIL\&GIN dataset macro minimum and maximum values. Macro mean [minimum, maximum] values over test subjects.}
\centering
\begin{tabular}{cccccccc}
\hline
& \textbf{Method} & \textbf{Accuracy\textsubscript{m}} & \textbf{Sensitivity\textsubscript{m}} & \textbf{Precision\textsubscript{m}} & \textbf{F1-score\textsubscript{m}} & \textbf{VGR} \\
\hline
\#1 & \textbf{length} & [0.12, 0.16] & [0.5, 0.52] & [0.55, 0.57] & [0.11, 0.15] & [0.94, 1] \\
\#2 & \textbf{no\_loops} & [0.15, 0.21] & [0.53, 0.55] & [0.55, 0.58] & [0.15, 0.21] & [0.94, 1] \\
\#3 & \textbf{no\_end\_in\_csf} & [0.18, 0.24] & [0.54, 0.56] & [0.55, 0.57] & [0.18, 0.24] & [0.94, 1] \\
\#4 & \textbf{end\_in\_atlas} & [0.68, 0.75] & [0.56, 0.74] & [0.55, 0.62] & [0.53, 0.62] & [0.88, 1] \\
\#5 & \textbf{recobundles\_single} & [0.81, 0.87] & [0.79, 0.88] & [0.65, 0.72] & [0.68, 0.75] & [0.72, 1] \\
\#6 & \textbf{recobundles\_multi} & [0.78, 0.86] & [0.75, 0.85] & [0.64, 0.69] & [0.67, 0.72] & [0.89, 1] \\
\#7 & \textbf{FINTA} & [0.89, 0.91] & [0.89, 0.92] & [0.76, 0.81] & [0.81, 0.85] & [0.68, 0.94] \\
\end{tabular}
\end{table*}

\begin{table*}[!htbp]
\caption{\label{tab:bil_gin_dataset_results_weighted_min_max}Callosal BIL\&GIN dataset weighted minimum and maximum values. Weighted mean [minimum, maximum] values over test subjects. The valid gyrus-wise rate (VGR) is the same as the one reported in table \ref{tab:bil_gin_dataset_results_macro_min_max}.}
\centering
\begin{tabular}{cccccccc}
\hline
& \textbf{Method} & \textbf{Balanced Accuracy} & \textbf{Sensitivity\textsubscript{w}} & \textbf{Precision\textsubscript{w}} & \textbf{F1-score\textsubscript{w}} & \textbf{VGR} \\
\hline
\#1 & \textbf{length} & [0.5, 0.52] & [0.12, 0.16] & [0.87, 0.92] & [0.05, 0.08] & [0.94, 1] \\
\#2 & \textbf{no\_loops} & [0.53, 0.55] & [0.15, 0.21] & [0.88, 0.92] & [0.13, 0.18] & [0.94, 1] \\
\#3 & \textbf{no\_end\_in\_csf} & [0.54, 0.56] & [0.18, 0.24] & [0.88, 0.92] & [0.16, 0.22] & [0.94, 1] \\
\#4 & \textbf{end\_in\_atlas} & [0.56, 0.74] & [0.68, 0.75] & [0.79, 0.88] & [0.73, 0.79] & [0.88, 1] \\
\#5 & \textbf{recobundles\_single} & [0.79, 0.88] & [0.81, 0.87] & [0.89, 0.93] & [0.84, 0.89] & [0.72, 1] \\
\#6 & \textbf{recobundles\_multi} & [0.75, 0.85] & [0.78, 0.86] & [0.86, 0.93] & [0.8, 0.88] & [0.89, 1] \\
\#7 & \textbf{FINTA} & [0.89, 0.93] & [0.89, 0.93] & [0.93, 0.95] & [0.9, 0.93] & [0.68, 0.94] \\
\end{tabular}
\end{table*}

\subsection{The callosal BIL\&GIN dataset}
\label{subsec:considered_bil_gin_cc_regions}
\begin{figure*}[!htbp]
\centering
\includegraphics[scale=0.5, trim=0.5in 0in 0.5in 0in, clip=true, width=0.9\linewidth, keepaspectratio=true]{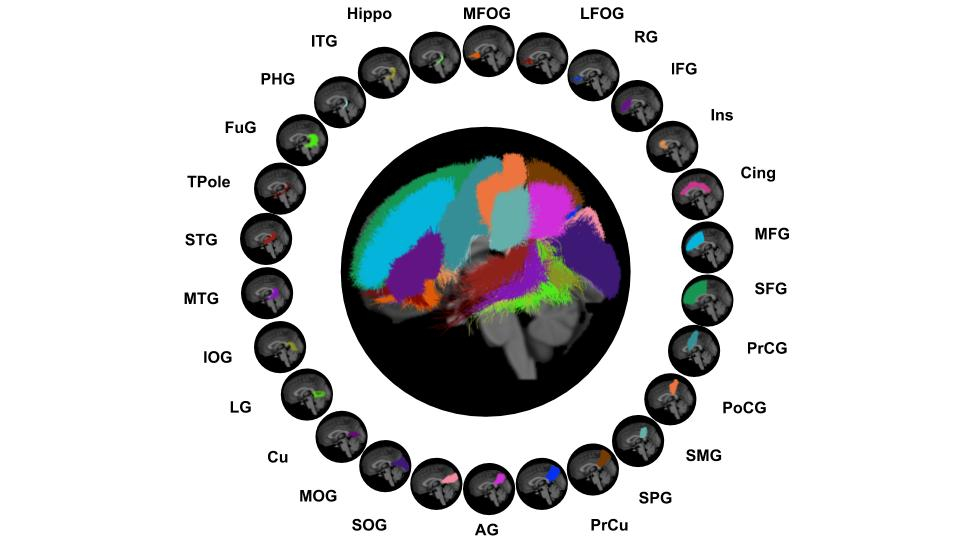} 
\caption{\label{fig:bil_gin_dataset_cc_regions}Callosal BIL\&GIN dataset gyrus-wise homotopic streamline models. All sagittal left views.}
\end{figure*}

The callosal BIL\&GIN dataset was composed of the streamlines passing through the corpus callosum in $39$ participants of the BIL\&GIN dataset. For each subject, callosal streamlines were extracted from a whole brain tractogram computed with the probabilistic setting of the Particle Filtering Tracking (PFT) method by \citet{Girard:Neuroimage:2014} as previously described in \citet{Chenot:BrainStructFunc:2019}. Each individual callosal tractogram was composed of anatomically plausible and implausible streamlines. The former were composed of gyral-based homotopic callosal streamlines, any other type of callosal streamline being considered as anatomically implausible. Twenty-six ($26$) different types of gyral-based homotopic streamlines were considered depending on the homotopic gyri they belonged to, namely the $26$ gyral regions of interest of the JHU template \citep{Zhang:Neuroimage:2010}. Each gyral ROI was composed of both cortical and superficial white matter parts of the gyrus as defined in the JHU template excepted for the insula, hippocampus and parahippocampal gyrus that have no superficial white matter part. The $26$ gyral-based ROIs were therefore: Angular Gyrus (AG); Cingulum (Cing); Cuneus (Cu); Fusiform Gyrus (FuG); Hippocampus (Hippo); Inferior Frontal Gyrus (IFG); Inferior Occipital Gyrus (IOG); Inferior Temporal Gyrus (ITG); Insula (Ins); Lateral Fronto-Orbital Gyrus (LFOG); Lingual Gyrus (LG); Middle Frontal Gyrus (MFG); Middle Fronto-Orbital Gyrus (MFOG); Middle Occipital Gyrus (MOG); Middle Temporal Gyrus (MTG); Parahippocampal Gyrus (PHG); Post-Central Gyrus (PoCG); Pre-Central Gyrus (PrCG); Pre-Cuneus (PrCu); Rectus Gyrus (RG); Superior Frontal Gyrus (SFG); Supramarginal Gyrus (SMG); Superior Occipital Gyrus (SOG); Superior Parietal Gyrus (SPG); Superior Temporal Gyrus (STG); and Temporal Pole (TPole). The gyrus-wise homotopic callosal streamline models, used in the RecoBundles filtering experiments, were built using the aforementioned procedure and according to the method described in \citet{Rheault:PhD:2020} over the entire BIL\&GIN population ($411$ subjects). Figure \ref{fig:bil_gin_dataset_cc_regions} provides an overview of the considered models.

As confirmed by diffusion-weighted tractography studies, the general research on the anatomy of the corpus callosum has traditionally been limited to its homotopic feature \citep{Hofer:Neuroimage:2006, Chao:HBM:2009}. Although recent developments \citep{DeBenedictis:HBM:2016} have proposed solutions to reconstruct the most lateral homotopic connectivity, and, to a lesser extent, some heterotopic callosal connectivity in the frontal cortex, these still require further investigation with anatomical validation from dissection to be considered for the whole brain. Hence, in this work we focus on the most reliable representation of the callosal streamlines at the whole-brain level, namely the homotopic ones.

\subsection{Callosal BIL\&GIN gyrus-wise analysis}
\label{subsec:bil_gin_regionwise_analysis}
The gyral segment pairs defined in section \ref{subsec:considered_bil_gin_cc_regions} for the callosal BIL\&GIN dataset solely allow the extraction of homotopic streamlines. Hence, by definition, no implausible streamline can be recovered using the constraints established by these criteria. As a consequence, the only measure that can be reported for the predictions over the gyral segments is the sensitivity (see table \ref{tab:bil_gin_dataset_classification_performance_regionwise}). The mean number of segments with a non-zero streamline count in the reference data was $17$ for the test subjects. Similarly, there was a non-negligible number of segments whose streamline density was very low compared to the most heavily populated ones. Test subjects averaged $5287$ (standard deviation: $1450$) homotopic streamlines across all segments, and the number of segments that had a mean streamline count larger than $264$ ($5\%$ of the mean homotopic streamline count) was $3$: MFG; PrCu; and SFG.

\subsection{Connectivity analysis}
\label{subsec:connectivity_analysis}
Additional experiments were performed to quantify the effect of FINTA into downstream tractography tasks. Structural connectivity matrices in terms of the streamline count were computed and analyzed for the ``Fiber Cup'' and ISMRM 2015 Tractography Challenge datasets. The following measures were computed on the binarized connectivity matrices to quantitatively analyze and compare them:
\begin{itemize}
\item \textit{Density}: occupation index of the given connectivity matrix, an empty matrix having a value of $0$, and $1$ corresponding to a fully occupied matrix.
\item \textit{Chi-square}: defined as the squared element-wise differences across the matrices divided by their addition as in \citet{StOnge:Neuroimage:2018}. Lower is better.
\item \textit{Total variation}: computed as the absolute difference between matrices and normalized to the $[0, 1]$ range. Lower is better.
\item \textit{Correlation}: as defined by \citet{Deslauriers-Gauthier:MIA:2020} and modified to be normalized to the $[0, 1]$ range. Higher is better.
\end{itemize}

Considered together, the presented results in terms of the density of the matrices and the distances across the matrices show that the connectomes issued by FINTA are populated as expected.

\subsubsection{``Fiber Cup'' synthetic dataset}
\label{subsec:synthetic_data_connectivity_analysis}
\begin{figure*}[!htb]
\centering
\includegraphics[scale=0.9, trim=6in 1.25in 6in 2in, clip=true, width=0.35\linewidth, keepaspectratio=true]{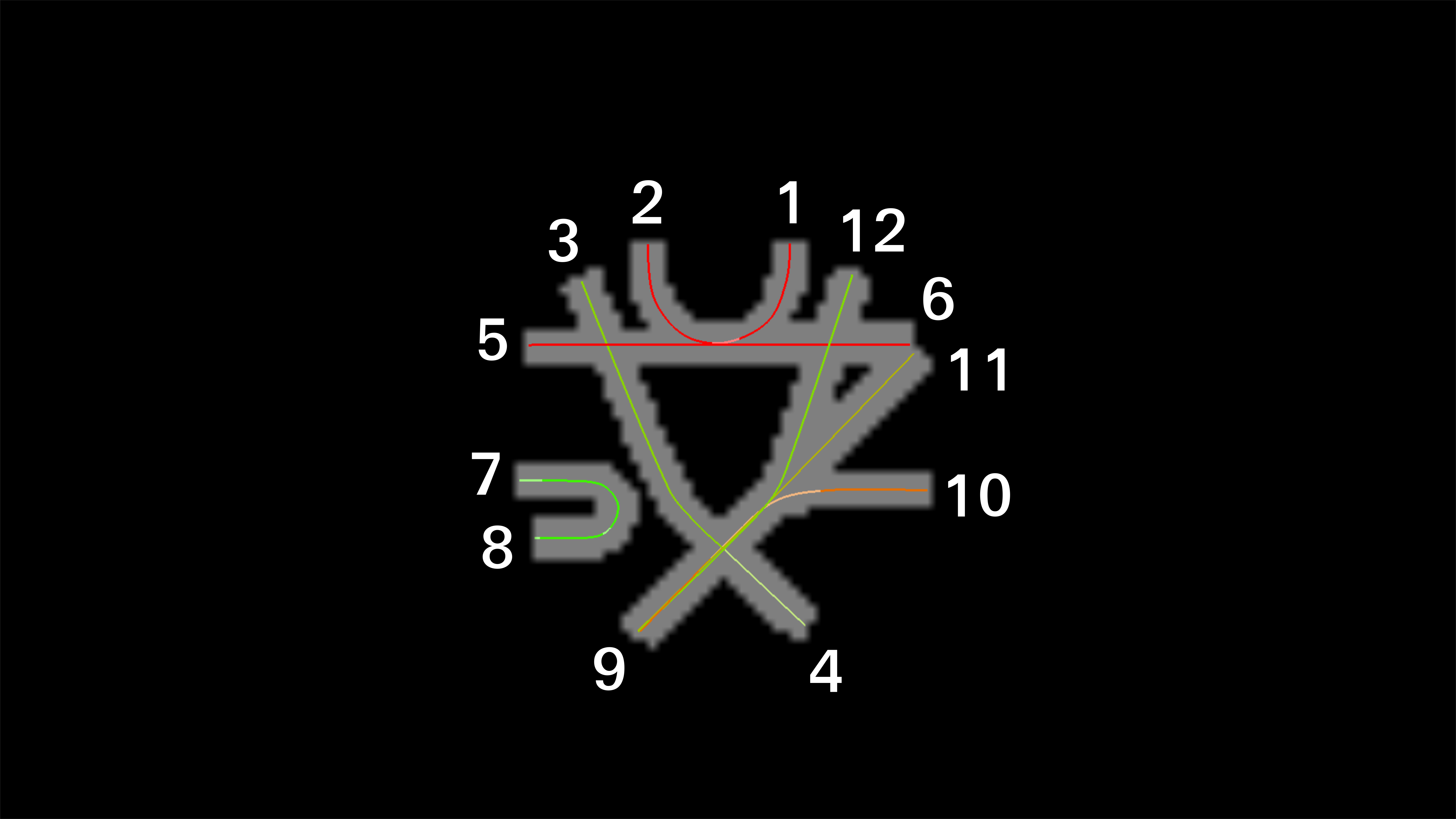} 
\caption{\label{fig:fibercup_dataset_connectivity_regions}``Fiber Cup'' dataset connectivity region definition.}
\end{figure*}

\begin{figure*}[!ht]
\centering
\setlength{\tabcolsep}{1pt}
\begin{tabular}{ccccc}
\includegraphics[scale=0.95, trim=0in 0in 0.75in 0in, clip=true, width=0.2275\linewidth, keepaspectratio=true]{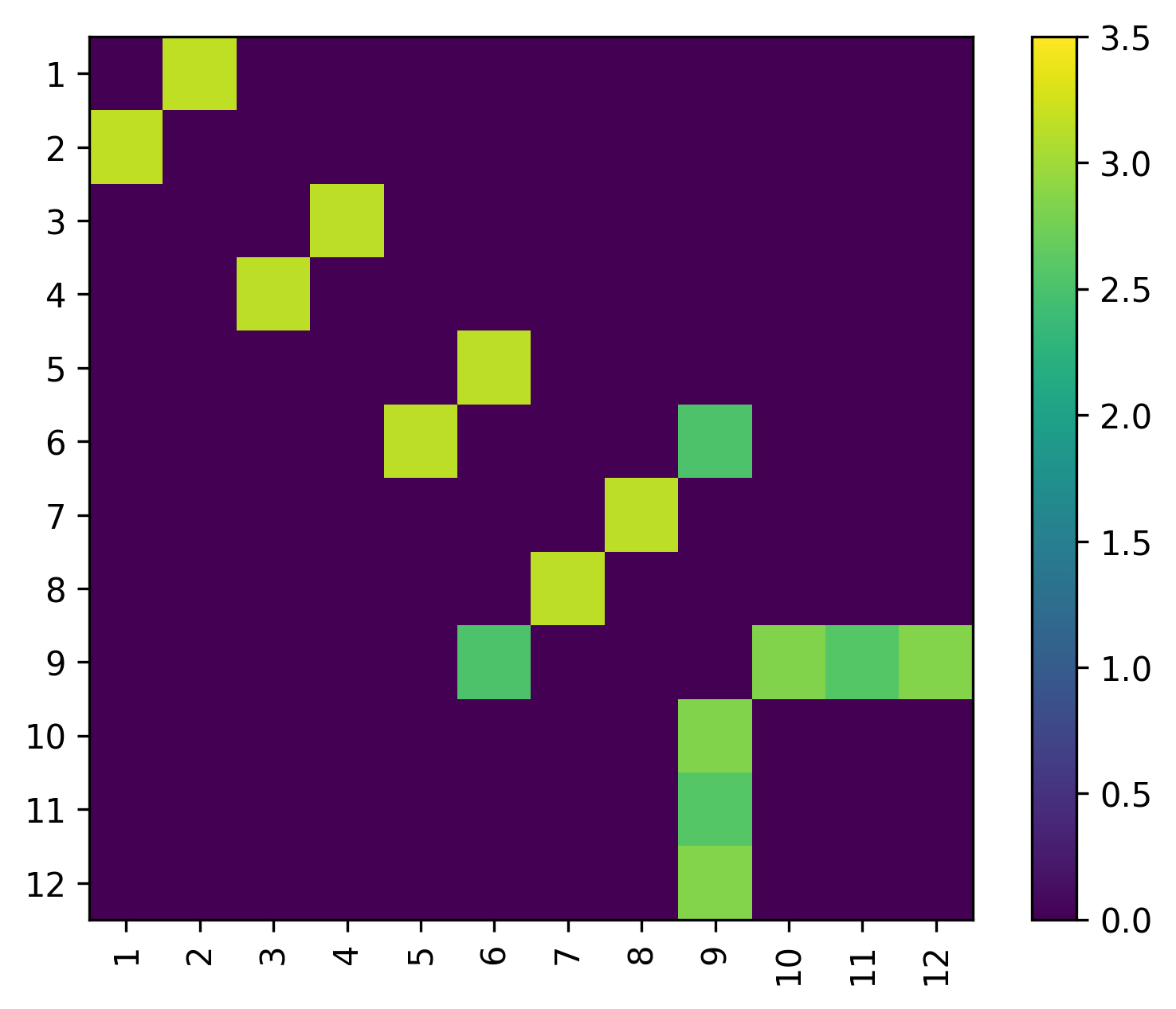} &
\includegraphics[scale=0.95, trim=0in 0in 0.75in 0in, clip=true, width=0.2275\linewidth, keepaspectratio=true]{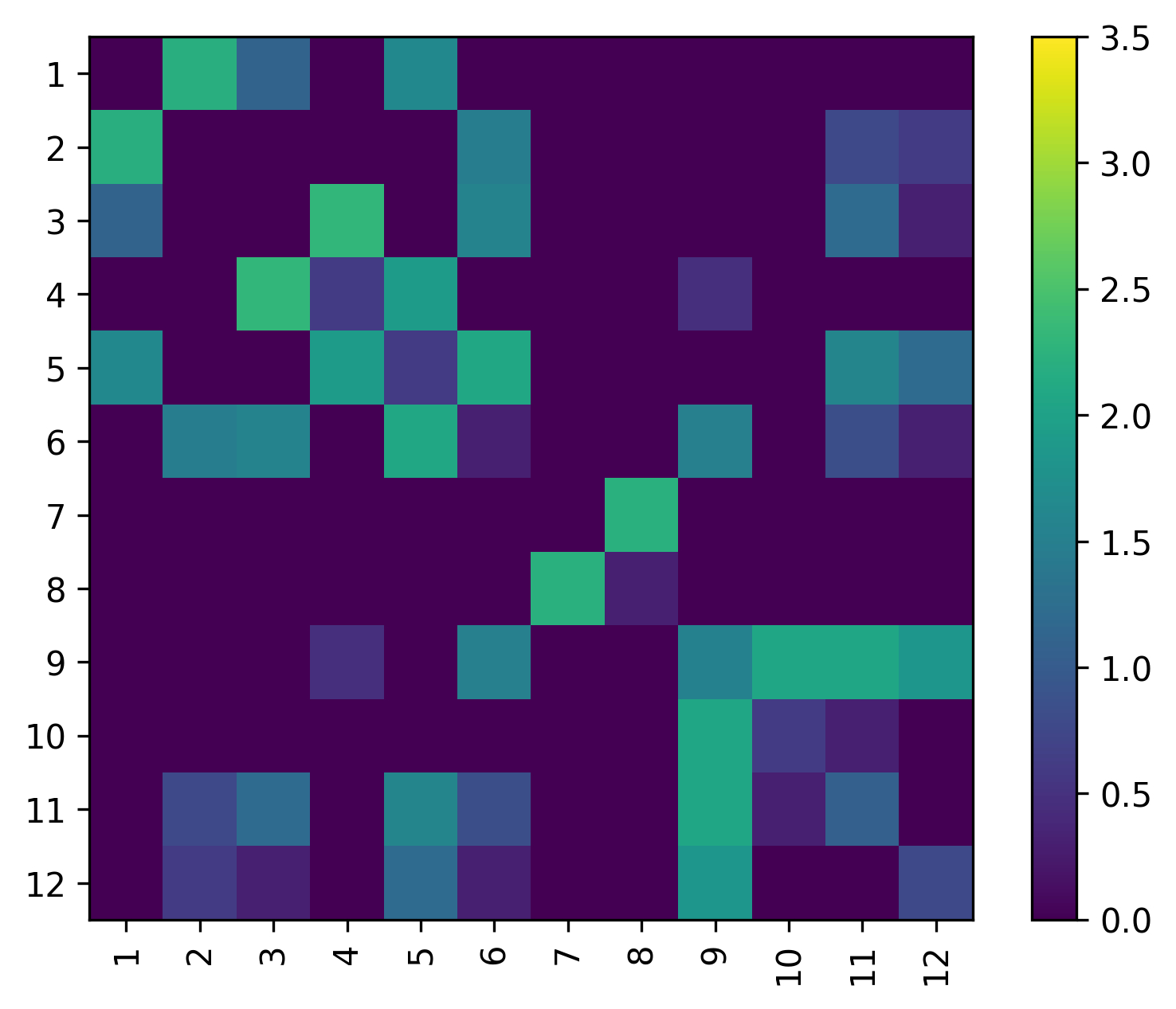} &
\includegraphics[scale=0.95, trim=0in 0in 0.75in 0in, clip=true, width=0.2275\linewidth, keepaspectratio=true]{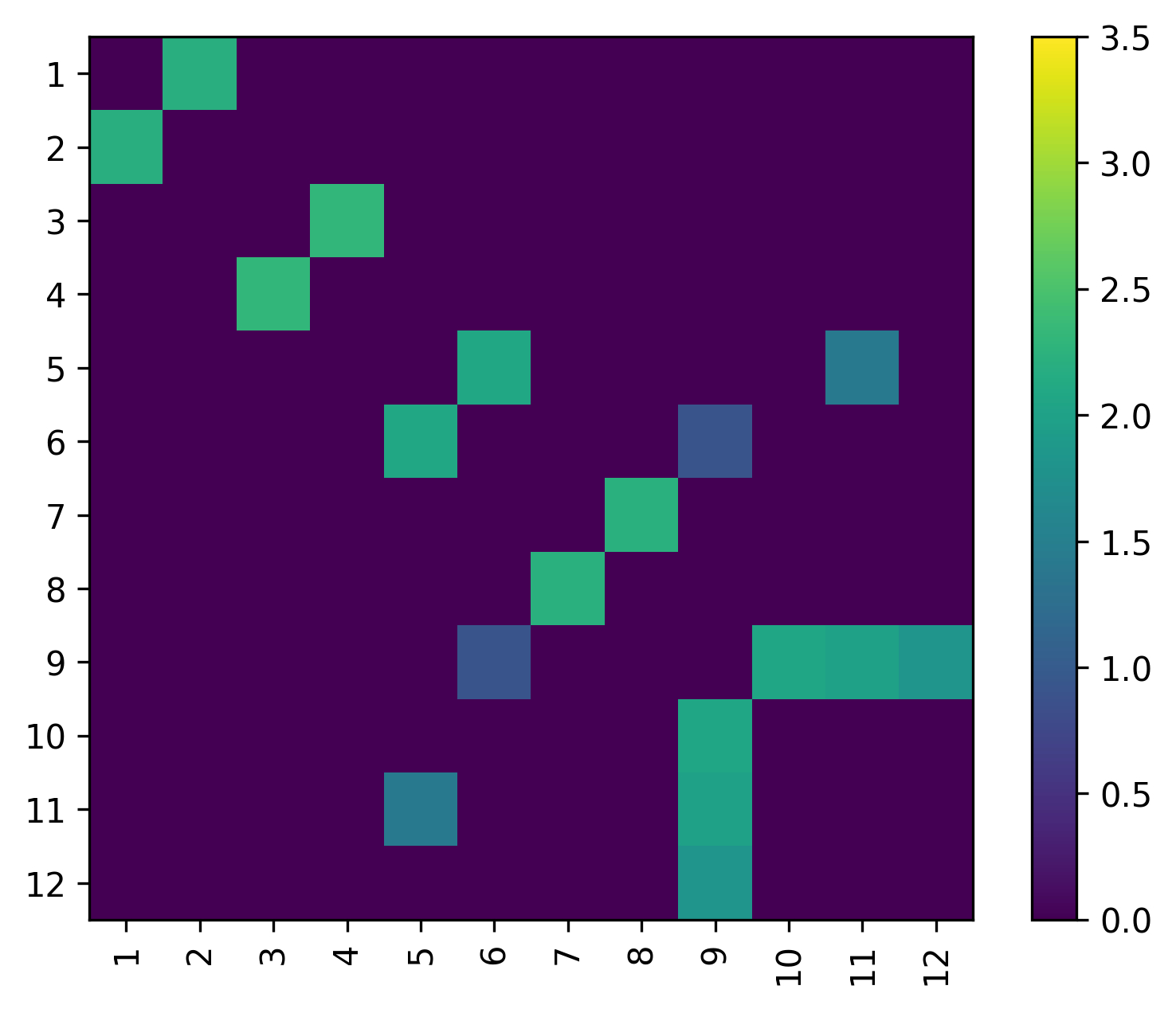} &
\includegraphics[scale=0.95, trim=0in 0in 0.75in 0in, clip=true, width=0.2275\linewidth, keepaspectratio=true]{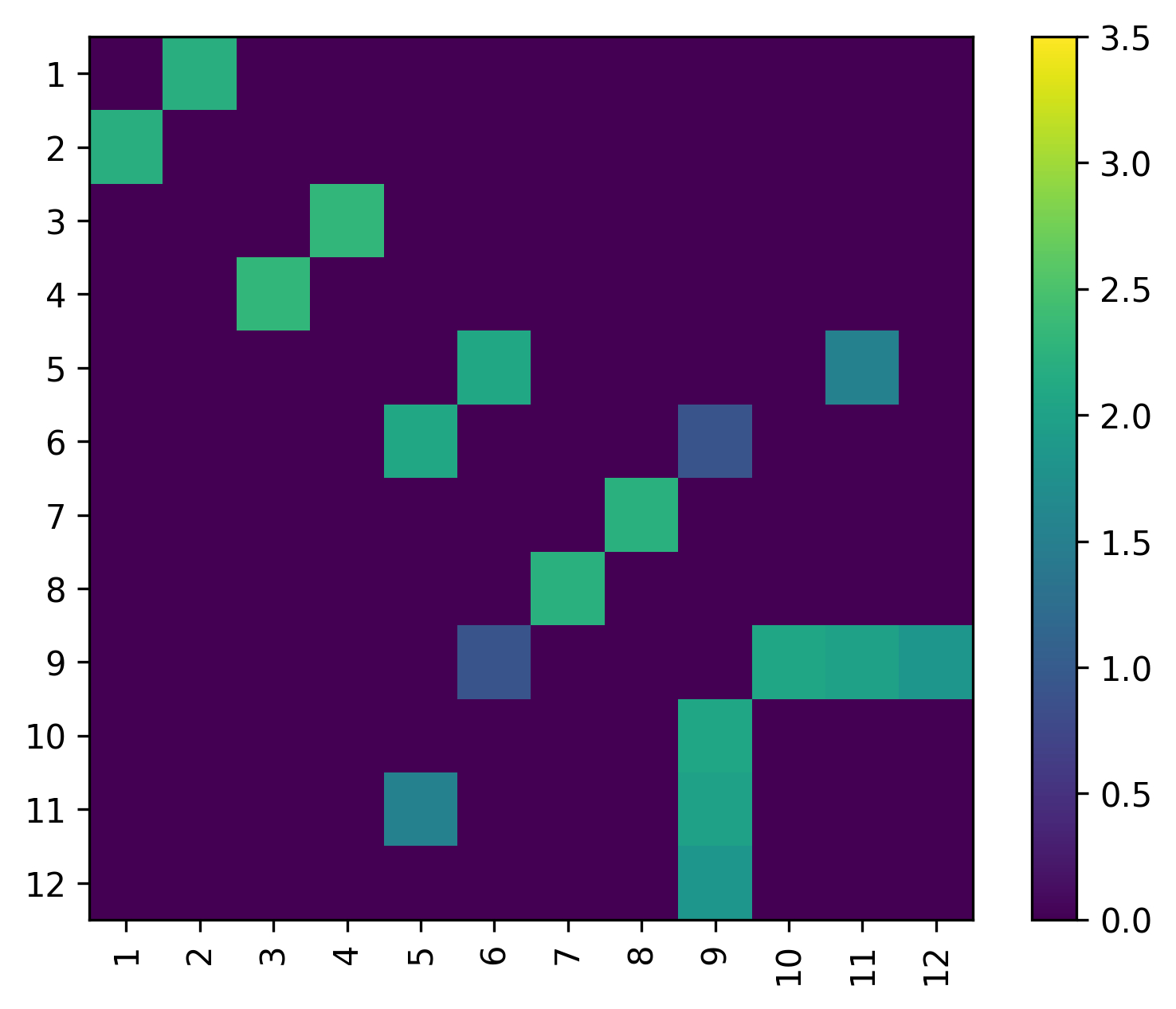} &
\includegraphics[scale=0.95, clip=true, width=0.0455\linewidth, keepaspectratio=true]{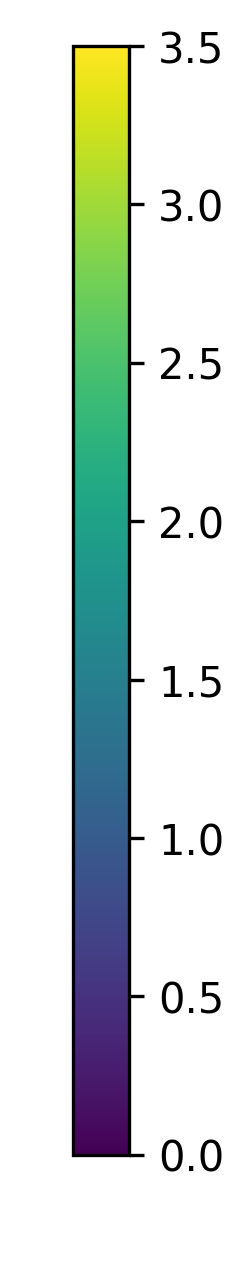} \\
\textbf{(a)} & \textbf{(b)} & \textbf{(c)} & \textbf{(d)} & \\
\includegraphics[scale=0.95, trim=0in 0in 0.75in 0in, clip=true, width=0.2275\linewidth, keepaspectratio=true]{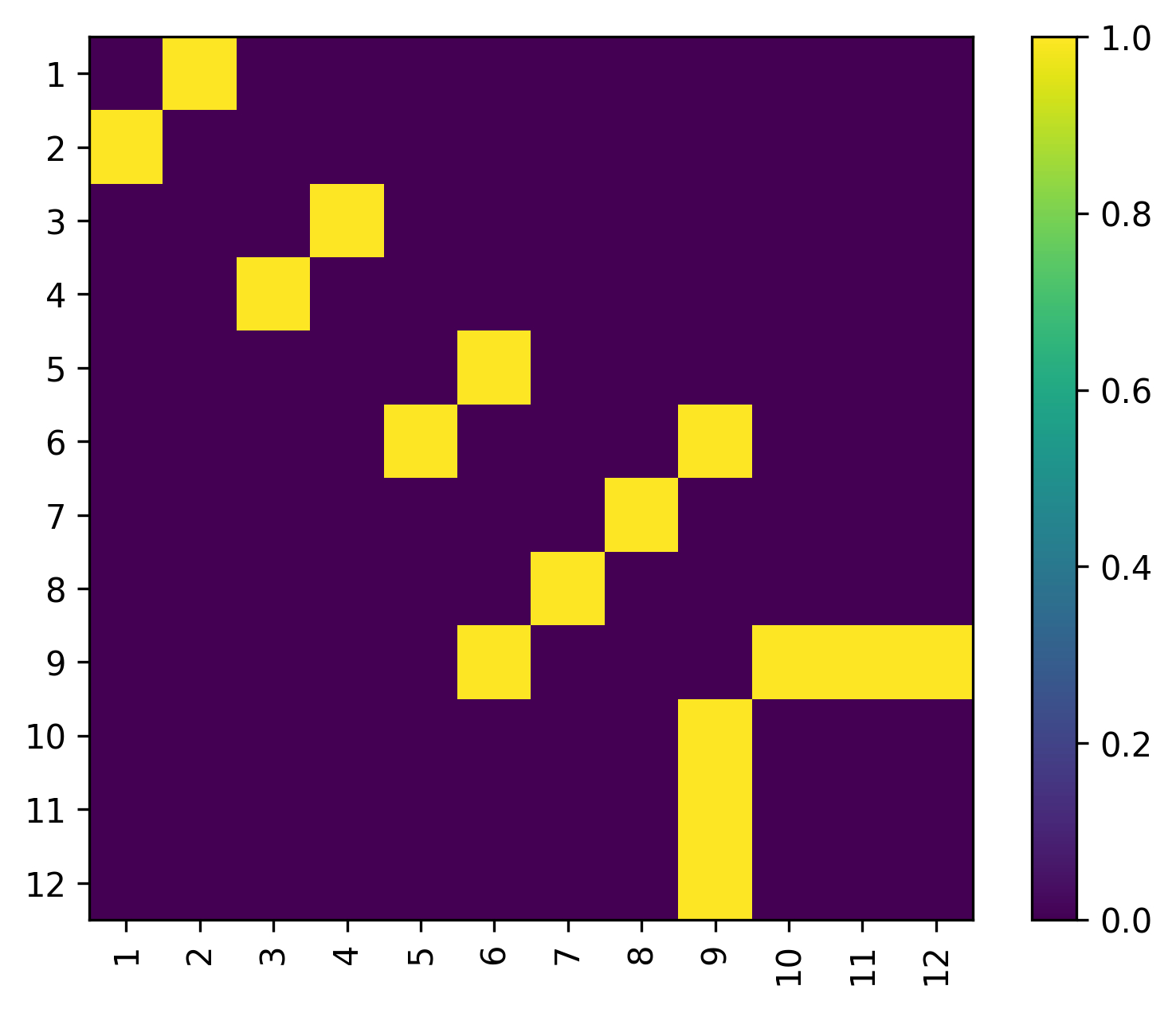} &
\includegraphics[scale=0.95, trim=0in 0in 0.75in 0in, clip=true, width=0.2275\linewidth, keepaspectratio=true]{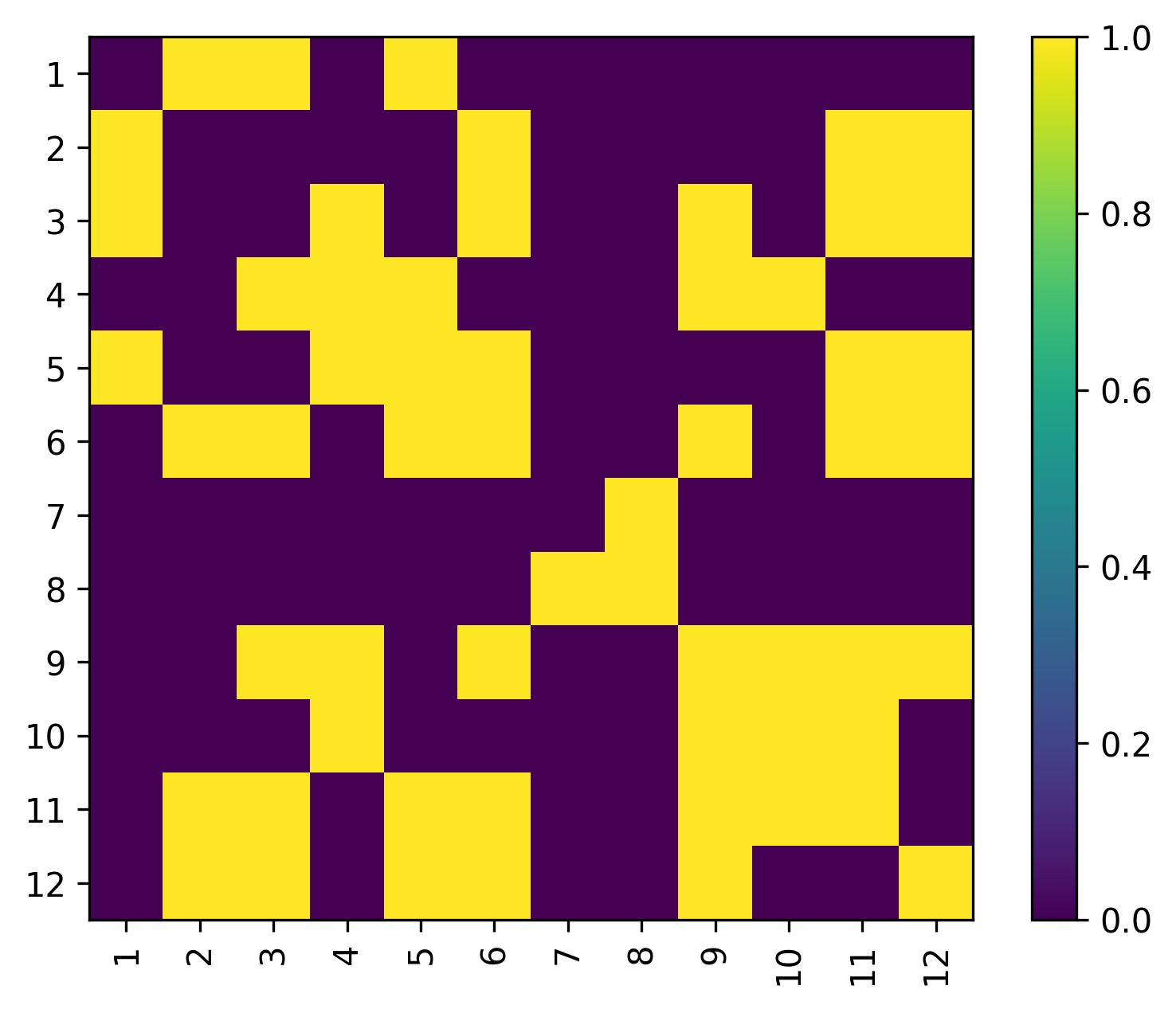} &
\includegraphics[scale=0.95, trim=0in 0in 0.75in 0in, clip=true, width=0.2275\linewidth, keepaspectratio=true]{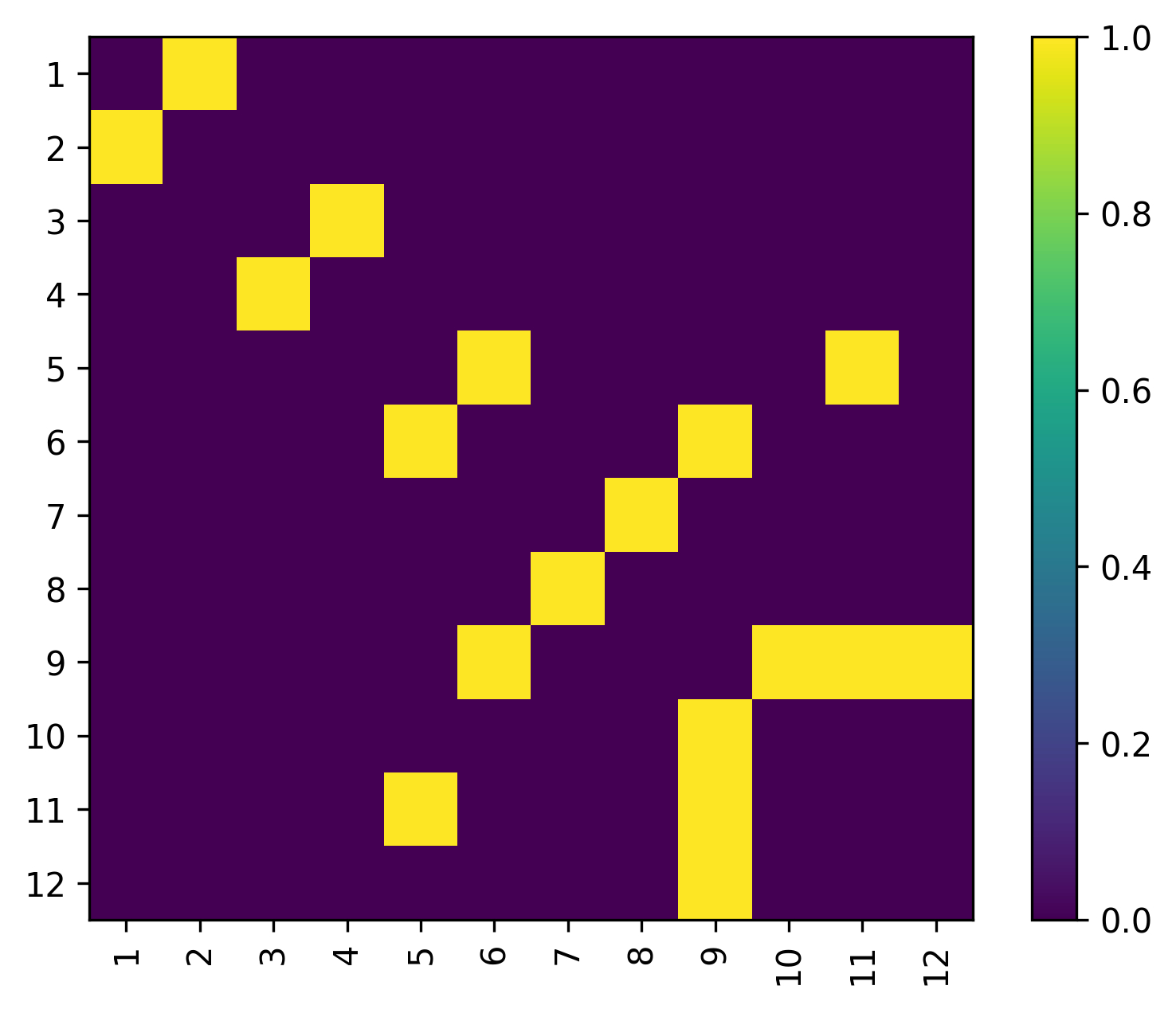} &
\includegraphics[scale=0.95, trim=0in 0in 0.75in 0in, clip=true, width=0.2275\linewidth, keepaspectratio=true]{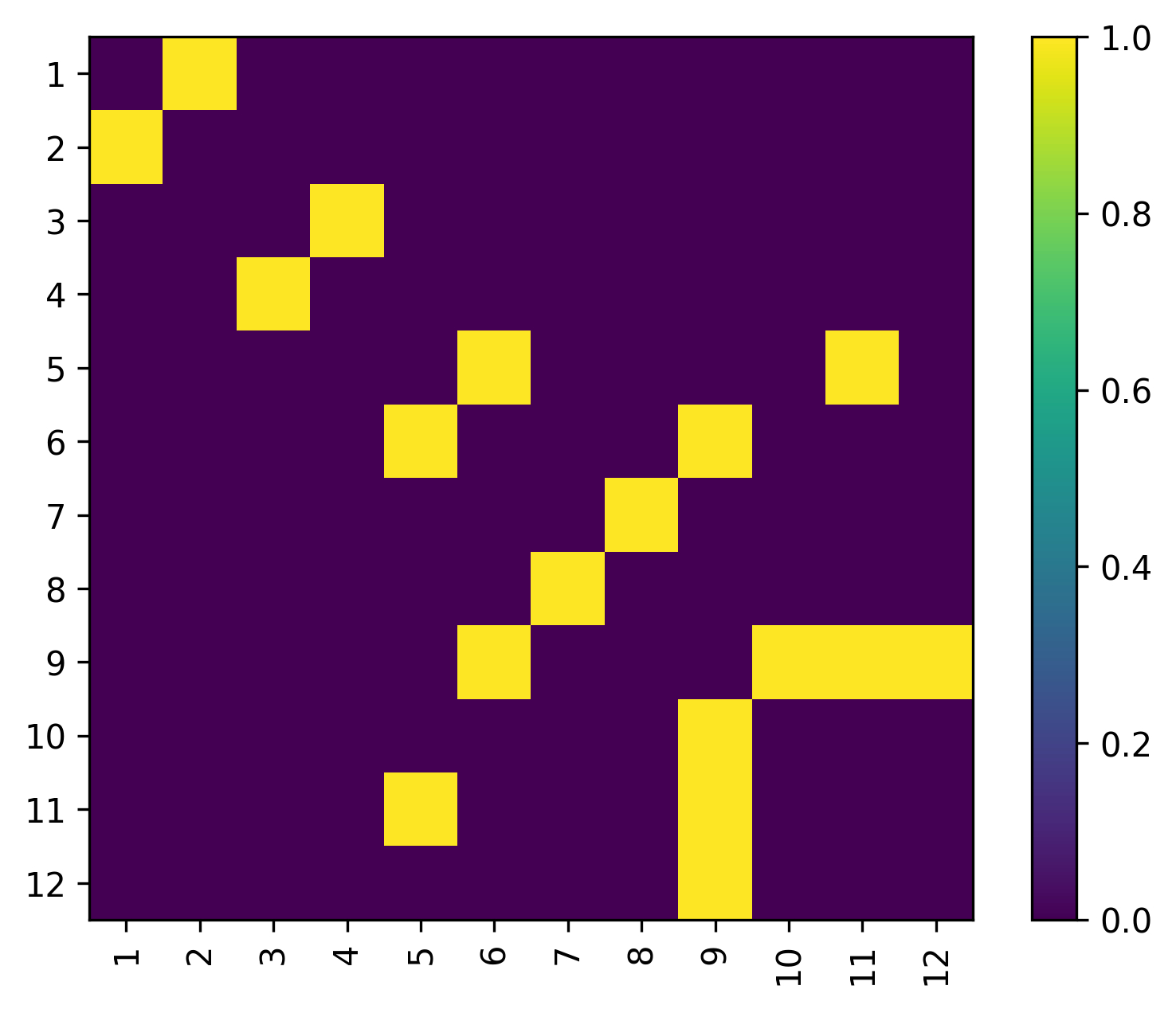} & \\
\textbf{(e)} & \textbf{(f)} & \textbf{(g)} & \textbf{(h)} & \\
\end{tabular}
\caption{\label{fig:fibercup_dataset_connectivity_matrices}Streamline count connectivity matrices for the ``Fiber Cup''. Upper row: log-scale; lower row: binarized. (a) and (e): ground truth; (b) and (f): unfiltered tractogram; (c) and (g): \textit{Tractometer\_basic}-scored tractogram; (d) and (h): FINTA-filtered tractogram.}
\end{figure*}

Figure \ref{fig:fibercup_dataset_connectivity_regions} shows the defined regions used to compute the connectivity on the ``Fiber Cup'' dataset. Figure \ref{fig:fibercup_dataset_connectivity_matrices} groups the log-scale and binarized streamline count connectivity matrices. It detaches from the figures that FINTA is in agreement with the \textit{Tractometer\_basic}-scored data, which is at the same time, close to the ground truth.

Table \ref{tab:fibercup_dataset_connectome_density} displays the density of the binarized connectivity matrices. The results reveal that FINTA is able to remove the undesired connections from the unfiltered dataset, and follows closely the value expected from the ground truth.
\begin{table}[!htb]
\caption{\label{tab:fibercup_dataset_connectome_density}Binarized streamline count connectivity matrix density for the ``Fiber Cup'' dataset. Note that the data corresponds to the test data (except for the ground truth).}
\centering
\begin{tabular}{ccccc}
\textbf{Data} & \textbf{Ground truth} & \textbf{Unfiltered} & \textbf{Tractometer\_basic} & \textbf{FINTA} \\
\hline
\textbf{Density} & 0.12 & 0.5 & 0.14 & 0.14
\end{tabular}
\end{table}

Table \ref{tab:fibercup_dataset_connectome_distance} shows that FINTA is identically correlated to the ground truth when compared to the scored data (i.e. \textit{Tractometer\_basic}), meaning that the proposed filtering method matches the expectations from the scoring.

\begin{table}[!htb]
\caption{\label{tab:fibercup_dataset_connectome_distance}Distance measures on the binarized connectivity matrices for the ``Fiber Cup'' dataset.}
\centering
\begin{tabular}{cccc}
& \textbf{Chi-square} & \textbf{Total variation} & \textbf{Correlation} \\
\hline
\textbf{Ground truth vs. Unfiltered} & 21.0 & 0.29 & 0.48 \\
\textbf{Ground truth vs. Tractometer\_basic} & 1.0 & 0.01 & 0.93 \\
\textbf{Ground truth vs. FINTA} & 1.0 & 0.01 & 0.93 \\
\textbf{Tractometer\_basic vs. FINTA} & 0 & 0 & 1.0
\end{tabular}
\end{table}

\subsubsection{ISMRM 2015 Tractography Challenge human-based synthetic data}
\label{subsec:human_based_synthetic_data_connectivity_analysis}
\begin{figure*}[!htb]
\centering
\setlength{\tabcolsep}{1pt}
\begin{tabular}{ccccc}
\includegraphics[scale=0.95, trim=0in 0in 0.75in 0in, clip=true, width=0.2275\linewidth, keepaspectratio=true]{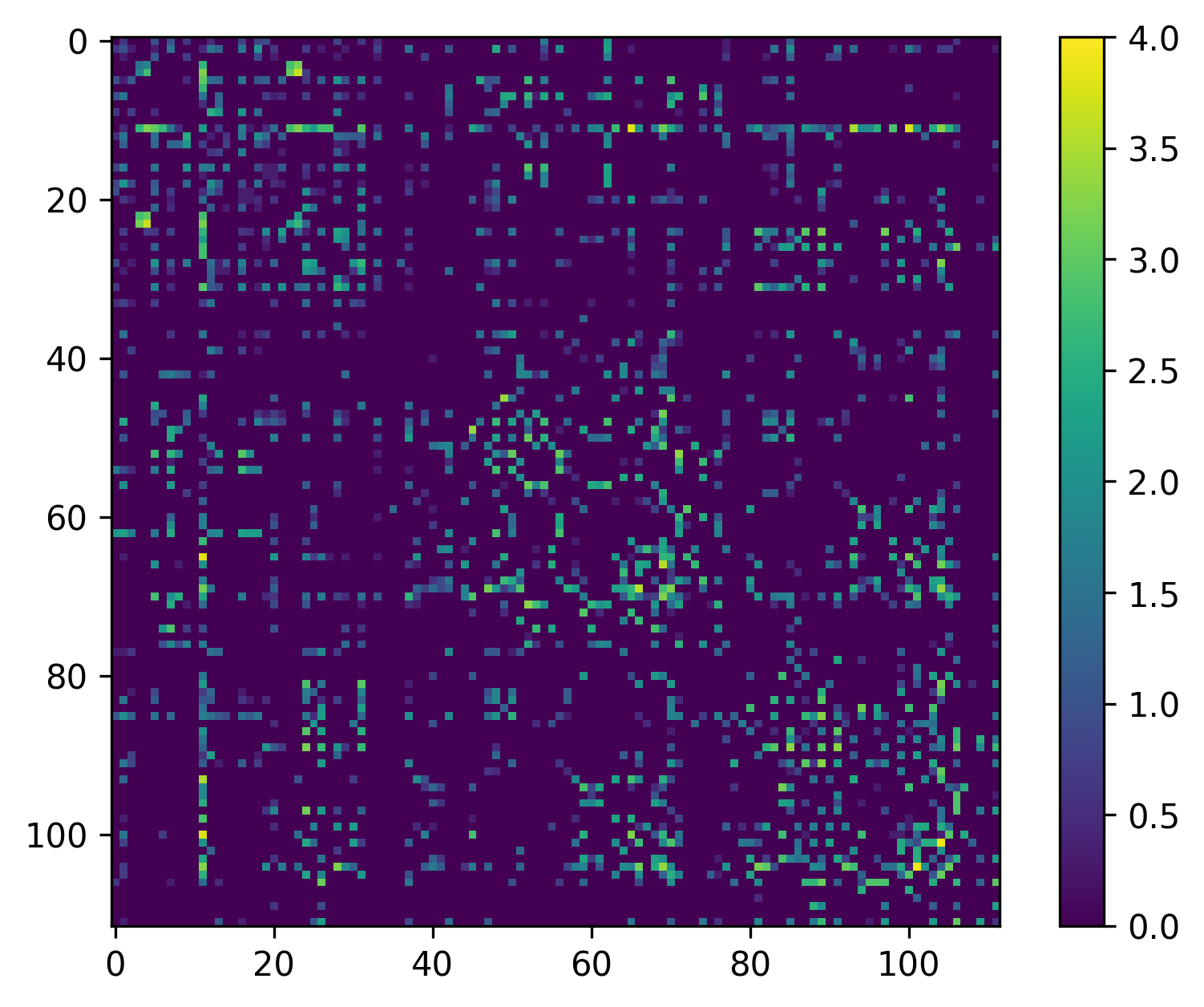} &
\includegraphics[scale=0.95, trim=0in 0in 0.75in 0in, clip=true, width=0.2275\linewidth, keepaspectratio=true]{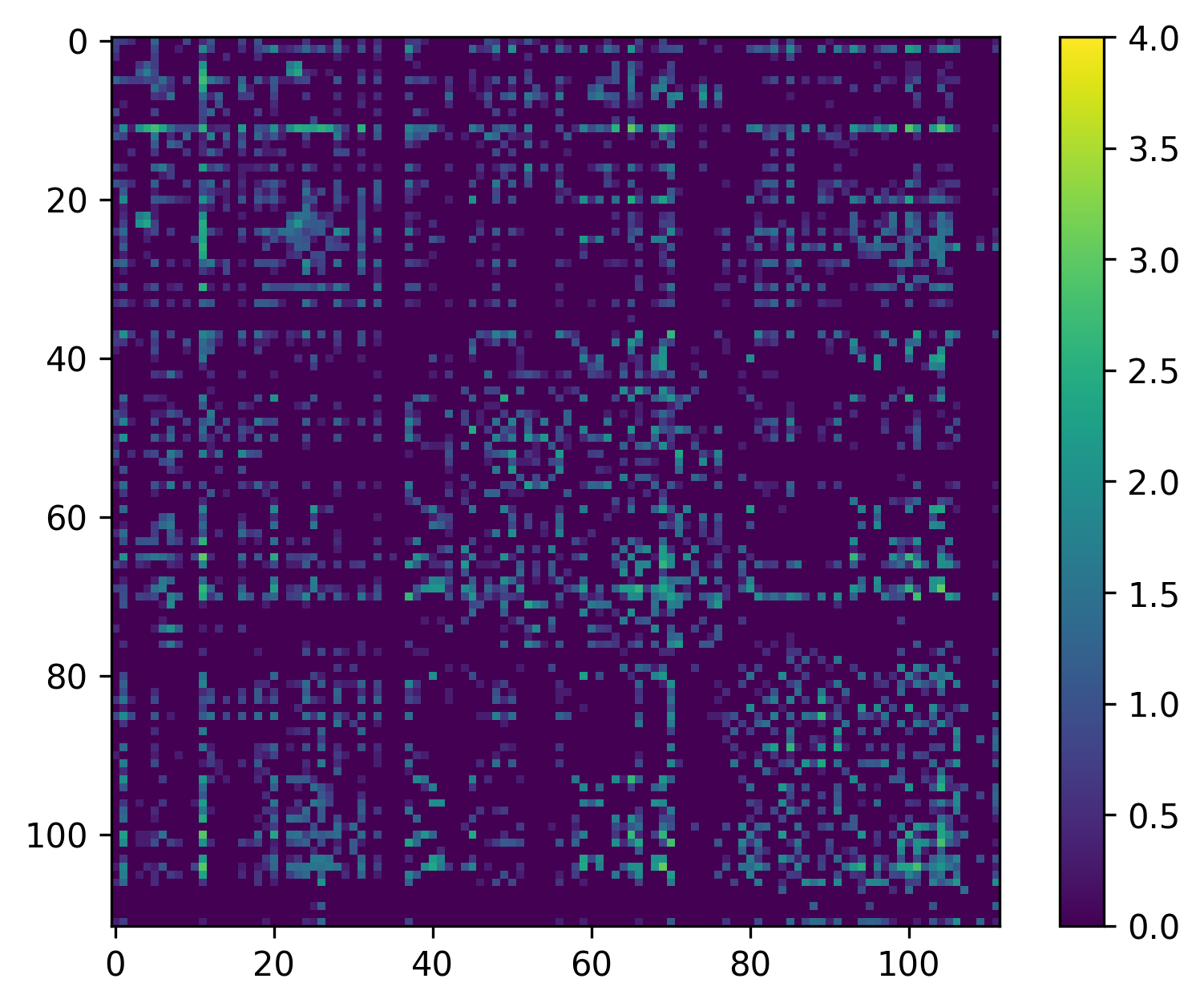} &
\includegraphics[scale=0.95, trim=0in 0in 0.75in 0in, clip=true, width=0.2275\linewidth, keepaspectratio=true]{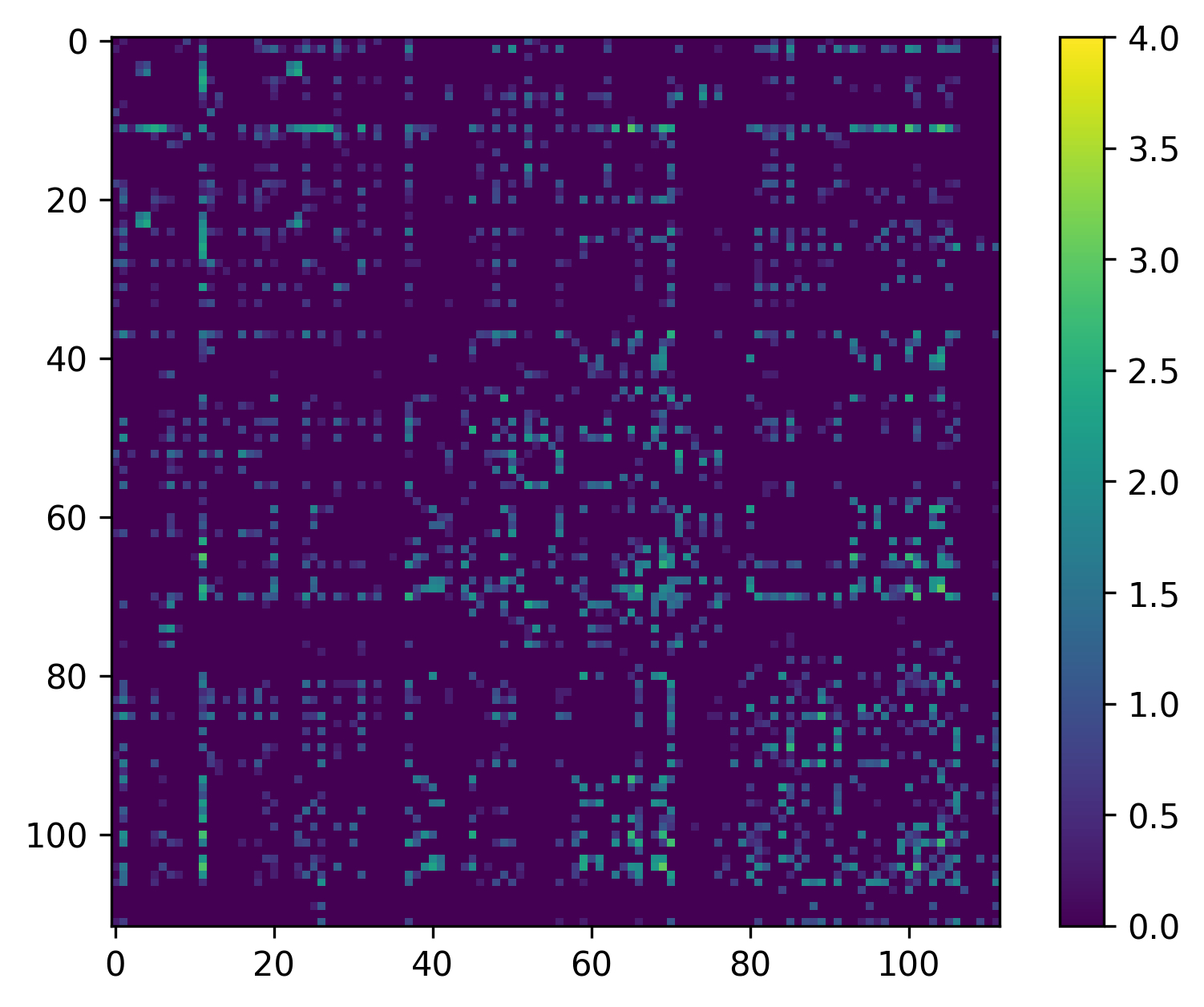}  &
\includegraphics[scale=0.95, trim=0in 0in 0.75in 0in, clip=true, width=0.2275\linewidth, keepaspectratio=true]{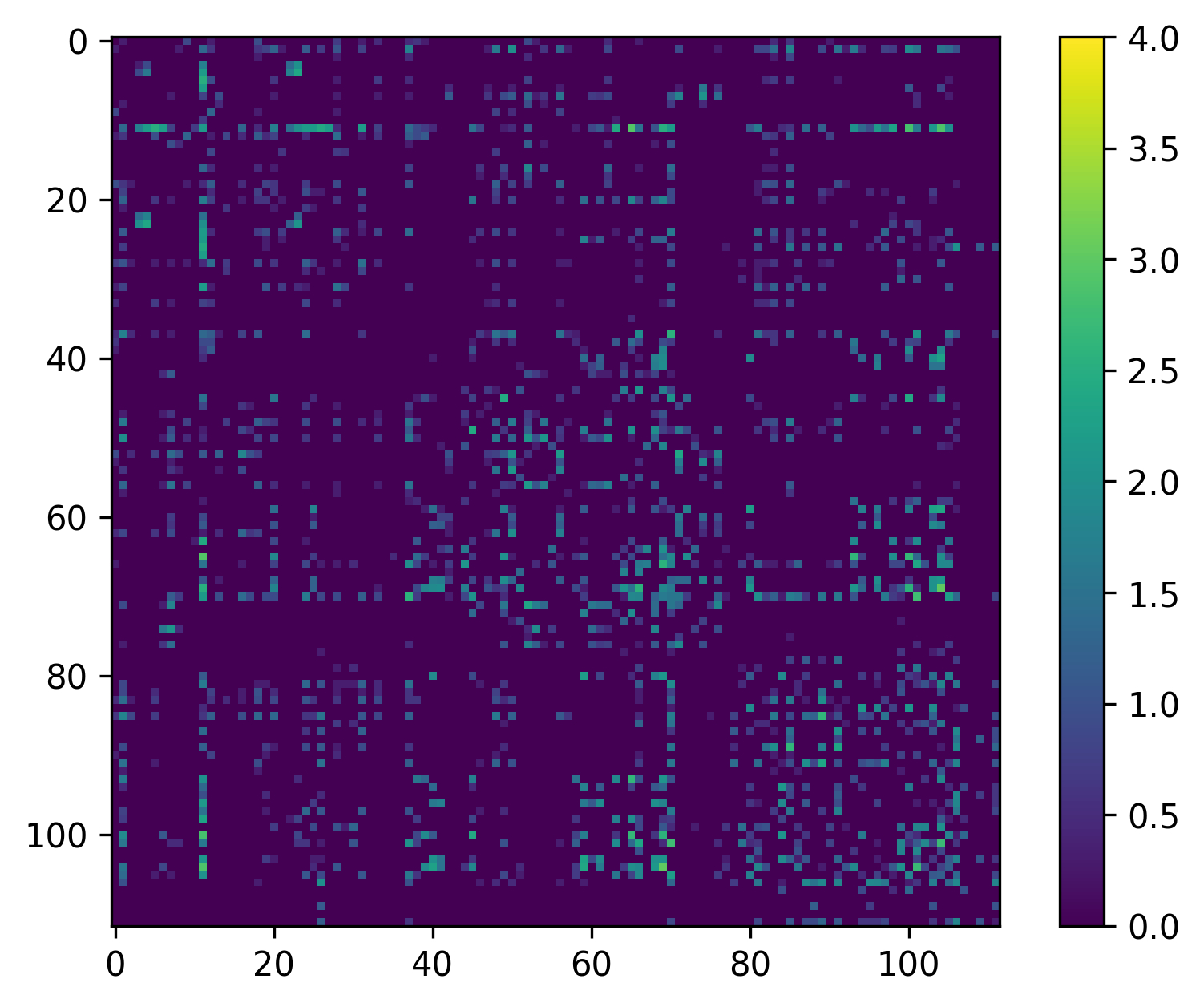} &
\includegraphics[scale=0.95, clip=true, width=0.0445\linewidth, keepaspectratio=true]{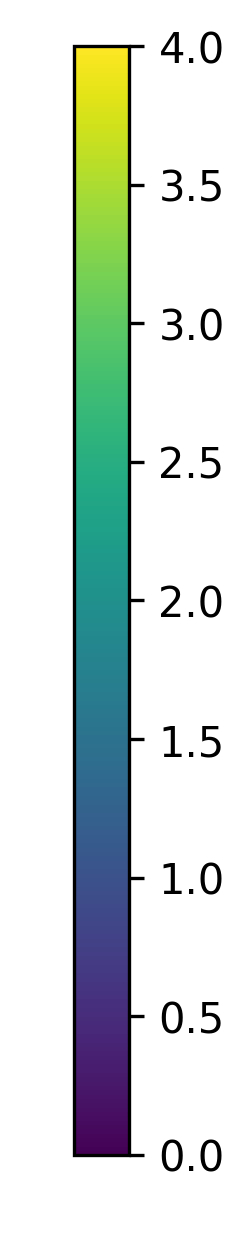} \\
\textbf{(a)} & \textbf{(b)} & \textbf{(c)} & \textbf{(d)} & \\
\includegraphics[scale=0.95, trim=0in 0in 0.75in 0in, clip=true, width=0.2275\linewidth, keepaspectratio=true]{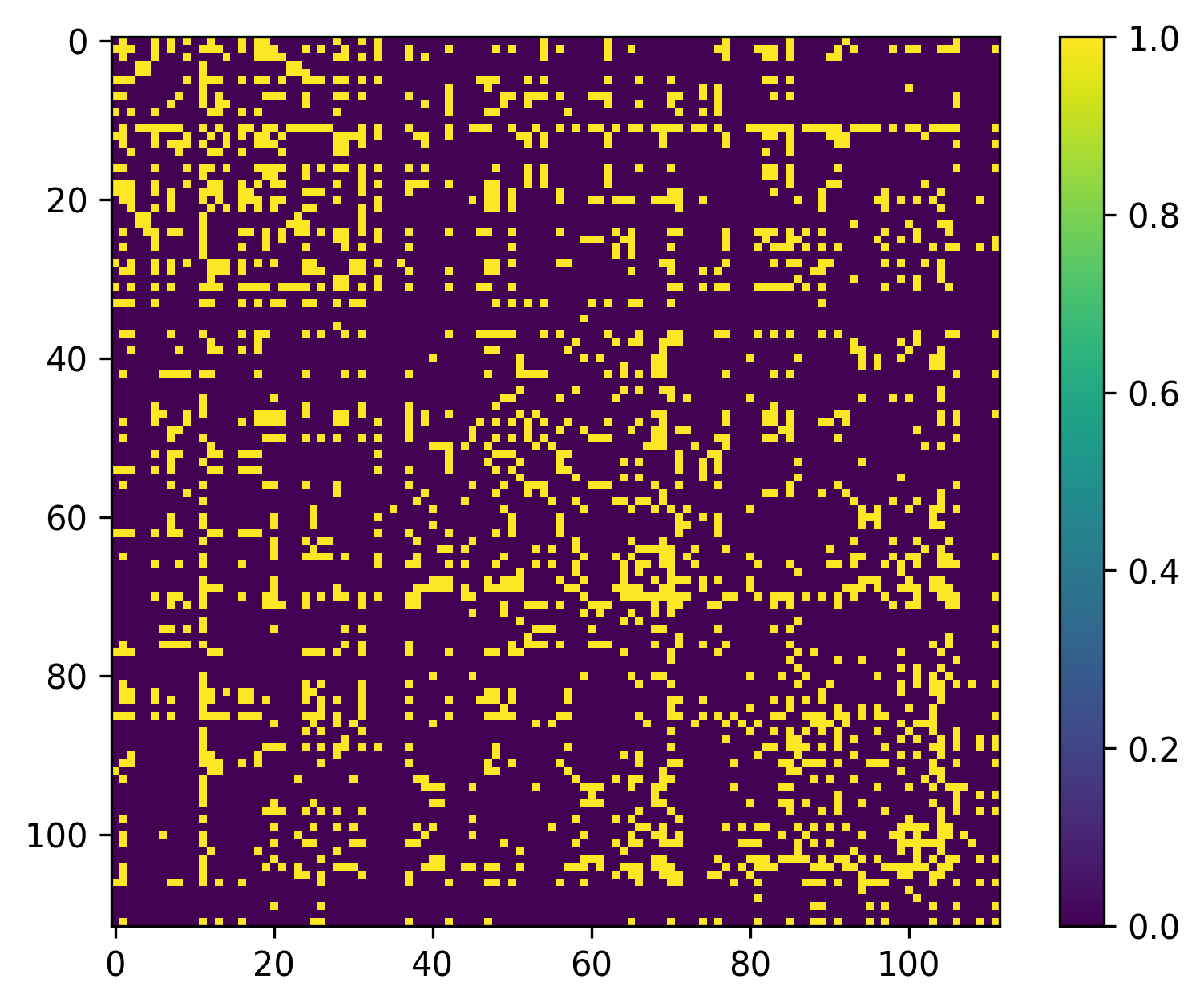} &
\includegraphics[scale=0.95, trim=0in 0in 0.75in 0in, clip=true, width=0.2275\linewidth, keepaspectratio=true]{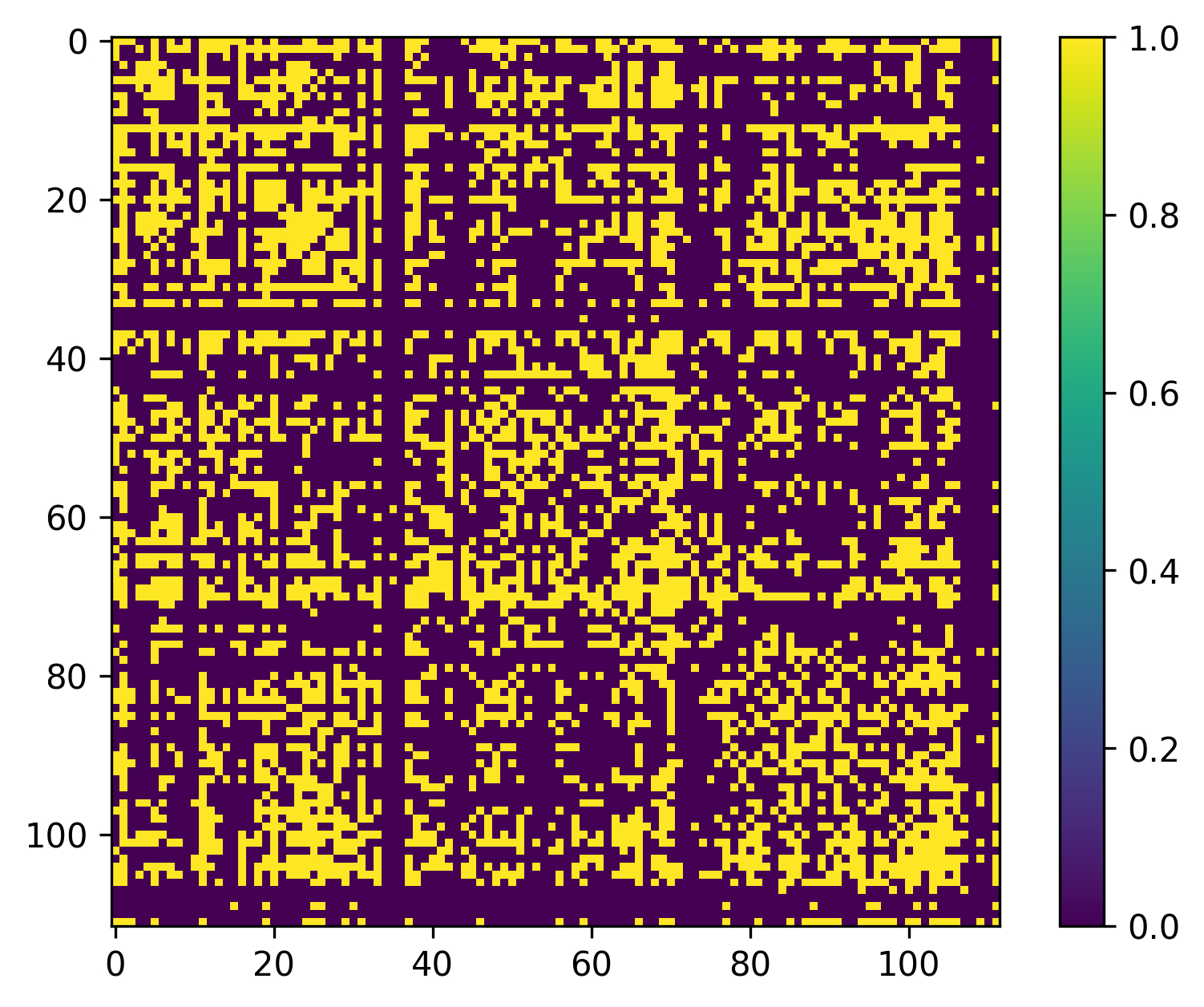} &
\includegraphics[scale=0.95, trim=0in 0in 0.75in 0in, clip=true, width=0.2275\linewidth, keepaspectratio=true]{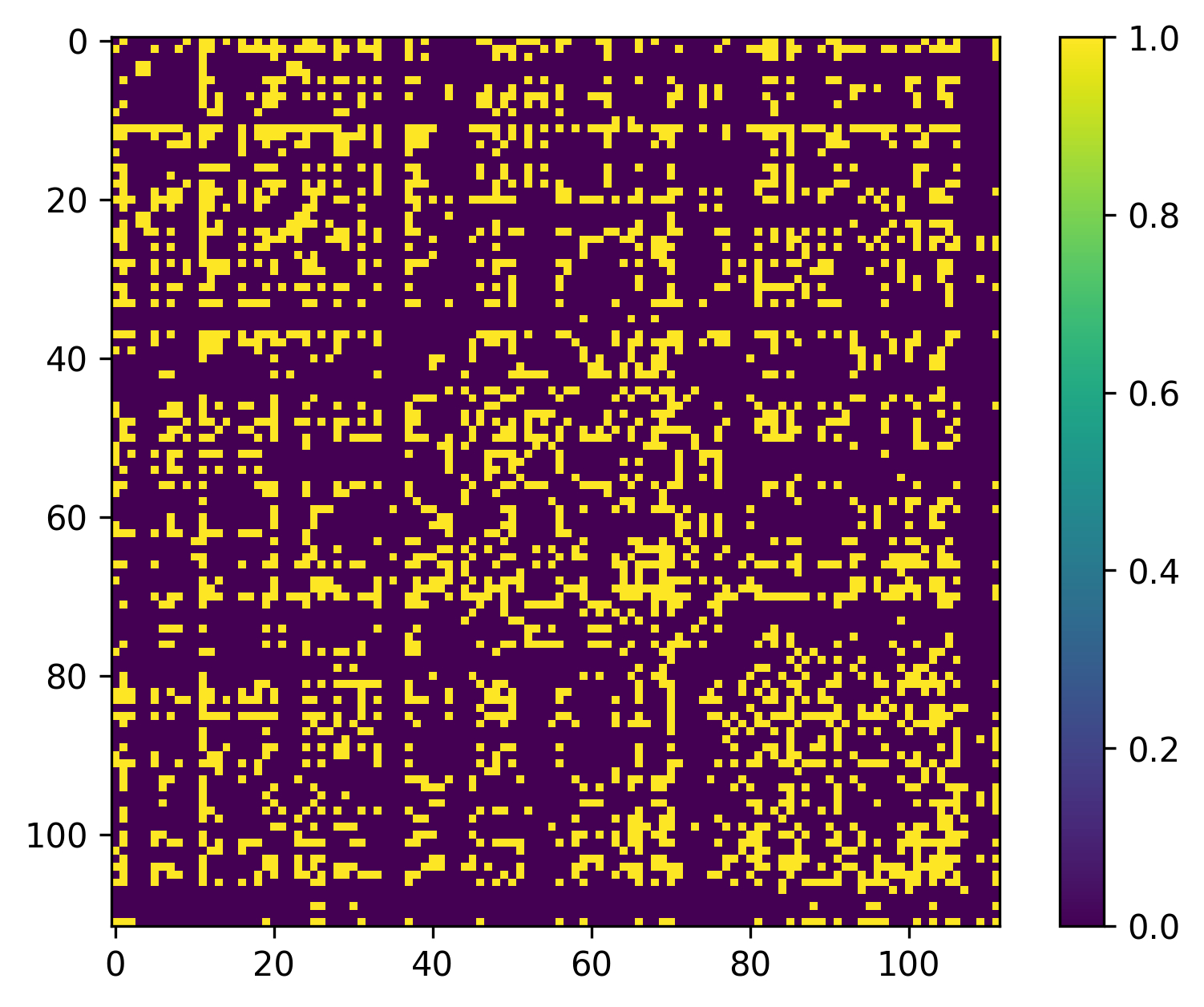} &
\includegraphics[scale=0.95, trim=0in 0in 0.75in 0in, clip=true, width=0.2275\linewidth, keepaspectratio=true]{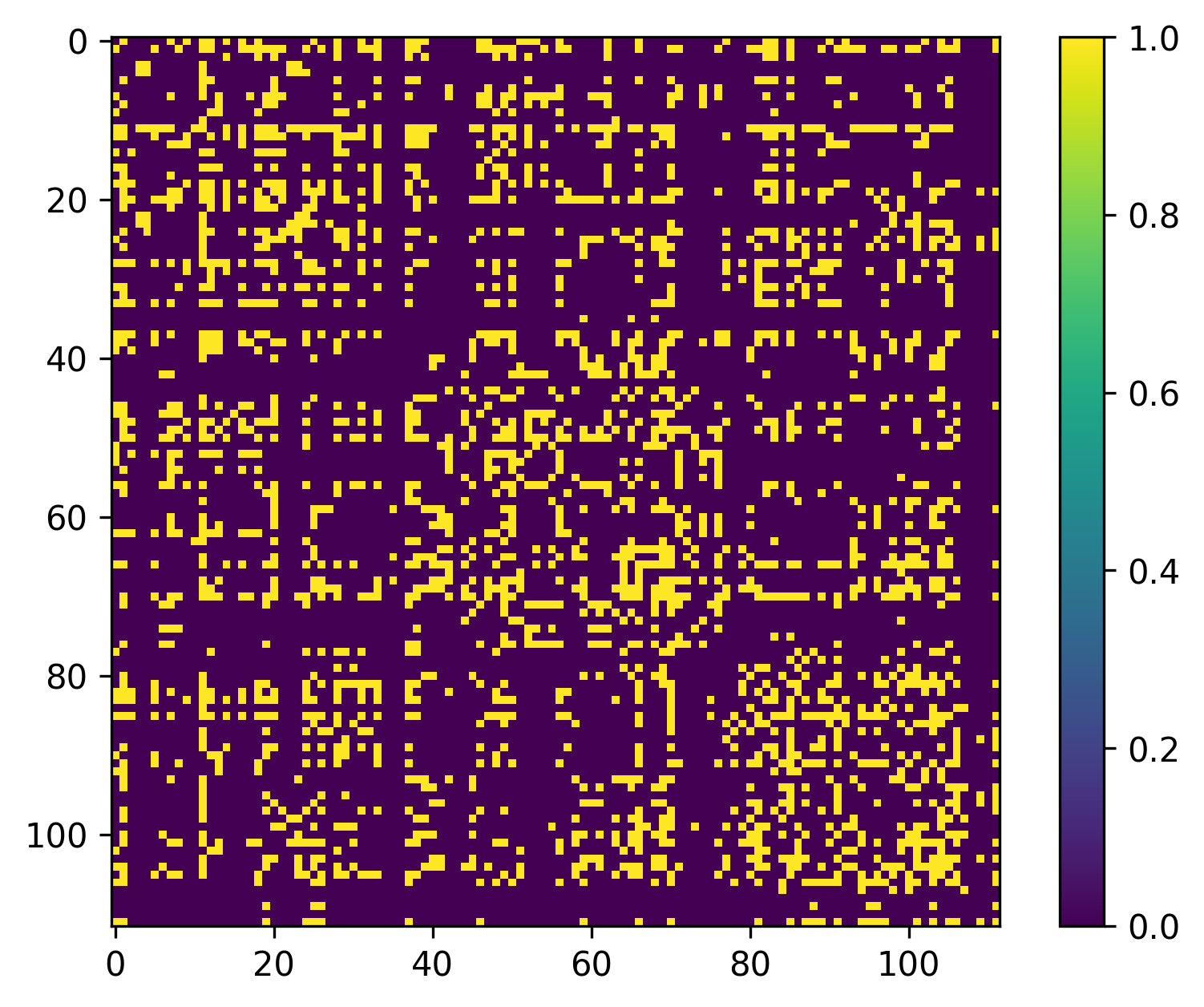} & \\
\textbf{(e)} & \textbf{(f)} & \textbf{(g)} & \textbf{(h)} & \\
\end{tabular}
\caption{\label{fig:ismrm_2015_tractography_challenge_dataset_connectivity_matrices}Streamline count connectivity matrices for the ISMRM 2015 Tractography Challenge dataset (local probabilistic tracking). Upper row: log-scale; lower row: binarized. (a) and (e): ground truth; (b) and (f): unfiltered tractogram; (c) and (g): \textit{Tractometer\_adv}-scored tractogram; (d) and (h): FINTA-filtered tractogram.}
\end{figure*}

The connectivity analysis was performed on the local probabilistic tracking data (see section \ref{sec:experiment_design}) using the $112$ endpoint regions corresponding to those originally used in the Challenge data preparation. Figure \ref{fig:ismrm_2015_tractography_challenge_dataset_connectivity_matrices} shows the log-scale and binarized streamline count connectivity matrices. A qualitative assessment evidences FINTA's ability to coherently dismiss the undesired associations from the connectome of a raw, unfiltered tractogram.

The values in table \ref {tab:ismrm_2015_tractography_challenge_dataset_connectome_density} evidence that the connectivity matrix density corresponding to the tractogram filtered with FINTA matches closely the one obtained for the \textit{Tractometer\_adv}-scored data, being both slightly above the ground truth value.

\begin{table}[!htb]
\caption{\label{tab:ismrm_2015_tractography_challenge_dataset_connectome_density}Binarized streamline count connectivity matrix density for the local probabilistic tracking ISMRM 2015 Tractography Challenge dataset. Note that the data corresponds to the test data (except for the ground truth).}
\centering
\begin{tabular}{ccccc}
\textbf{Data} & \textbf{Ground truth} & \textbf{Unfiltered} & \textbf{Tractometer\_adv} & \textbf{FINTA} \\
\hline
\textbf{Density} & 0.17 & 0.34 & 0.21 & 0.2
\end{tabular}
\end{table}

As it is the case with the ``Fiber Cup'' dataset, table \ref{tab:ismrm_2015_tractography_challenge_dataset_connectome_distance} shows that FINTA exhibits a correlation similar to the \textit{Tractometer\_adv}-scored data with respect to the ground truth on the realistic whole-brain synthetic dataset considered. Overall these results reveal that FINTA successfully filtered the tractogram without introducing significant changes to the structural connectivity.

\begin{table}[!htb]
\caption{\label{tab:ismrm_2015_tractography_challenge_dataset_connectome_distance}Distance measures on the binarized connectivity matrices for the ISMRM 2015 Tractography Challenge dataset.}
\centering
\begin{tabular}{cccc}
& \textbf{Chi-square} & \textbf{Total variation} & \textbf{Correlation} \\
\hline
\textbf{Ground truth vs. Unfiltered} & 1353.5 & 0.22 & 0.49 \\
\textbf{Ground truth vs. Tractometer\_adv} & 744.5 & 0.12 & 0.62 \\
\textbf{Ground truth vs. FINTA} & 726.0 & 0.12 & 0.61 \\
\textbf{Tractometer\_adv vs. FINTA} & 242.5 & 0.04 & 0.88
\end{tabular}
\end{table}

\end{document}